\documentclass{jfm}

\usepackage[]{amsmath}
\usepackage[dvipsnames]{xcolor}
\usepackage[]{subfigure}
\usepackage{upgreek}
\usepackage{overpic}
\usepackage[]{multirow}
\usepackage[]{url}
\usepackage[]{nicefrac}
\usepackage{bm}
\usepackage{enumerate}
\usepackage[utf8]{inputenc}

\begin{document}

\newtheorem{lemma}{Lemma}
\newtheorem{corollary}{Corollary}

\shorttitle{Low-order model for successive bifurcations} %for header on odd pages
\shortauthor{N.~Deng, B.~R.~Noack and M.~Morzy\'nski, L.~R.~Pastur} %for header on even pages

\title{Low-order model for successive bifurcations of the fluidic pinball}

\author{Nan Deng\aff{1,2}\corresp{\email{nan.deng@ensta-paristech.fr}}, 
	Bernd R.~Noack\aff{2,3,4},
        Marek Morzy\'nski\aff{5}
        \and Luc~R.~Pastur\aff{1} 
}
\affiliation
{
\aff{1}
Institute of Mechanical Sciences and Industrial Applications, 
ENSTA-ParisTech, 
828 Bd des Mar\'echaux, 
F-91120 Palaiseau, France. 
\aff{2}
LIMSI,
CNRS, Universit\'e Paris-Saclay, B{\^a}t 507, rue du Belv\'ed\`ere, Campus Universitaire,
F-91403 Orsay, France
\aff{3}
Institute for Turbulence-Noise-Vibration Interaction and Control,
Harbin Institute of Technology,
Shenzhen Graduate School, 
University Town, Xili, Shenzhen 518058, 
People's Republic of China
\aff{4}
Institut f\"ur Str\"omungsmechanik und Technische Akustik (ISTA),
Technische Universit\"at Berlin,
M\"uller-Breslau-Stra{\ss}e 8,
D-10623 Berlin, Germany
\aff{5}
Chair of Virtual Engineering, 
Pozna\'n University of Technology,
Jana Pawla II 24, PL 60-965 Pozna\'n, Poland
}

\maketitle

%***********************************************************************
\begin{abstract} 
We propose the first least-order Galerkin model 
of an incompressible flow undergoing 
two successive supercritical bifurcations
of Hopf and pitchfork type.
A key enabler is a mean-field consideration 
exploiting the symmetry of the mean flow and the asymmetry of the fluctuation.
These symmetries generalize mean-field theory,
e.g.\ no assumption of slow growth-rate is needed.
The resulting 5-dimensional Galerkin model successfully describes
the phenomenogram of the fluidic pinball, 
a two-dimensional wake flow 
around a cluster of three equidistantly spaced cylinders. 
The corresponding transition scenario
is shown to undergo two successive supercritical bifurcations, 
namely a Hopf and a pitchfork bifurcations on the way to chaos. 
The generalized mean-field Galerkin methodology 
may be employed to describe other transition scenarios.
\end{abstract}

\section{Introduction}
\label{Sec:Introduction}

This study advances mean-field modeling
for successive symmetry breaking due to Hopf and pitchfork bifurcations.
The theoretical framework is applied to the transition 
of the flow around a cluster of circular cylinders, 
termed the \emph{fluidic pinball} 
for the possibility to control fluid particle by cylinder rotation
\citep{Noack2017put,Ishar2019jfm}.

Mean-field theory was pioneered by \citet{Landau1944} and \citet{Stuart1958jfm}
and is a singular triumph of nonlinear reduced-order modeling in fluid mechanics.
Already the most simple mean-field model for the supercritical Hopf bifurcation
reveals deep insights into the coupling between the fluctuations and the mean flow, 
e.g. the damping mechanism of unstable modes by Reynolds stress. 
In addition, 
\citet{Malkus1956jfm} principle of marginal stability for time-averaged flows,
the square root growth law of fluctuation level with increasing Reynolds number,
the cubic damping term from a linear-quadratic dynamics,
the energetic explanation  of this amplitude dynamics,
and the slaving principle leading to manifolds driven by ensemble-averaged Reynolds stress
are easily derived.
Also, the idea of center manifold theory 
and the surprising success of linear parameter-varying models
are analytically illustrated.
Historically, Landau was the first to derive the normal form of the dynamics
with Krylov-Bogoliubov approximation \citep[an averaging method for spiral phase paths, see  \textit{e.g.}][]{jordan1999nonlinear}
while Stuart could explain how the cubic damping term arises from the distorted mean flow.

Mean-field models for a supercritical Hopf bifurcation with an unstable oscillatory eigenmode
have been applied and validated for numerous configurations.
The onset of vortex shedding behind a cylinder wake has been thoroughly investigated \citep{Strykowski1990jfm,Schumm1994jfm,Noack2003jfm}. Even high-Reynolds number turbulent wake flow can display a distinct mean-field manifold  and modeled by a noise-driven mean-field model \citep{Bourgeois2013jfm}.

A supercritical pitchfork bifurcation similarly arises by an unstable eigenmode with a real eigenvalue. 
The onset of convection rolls in the Rayleigh-B\'enard problem 
is a famous example \citep{Zaitsev1971,Swift1977pra,Cross1993rpm}.
The features of a pitchfork bifurcation are observed for the sidewise symmetry breaking
of the time-averaged Ahmed body wake \citep{grandemange_PRE2012,Grandemange2013jfm,cadot_PRE2015,bonnavion_JFM2018} and more generally in three-dimensional wake flows \citep{mittal_AIAA1999,gumowski_PRE2008,szaltys_JFS2012,Grandemange_EXIF2014,rigas_JFM2014}. 
In contrast, the drag crisis of circular cylinder is associated with a subcritical bifurcation
into two asymmetric sheddings with opposite mean lift values \citep{Schewe1983jfm}.

Not surprisingly, numerous generalizations of mean-field models have been proposed.
\citet{Landau1944} and \citet{Hopf1948} 
have conjectured that high-dimensional fully developed turbulence
may be explained by an increasingly rapid succession of Hopf bifurcations.
This idea has been discarded as unlikely \citep[see, for instance,][]{Landau1987book}. 
The seminal paper by \citet{Ruelle1971nature} 
showed that turbulence does not arise as a successive superposition of oscillators, 
but irregular chaotic behavior can already appear after few bifurcations. 
A second direction is the explanation of nonlinear coupling
between two incommensurable shedding frequencies \citep{Luchtenburg2009jfm},
also referred to as \emph{frequency crosstalk} in the following.
This amplitude coupling over the mean flow has been termed \emph{quasi-laminar}
in \citet{Reynolds1972jfm} pioneering theoretical foundation of the triple decomposition. 
The advancements also include subcritical bifurcations \citep{Watson1960jfm}. 
More specifically, the case of a codimension two bifurcation, 
involving both a pitchfork and a Hopf bifurcation, was addressed in  \citet{Meliga2009jfm}, 
who derived the amplitude equation based on the weakly nonlinear analysis of the wake of a disk. 
\citet{Fabre2008pof} derived the same equation solely based on symmetry arguments for the wake of axisymmetric bodies.
A resolvent analysis follows mean-field considerations in decomposing the flow in a time-resolved linear dynamics
and a feedback-term with the quadratic nonlinearity \citep{Gomez2016jfm,Rigas2017aiaa}. 

Our study develops a generalized mean-field Galerkin model 
for the first two bifurcations of the fluidic pinball with increasing Reynolds number.
The primary supercritical bifurcation leads to the periodic vortex shedding which is statistically symmetric.
At higher Reynolds numbers, the resulting limit cycle undergoes a pitchfork bifurcation 
into a stable, asymmetric, mirror-symmetric pair of periodic solutions. This local bifurcation has a transverse effect resulting from the decoupling of these two bifurcations (see appendix~\ref{Sec:ODE}), which simultaneously leads to an identical local pitchfork bifurcation of the steady solution, into an unstable, asymmetric, mirror-symmetric pair of steady solutions. 
The underlying dynamics is modeled with a small number of assumptions.
The key simplification results from exploiting the symmetry
of the mean flow and the antisymmetry of the fluctuation. The generalized mean-field Galerkin methodology can be expected to be useful for describing other transition scenarios.

The manuscript is organized as follows.
In \S~\ref{Sec:FlowConfig}, the numerical plant is introduced
and the Reynolds-number dependent flow behavior described.
This phenomenology drives the mean-field modeling of the first two bifurcations in \S~\ref{Sec:Methodology}.
The resulting models for the Hopf 
and subsequent pitchfork bifurcation
are present in \S~\ref{Sec:Primary} and \S~\ref{Sec:Secondary}, respectively.
\S~\ref{Sec:Conclusions} summarizes the results
and outline future directions of research.

\section{Flow configuration}
\label{Sec:FlowConfig}

In this section we describe the numerical toolkit and the flow features as the Reynolds number is increased. The direct Navier-Stokes solver with MATLAB interfaces, used for the simulation, is described in \S~\ref{Sec:DNS}. The fluidic pinball configuration and the flow features and route to chaos are described in  \S~\ref{Sec:Configuration}  and  \S~\ref{Sec:Features}, respectively.

\subsection{Direct Navier-Stokes solver}
\label{Sec:DNS}

\begin{figure}
\begin{centering}
\begin{tabular}{cc}
(a) & (b) \\
\includegraphics[width=0.48\textwidth]{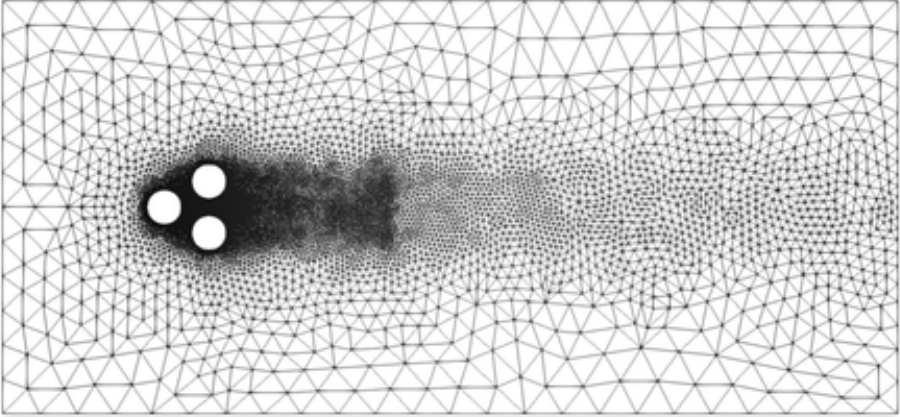}  &
\includegraphics[width=0.48\textwidth]{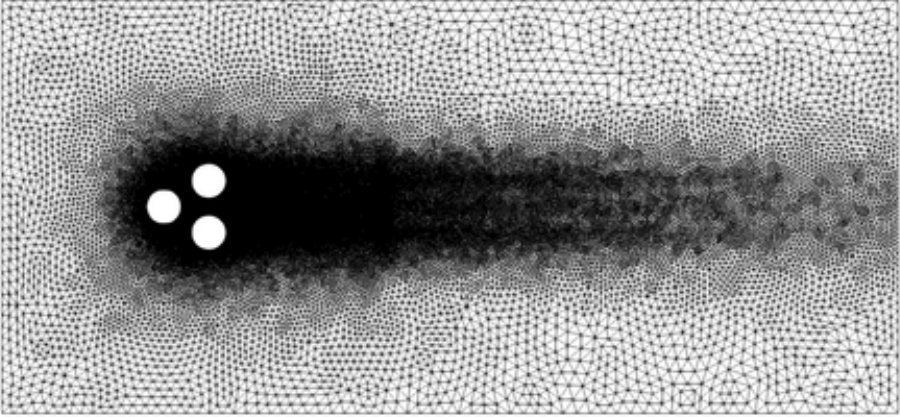}  \\
\end{tabular}
\end{centering}
\caption{Computational grid for the fluidic pinball, with 8633 (a) and 54195 vertices (b).
}
\label{fig:Pinball_Grid}
\end{figure}

\begin{figure}
\begin{centering}
\begin{tabular}{cc}
(a) & (b) \\
\includegraphics[width=0.48\textwidth]{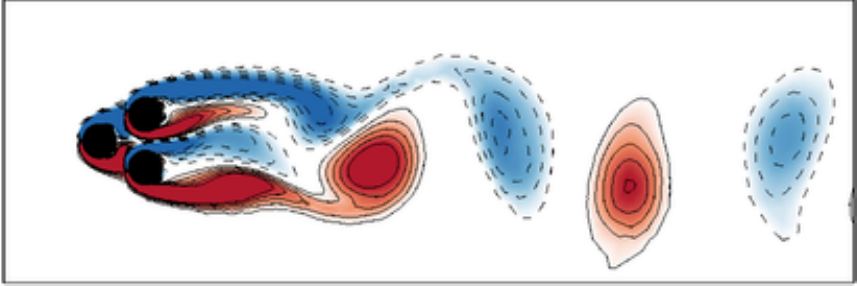}  &
\includegraphics[width=0.48\textwidth]{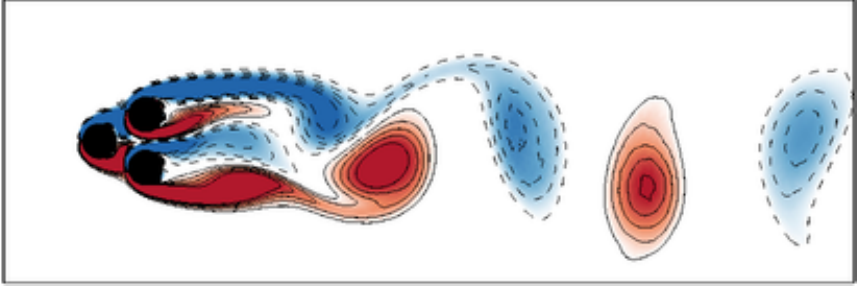}  \\
\end{tabular}
\end{centering}
\caption{DNS computations on a grid with 8\,633 nodes (a) and 54\,195 nodes (b). Vorticity depicted with color is $[-1.5, 1.5]$, $t=200$.
The initial kick is provided by a rotation of all three cylinders at $t=0.2$, see text for details. 
}
\label{fig:Comparison}
\end{figure}

The unsteady Navier-Stokes solver is based on fully implicit time integration and Finite-Element Method discretization~\citep{Noack2017put,Noack2003jfm,Noack2016jfm}. The time integration is third-order accurate while FEM discretization employs a second--order Taylor-Hood finite elements~\citep{Taylor1973ef}. 
The solution is obtained iteratively, with the Newton-Raphson type approach. The tangent matrix is updated on each iteration and computations are carried out until the residual is under a prescribed tolerance.
The steady solution is obtained in a similar Newton-Raphson iteration for the steady Navier-Stokes equations. The convergence to one of the three steady solutions with different states of the base-bleeding jet is triggered by appropriate ``initial'' conditions in the iteration, see appendix~\ref{Sec:AsymSS}. The solver quickly converges to one of the steady states and a final, near-zero residual confirms that this is indeed the steady flow solution sought.
The computational domain is discretized on an unstructured grid. Pinball configuration uses a grid with 4225 triangles and 8633 vertices (see figure~\ref{fig:Pinball_Grid}(a)). To test the grid dependency of the solution we increased the number of triangles by nearly a factor 4 (26849 elements and 54195 nodes, see figure~\ref{fig:Pinball_Grid}(b)). 
The flow patterns shown in figure~\ref{fig:Comparison} develop from a steady solution at $Re=100$ subjected to an instantaneous rotation of cylinders at $T=0.2$. The upper cylinder rotates counterclockwise, 
the lower one clockwise and the center cylinder also in a clockwise direction --- all with unit circumferential velocity, i.e.~the velocity of oncoming flow $U_\infty$. 
This configuration and boundary conditions result in a vortex shedding shown in  figure~\ref{fig:Comparison} for the time instance $t=200$. Both simulations prove grid independence and yield dynamically consistent results (see figure~\ref{fig:Comparison}).

\subsection{Pinball configuration}
\label{Sec:Configuration}

We refer to the configuration shown in figure~\ref{fig:Pinball_Grid} as the \emph{fluidic pinball} 
as the rotation speeds allow one to change the paths 
of the incoming fluid just as flippers manipulate the ball of a conventional pinball machine.
The fluidic pinball is a set of three equal circular cylinders with radius $R$  
placed parallel to each other in a viscous incompressible uniform flow at speed $U_\infty$. The flow over a cluster of three parallel cylinders has been experimentally studied involving heat transfer,  fluid-structure interactions and multiple frequencies interactions over the past few decades~\citep{price1984aerodynamic,sayers1987flow, lam1988phenomena, tatsuno1998effects, bansal2017experimental}. For the fluidic pinball, the cylinders can rotate at different speeds creating a kaleidoscope of vortical structures or variety of steady flow solutions. 
The configuration is used for evaluation of flow controllers~\citep{Cornejo2017limsi} 
as this problem is  a challenging task for control methods comprising several frequency crosstalk mechanisms~\citep{Noack2017put}. 
The centers of the cylinders form an equilateral triangle with side length $3R$, 
symmetrically positioned with respect to the flow. 
The leftmost triangle vertex points upstream, while the rightmost side is orthogonal to the oncoming flow. 
The origin of the Cartesian coordinate system is placed in the middle of the top and bottom cylinder.
The fluidic pinball computational domain, shown in figure~\ref{fig:Pinball_Grid} is bounded 
by the rectangle $[-6,20] \times [-6,6]$.

Without forcing, the boundary conditions comprise a no-slip condition 
on the cylinders and a unit velocity in the far field:
\begin{equation}
\label{Eqn:BoundaryCondition}
U_r = 0 \hbox{\ on the cylinders and }
U_\infty = \bm{e}_x \hbox{\ at infinity.}
\end{equation}
The far field boundary conditions are exerted on the inflow, upper and lower boundaries while the outflow boundary is assumed to be a stress--free one, transparent for the outgoing fluid structures. A typical initial condition is the unstable steady Navier-Stokes solution $\bm{u}_s(\bm{x})$.

In this study, all three cylinders remain static as we are interested in the natural dynamics of the flow as the Reynolds number is increased. 

\subsection{Flow features}
\label{Sec:Features}

\begin{figure}
\begin{center}
\includegraphics[width=\textwidth]{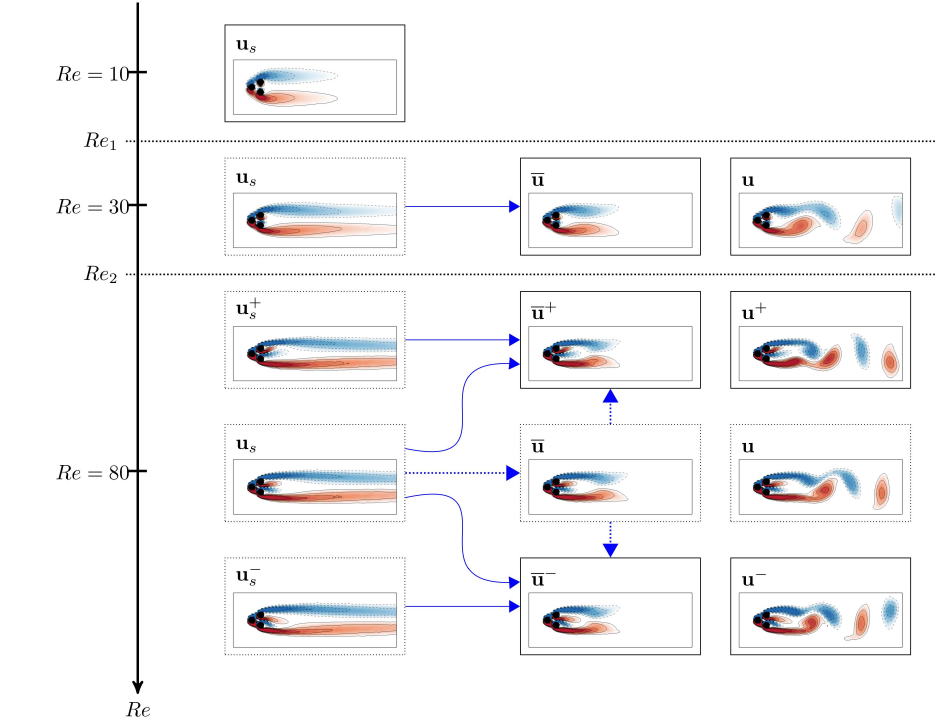}  \\
\caption{Flow states at different values of the Reynolds number: the stable states are labeled with a solid-line box and the unstable states with a dashed-line box. Steady solutions $\bm{u}_s(\bm{x})$ (on the left side): three steady solutions $\bm{u}_s(\bm{x})$ (symmetric), $\bm{u}_s^\pm(\bm{x})$ (asymmetric), exist at $Re=80$. For $Re=30$ and $Re=80$, the steady solutions are unstable: the permanent regime $\bm{u}(\bm{x},t) = \overline{\bm{u}}(\bm{x}) + \bm{u}^\prime (\bm{x},t)$ is unsteady with mean flow field $\overline{\bm{u}}(\bm{x})$. At $Re=80$, the instantaneous flow field $\bm{u}(\bm{x},t) = \overline{\bm{u}}(\bm{x}) + \bm{u}^\prime (\bm{x},t)$, transiently explored when starting close to $\bm{u}_s(\bm{x})$, is unstable with respect to either %asymptotic flow fields 
$\bm{u}^\pm(\bm{x},t) = \overline{\bm{u}}^\pm(\bm{x}) + \bm{u}^\prime (\bm{x},t)$. }
\label{Fig:FlowStates} 
\end{center}
\end{figure}

The steady solution $\bm{u}_s$, shown in figure~\ref{Fig:FlowStates} for different values of the Reynolds number $Re$, is stable up to the critical value $Re_{1}\approx 18$.
This value corresponds to $5/2\times Re_{1}\approx 45$ with respect to the actual body height $5R$, which is consistent with the critical value of the Reynolds number found for a single cylinder \citep{ding1999three,barkley2006europhys}. Beyond $Re_{1}$, the steady solution becomes unstable with respect to vortices periodically and alternately shed at the top and bottom of the two right-most cylinders, following a Hopf bifurcation (instability of the fixed point \textit{via} a pair of complex-conjugated eigenvalues, see e.g. \citet{strogatz1994book}). In addition to the resulting von K\'arm\'an street of vortices, the gap between the cylinders makes possible the formation of a jet at the base of the two outer cylinders. The steady solution $\bm{u}_s$, the mean flow $\overline{\bm{u}}$ and the instantaneous flow field $\bm{u}$, are shown in figure~\ref{Fig:FlowStates}, for $Re=30$. 

The flow passing between the two rearward cylinders, the base-bleeding flow, has a critical impact on the successive bifurcations undergone by the system on the route to chaos. Indeed, beyond a secondary critical value $Re_{2}\approx 68$ of the Reynolds number, the system undergoes a pitchfork bifurcation, which affects both the fixed point or the limit cycle, \textit{via} a real eigenvalue, see e.g. \citet{strogatz1994book}. As a result, the symmetry of both the steady solution and the mean flow is broken with respect to the symmetry plane defined by $y=0$. This is illustrated by the two mirror-conjugated steady solutions $\bm{u}_s^\pm$ and the two associated mean flows $\overline{\bm{u}}^\pm$, shown in figure~\ref{Fig:FlowStates} for $Re=80$, where the base-bleeding jet appears deflected to either upward or downward with respect to the symmetry plane. Note, however, that a symmetry-preserving mean flow ($\overline{\bm{u}}$ in figure~\ref{Fig:FlowStates} for $Re=80$) still exists beyond the secondary bifurcation, so that three mean flows exist beyond $Re_{2}$: two of them, $\overline{\bm{u}}^\pm$, are mirror-conjugated and break the symmetry, while the last one, $\overline{\bm{u}}$ preserves the symmetry. This bifurcation of the limit cycle is coincident with the bifurcation of the fixed point, as three steady solutions can be found beyond $Re_{2}$: $\bm{u}_s^\pm$ are mirror-conjugated and break the symmetry, while the last one, $\bm{u}_s$ preserves the symmetry. Yet, all three of them are unstable with respect to the cyclic shedding of von K\'arm\'an vortices, in which symmetry properties of the steady solution are succeeded in the resulting mean flow. 
When the initial condition is close to the symmetric steady solution $\bm{u}_s$, the flow regime arrives after a long transient on a limit cycle whose mean flow $\overline{\bm{u}}$ is symmetric, as illustrated in figure~\ref{Fig:FlowStates} for $Re=80$. However, the dynamics of this limit cycle is only transient, indicating that it is not a stable state. After a new transient, depending on the details of the initial condition, the flow regime eventually reaches one of the two mirror-conjugated limit cycles (centered on either $\overline{\bm{u}}^\pm$). When the initial condition already breaks the symmetry of the flow configuration, the unstable ``symmetry-centered'' limit cycle is not explored and the system reaches directly one of the two stable limit cycles. 
The transient dynamics between these six typical states beyond $Re_2$ illustrate a transverse action on the original state space resulting from the new active symmetric breaking mode decoupling with the primary Hopf bifurcation, as detailed in appendix~\ref{Sec:TranDyn}. The new active degree of freedom introduced by the pitchfork instability is responsible for the two simultaneous local pitchfork bifurcations, of both the steady solution and the periodic solution, as shown in the appendices~\ref{Sec:LSA} and \ref{Sec:FSA}. The simultaneous bifurcation of the steady and periodic solution has also been observed for the cylinder wake transition from stability analyses \citep{Noack1994jfm,Noack1994zamm} and from 3D Navier-Stokes simulations \citep{Zhang1994mpisf}. A further discussion about this non-generic situation is recorded in appendix~\ref{Sec:ODE}. Besides, the linear stability analysis (see appendix~\ref{Sec:LSA}) and the Floquet stability analysis (see appendix~\ref{Sec:FSA}) around $Re_2$ have been performed to prove these two simultaneous bifurcations.

\begin{figure}
\begin{center}
\includegraphics[width=\textwidth]{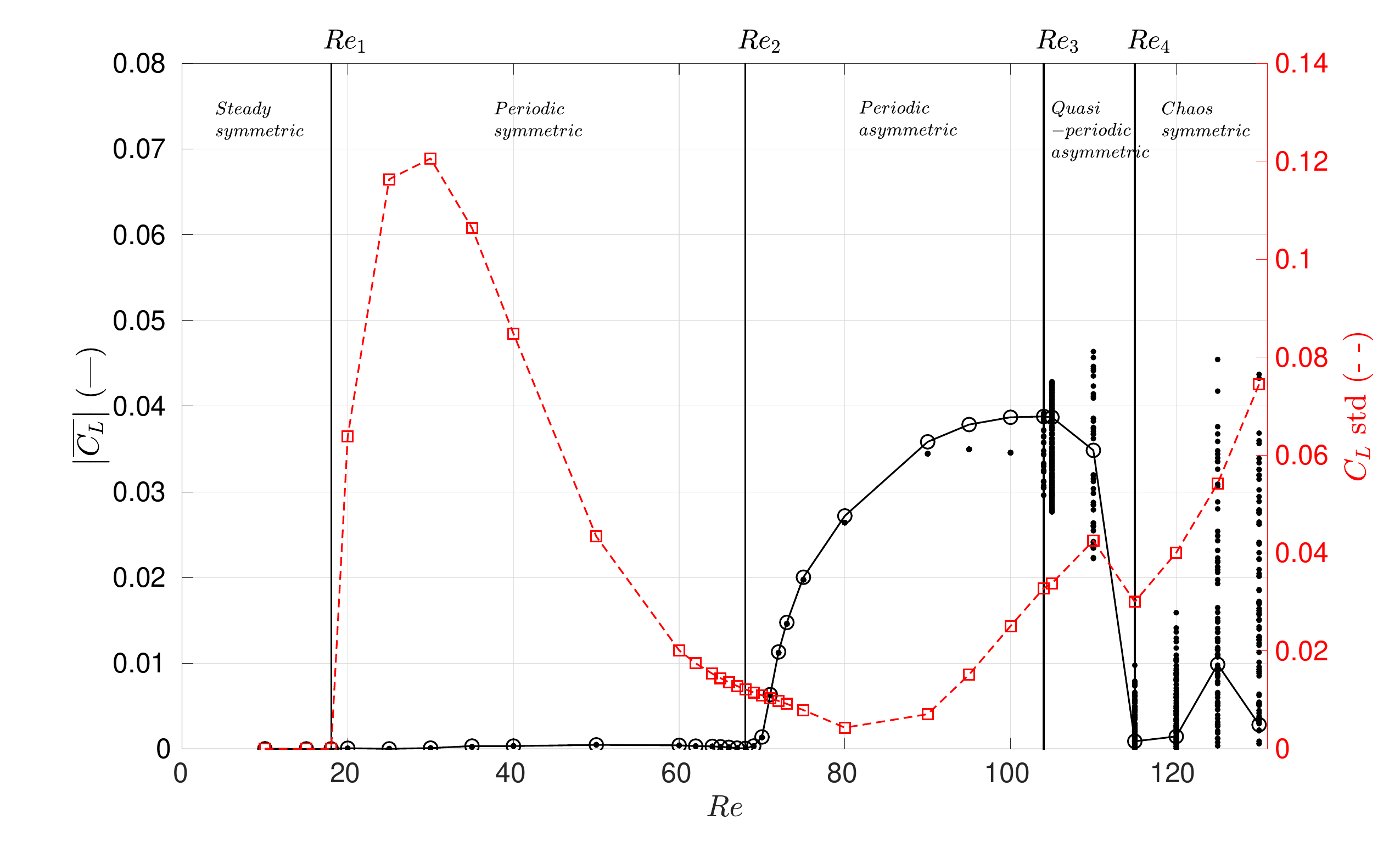}  \\
\caption{Bifurcation diagram based on the absolute value of the mean pressure lift coefficient $|\overline{C}_L|$ (circles + solid line) and its standard deviation (squares + dashed line). 
By symmetry, each non-vanishing mean lift value is associated with a positive and negative sign for the two attractors.
The vertically distributed black dots that are visible for $Re>Re_{3}$, are median values $\overline{c}_n=(C_n+C_{n+1})/2$ between successive local optima $C_n=C_L(t_n)$ of $C_L(t)$, at a given Reynolds number, where the $t_n$ are times at which $\dot{C}_L(t_n)=0$. Transition to unsteadiness occurs at $Re_{1}\approx 18$ (Hopf bifurcation), the average symmetry is broken beyond $Re_{2}\approx 68$ (pitchfork bifurcation), a secondary (incommensurable) frequency rises in the power spectrum at $Re_{3}\approx 104$ (Neimark-S\"acker bifurcation), and transition to chaos occurs at $Re_{4}\approx 115$. Note that the symmetry is statistically recovered in the chaotic regime ($\overline{C}_L\approx 0$). }
\label{Fig:CL} 
\end{center}
\end{figure}

As a result, when the symmetry of vortex shedding is broken, the mean value $\overline{C}_L$ (solid line) of the pressure lift coefficient $C_L = 2F_L/\rho U^2$, where $F_L$ is the total lift force from pressure, no longer vanishes, as shown in figure~\ref{Fig:CL}. At the precision of our investigation, both the Hopf and pitchfork bifurcations were found to be supercritical.

The fluctuation amplitude of the lift coefficient is minimum for $Re\approx 80>Re_{2}$, as shown in figure~\ref{Fig:CL} (dashed curve). It starts to decrease around $Re=30$, when the jet starts to grow at the base of the two outer cylinders. Henceforth, the growth of the base-bleeding jet, as the Reynolds number is increased, seems to be fed with the energy of the fluctuations. Transfers of energy between the dynamically dominant degrees of freedom will be made clear in \S~\ref{Sec:Methodology}.

\begin{figure}
\begin{center}
\begin{tabular}{ccc}
 (a) & (b) & (c) \\
 \includegraphics[width=0.32\textwidth]{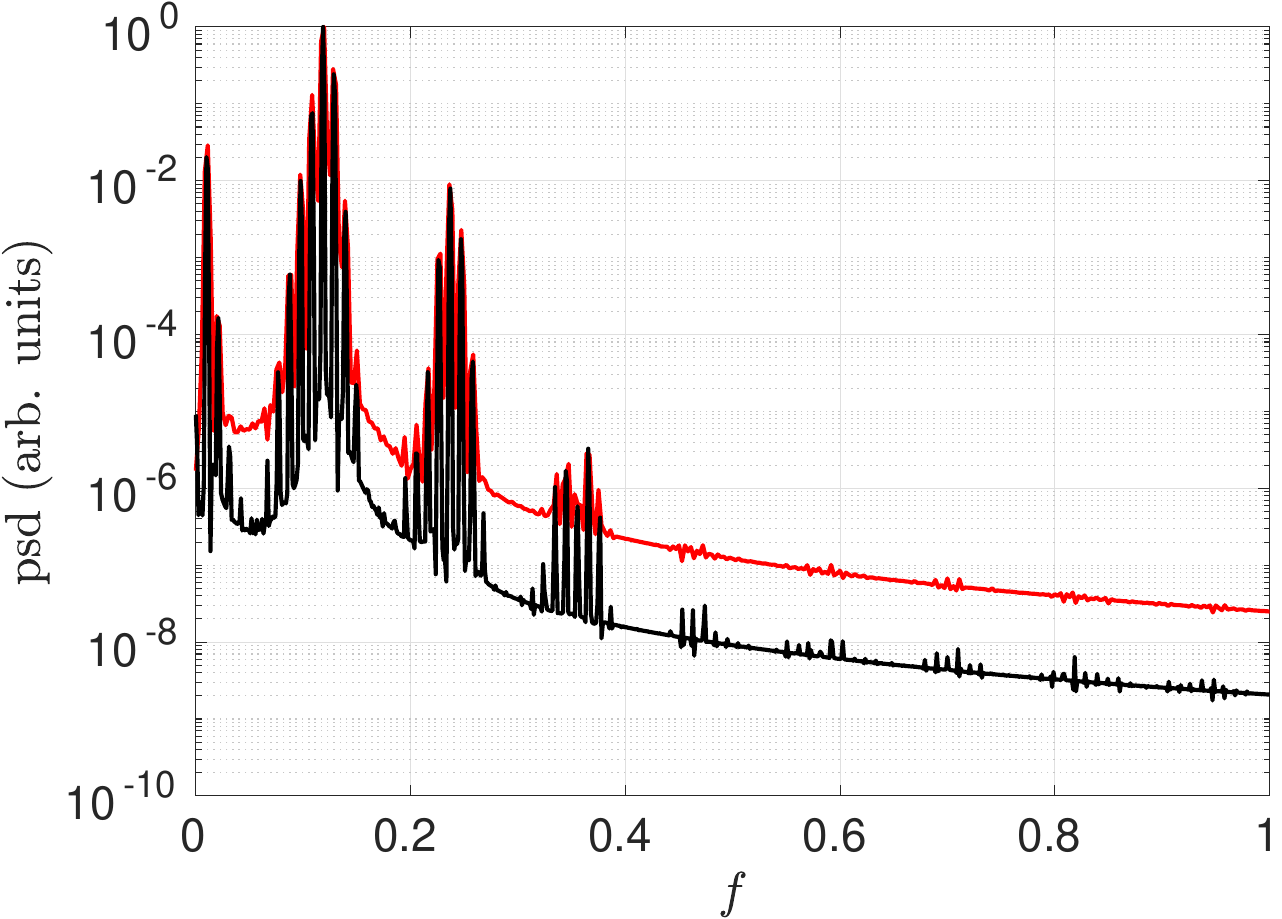}   &
\includegraphics[width=0.32\textwidth]{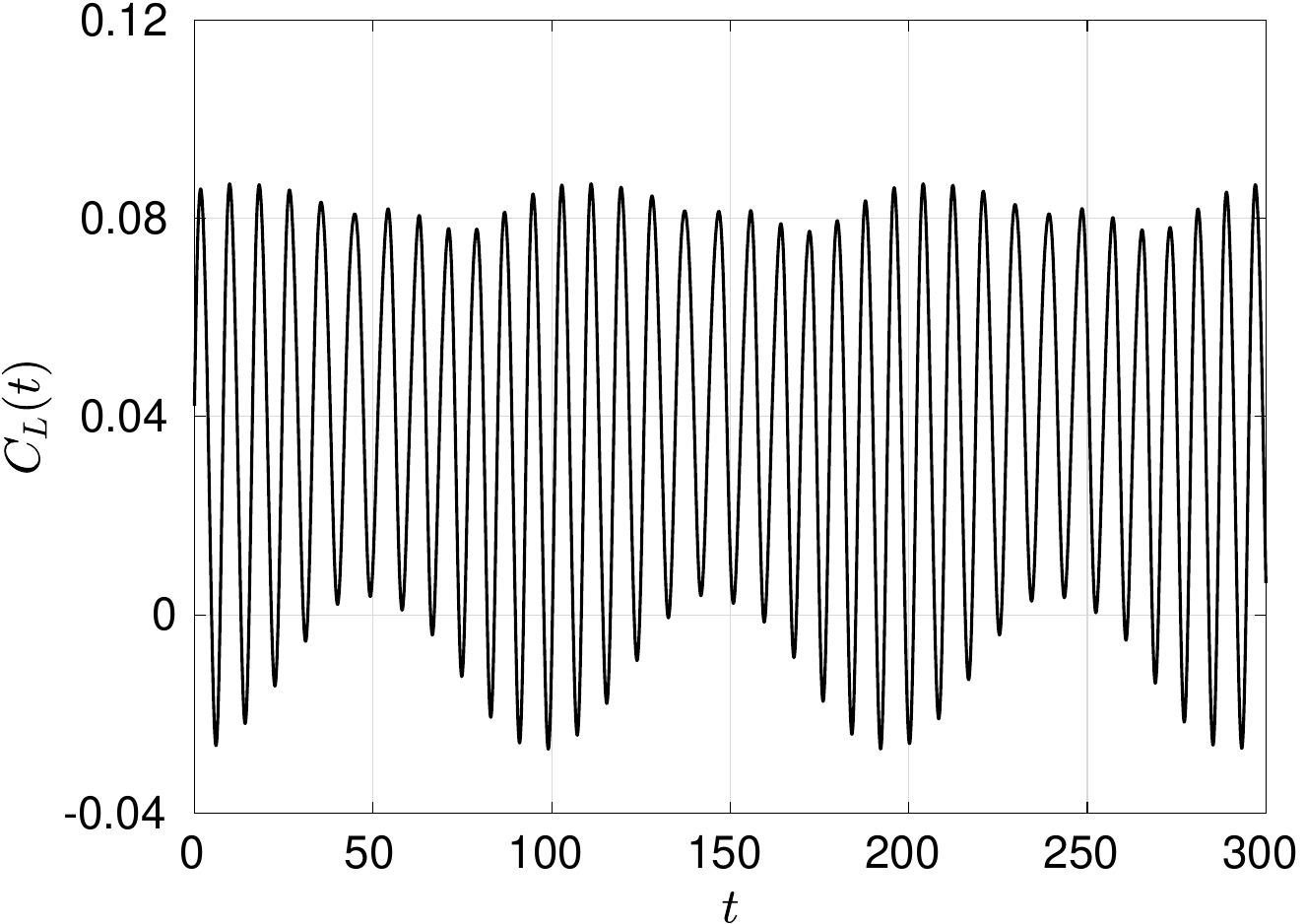}   &
\includegraphics[width=0.32\textwidth]{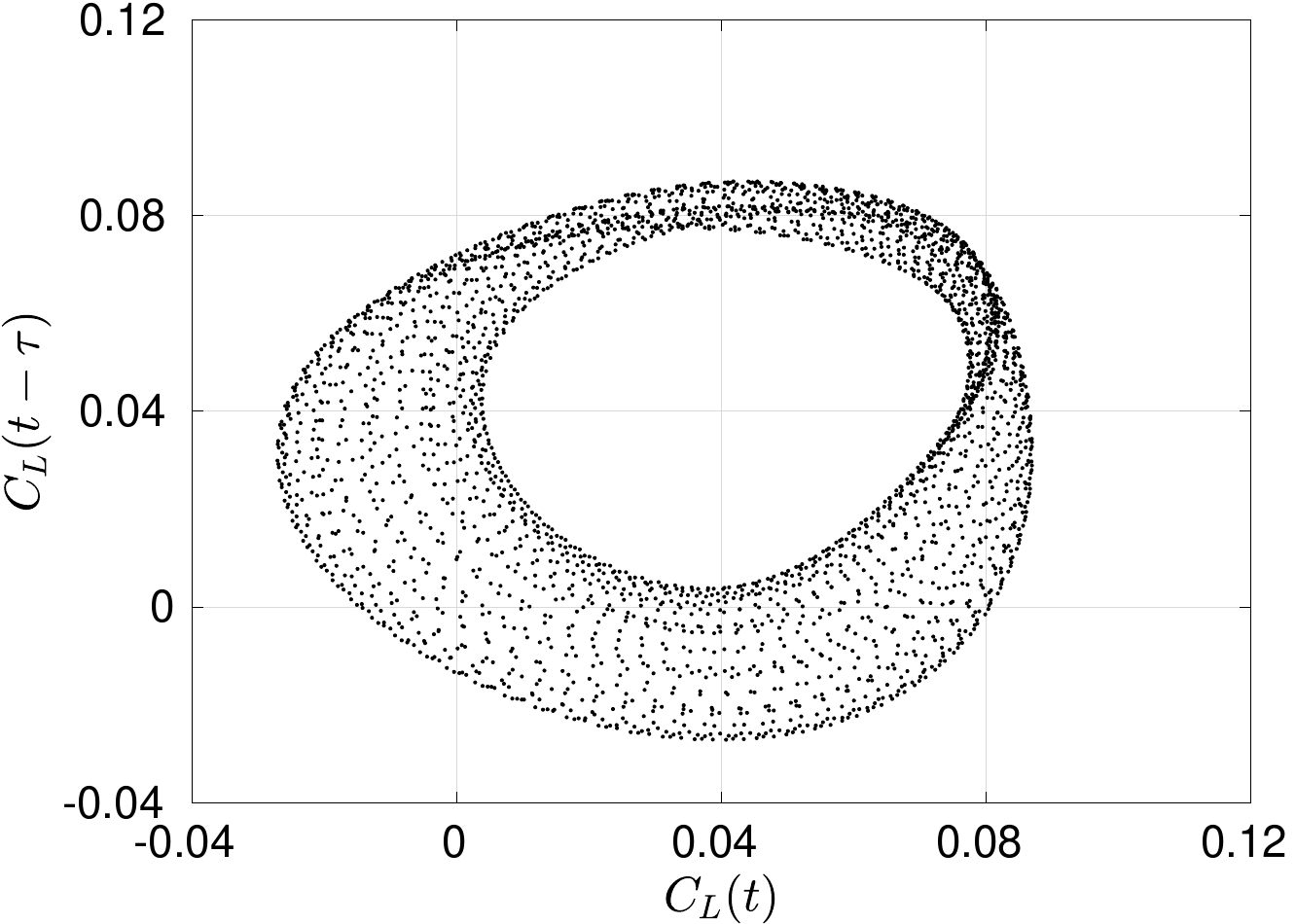}  \\ 
\end{tabular}
\caption{Quasi-periodic dynamics at $Re=105$ displayed by (a) the  power spectral density on time series of length  $T_{\rm data}=400$ (red curve), $T_{\rm data}=900$ (black curve),
(b) the time series and (c) the phase portrait of the pressure lift coefficient $C_L$.}
\label{Fig:Re105} 
\end{center}
\end{figure}

When the Reynolds number is further increased up to a critical value $Re_{3}\approx 104$, a new frequency rises in the power spectrum of the lift coefficient $C_L(t)$. This frequency is about one order of magnitude smaller than the natural frequency of the vortex shedding, as illustrated in figure~\ref{Fig:Re105}(a) for $Re=105$ and in the movie QP.MP4 of the additional materials. The low and natural frequencies couple to generate combs of sharp peaks in the power spectrum, while the background level depends on the length of the time series. The new frequency is associated with modulations of the base-bleeding jet around its deflected position. A visual inspection of both the time series and the phase portrait at $Re=105$ indicates that the new frequency also modulates the amplitude (figure~\ref{Fig:Re105}(b)) of the main oscillator and thickens the limit cycle associated with the main oscillator (figure~\ref{Fig:Re105}(c)). All these features are typical of a quasi-periodic dynamics, indicating that the system has most likely undergone a Neimark-S\"acker bifurcation, e.g. a secondary Hopf bifurcation, at $Re= Re_{3}$, after which two mirror-conjugated 2-tori exist in the state space of the system. 

\begin{figure}
\begin{center}
\begin{tabular}{ccc}
 (a) & (b) & (c) \\
\includegraphics[width=0.32\textwidth]{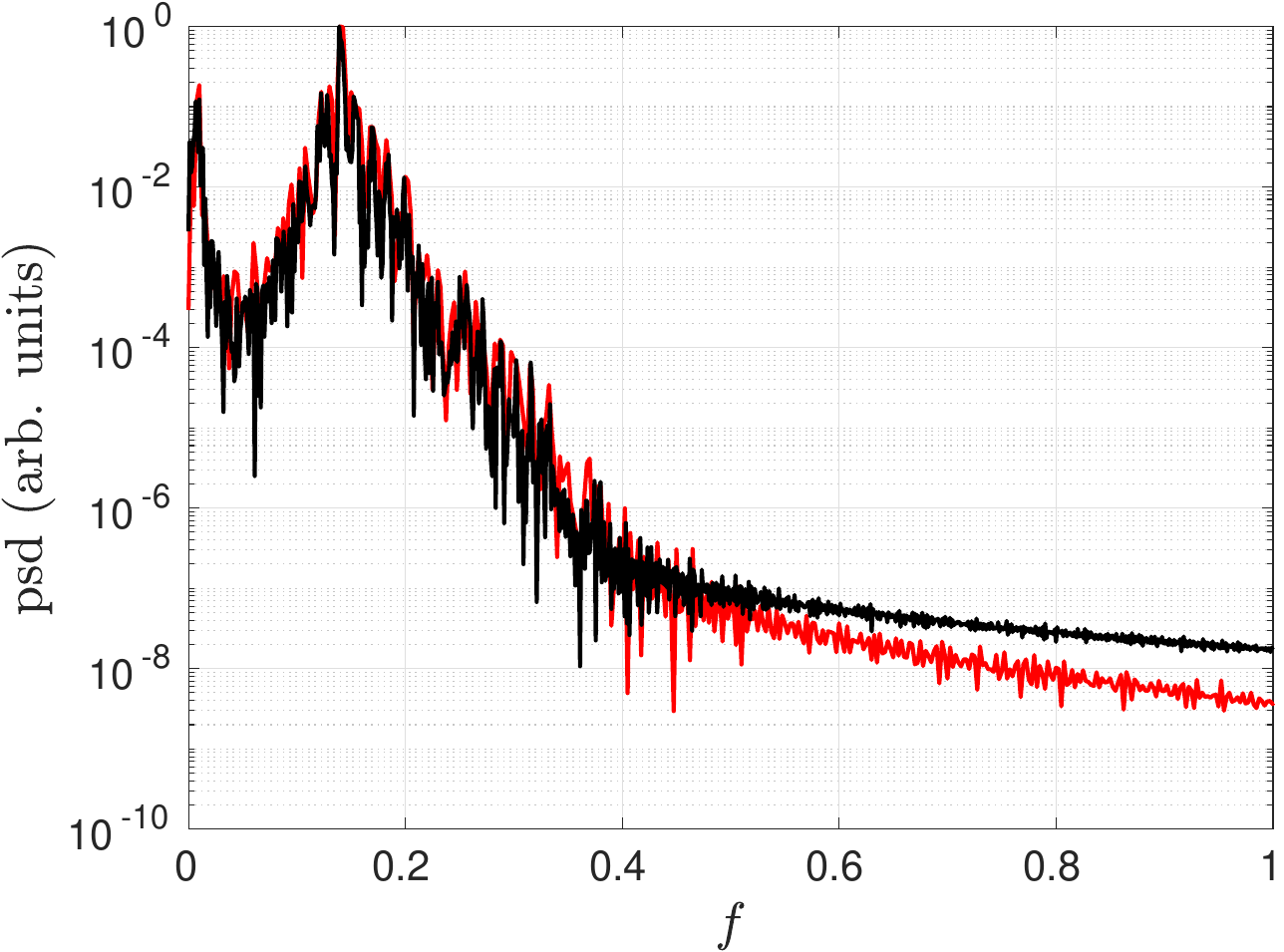}   &
\includegraphics[width=0.32\textwidth]{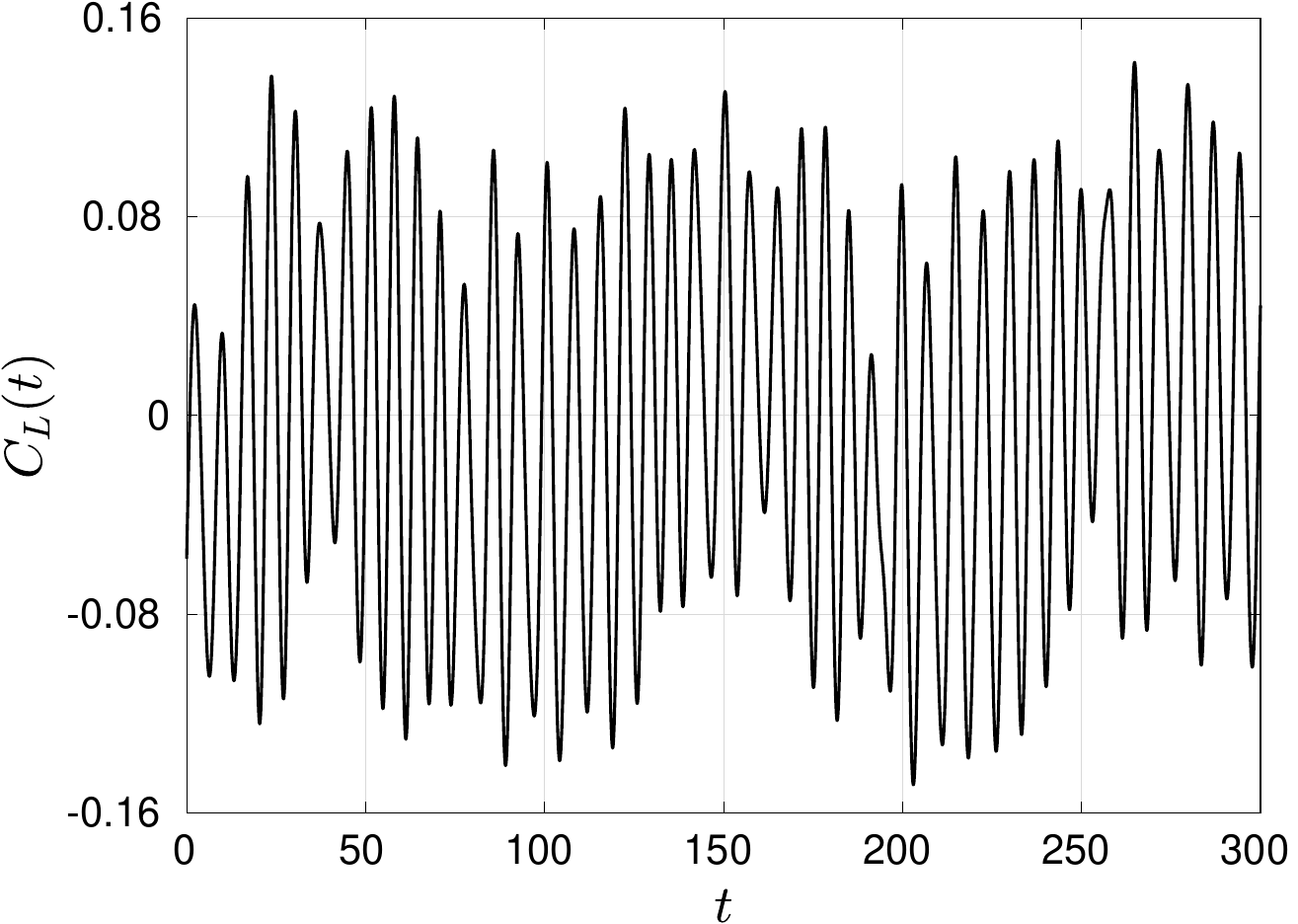}   &
\includegraphics[width=0.32\textwidth]{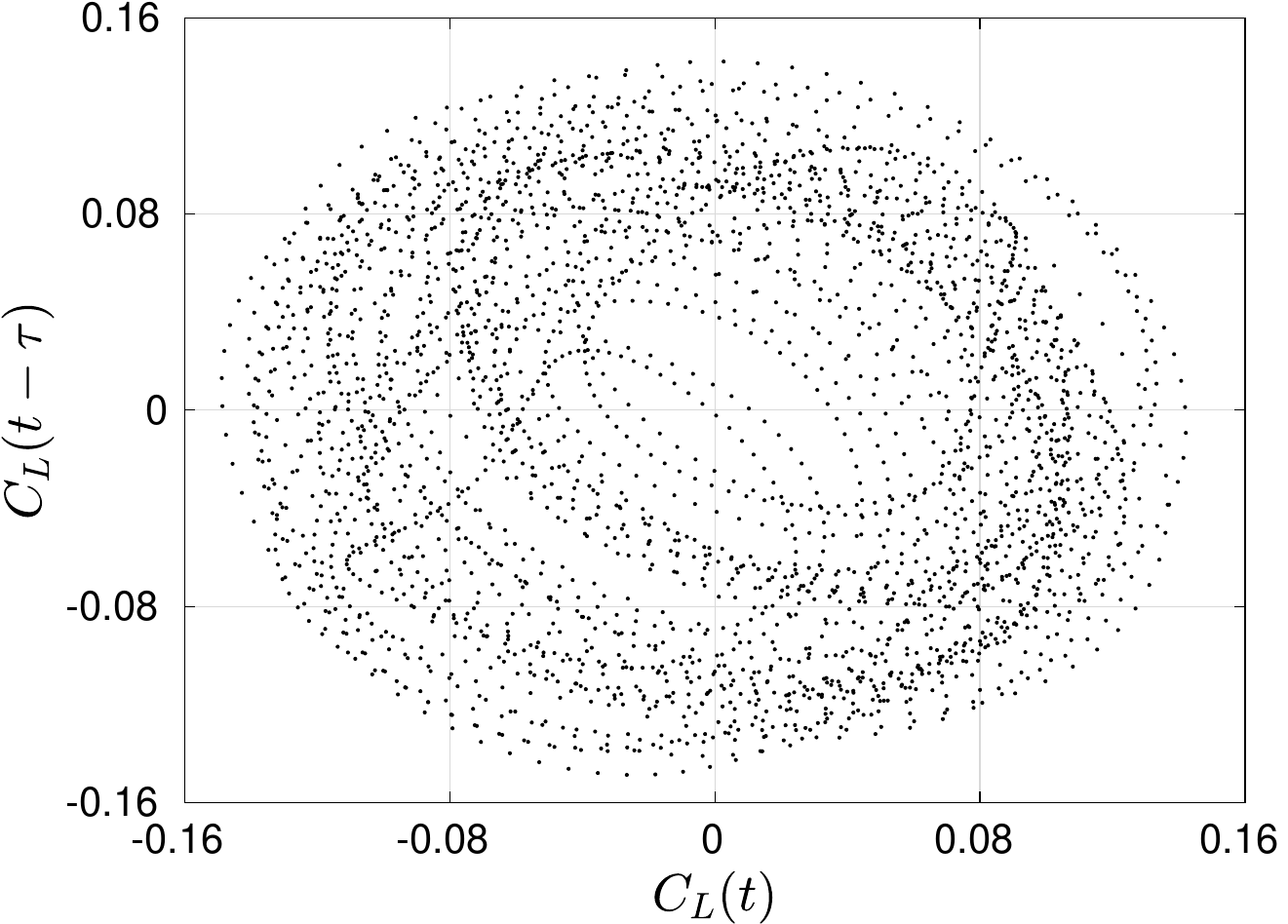}  \\ 
\end{tabular}
\caption{Chaotic dynamics at $Re=130$ displayed by (a) the power spectral density 
%(the maximum amplitude normalized)
on time series of length $T_{\rm data}=400$ (red curve), $T_{\rm data}=900$ (black curve), 
(b) the time series and (c) the phase portrait of the pressure lift coefficient $C_L$.}
\label{Fig:Re130} 
\end{center}
\end{figure}

At even larger values of the Reynolds number, $Re\geq Re_{4}\approx 115$, the main peak in the power spectral density of the lift coefficient widens significantly, as shown in figure~\ref{Fig:Re130}(a) for $Re=130$. In this new regime, the instantaneous flow field is characterized by random switches between an upward or downward base-bleeding jet, see the movie CHAOS.MP4 in the additional materials, and the mean flow $\overline{\bm{u}}$ is symmetric, as shown in figure~\ref{Fig:CL} for $Re>115$. The time series exhibits neither periodic nor quasi-periodic features anymore (see figure~\ref{Fig:Re130}(b)) and the phase portrait exhibits a much more complex dynamics (see figure~\ref{Fig:Re130}(c)). The dynamical regime henceforth exhibits many features of a chaotic regime, indicating that the system has most likely followed the Ruelle-Takens-Newhouse route to chaos \citep{newhouse1978}. 

%***********************************************************************
\section{Low-dimensional modeling}
\label{Sec:Methodology}

We derive a mean-field Galerkin model for the primary and secondary bifurcations of the fluidic pinball.
First (\S~\ref{Sec:Galerkin}), 
the Galerkin method is recapitulated as a very general approach to reduced-order models.
In \S~\ref{Sec:MeanField}, 
the constitutive equations of the mean-field model 
are derived from a minimal set of assumptions.
Then, mean-field Galerkin models are derived for the Hopf bifurcation (\S~\ref{Sec:Hopf}), 
the pitchfork bifurcation (\S~\ref{Sec:Pitchfork}) and the succession of both bifurcations (\S~\ref{Sec:MultipleBifurcation}).

%=======================================================================
\subsection{Galerkin method}
\label{Sec:Galerkin}

The starting point is the non-dimensionalized incompressible Navier-Stokes equations:
\begin{equation}
\label{Eqn:NSE}
\partial_t \bm{u} + \nabla \cdot \bm{u} \otimes \bm{u} = \nu \triangle \bm{u} - \nabla p,
\end{equation}
where $\nu = 1/Re$.
The velocity field satisfies the no-slip condition $\bm{u}=0$ on the cylinders,
the free-stream condition $\bm{u}=(1,0)$ at the inflow, 
a no-slip condition at the top and bottom boundary and the no-stress condition at the outflow.
The steady solution $\bm{u}_s$ satisfies the steady Navier-Stokes equations:
\begin{equation}
\label{Eqn:NSE:Steady}
 \nabla \cdot \bm{u}_s \otimes \bm{u}_s = \nu \triangle \bm{u}_s - \nabla p_s.
\end{equation}

The Galerkin method is based on an inner product in the space of square-integrable vector fields
$\mathcal{L}^2 ( \Omega ) $ in the observation domain $\Omega$.
The standard inner product between  $\bm{u} (\bm{x})$ and $\bm{v} (\bm{x})$ reads:
\begin{equation}
\label{Eqn:InnerProduct}
\left( \bm{u} , \bm{v} \right)_{\Omega} := \int\limits_{\Omega} \!\! d\bm{x} \> \bm{u}(\bm{x}) \cdot \bm{v}(\bm{x}).
\end{equation}
A traditional Galerkin approximation with a basic mode $\bm{u}_0$, for instance, the steady solution
and $N$ orthonormal expansion modes $\bm{u}_i ( \bm{x} )$, $i=1,\ldots,N$
with time-dependent amplitudes $a_i(t)$, reads:
\begin{equation} 
\label{Eqn:GalerkinExpansion}
\bm{u} (\bm{x},t ) = \bm{u}_0 ( \bm{x} ) + \sum\limits_{i=1}^N a_i(t) \bm{u}_i ( \bm{x} ).
\end{equation}
Orthonormality implies:
\begin{equation} 
\label{Eqn:Orthonormality}
\left( \bm{u}_i , \bm{u}_j \right)_{\Omega} = \delta_{ij}, \quad i,j \in \{1,\ldots,N\}.
\end{equation}
The projection of \eqref{Eqn:GalerkinExpansion} on \eqref{Eqn:NSE} 
leads to the linear-quadratic Galerkin system \citep{Fletcher1984book},
\begin{equation}
\label{Eqn:GalerkinSystem}
\frac{d}{dt} a_i = \nu \sum\limits_{j=0}^N l_{ij}^{\nu} a_j + \sum\limits_{j,k=0}^N q_{ijk}^{c} a_j a_k.
\end{equation}
Following \citet{Rempfer1994jfm},  $a_0=1$ is introduced.
The coefficients $l_{ij}^{\nu} = \left( \bm{u}_i, \triangle \bm{u}_j \right)_{\Omega}$ 
and  $q_{ijk}^{c} = \left( \bm{u}_i, \nabla \cdot \bm{u}_j \otimes \bm{u}_k \right)_{\Omega}$ 
parametrize the viscous and convective Navier-Stokes terms.
The pressure term vanishes for sufficiently large domains and is neglected in the following.

In the following, 
the steady solution is taken as the basic mode $\bm{u}_0 = \bm{u}_s$.
This implies that $\bm{a}=0$ is a fixed point of \eqref{Eqn:GalerkinSystem}
and the constant term $\nu l_{i0}^\nu + q_{i00}^{\nu}=0$ vanishes
as the projection of $\eqref{Eqn:NSE:Steady}$ onto the $i$th mode $\bm{u}_i$.
In this case, \eqref{Eqn:GalerkinSystem} can be re-written as a linear-quadratic system
of ordinary differential equations : 
\begin{equation}
\label{Eqn:DynamicalSystem}
\frac{d}{dt} a_i =  \sum\limits_{j=1}^N l_{ij} a_j + \sum\limits_{j,k=1}^N q_{ijk}  a_j a_k,
\end{equation}
where $l_{ij} = \nu l_{ij}^{\nu} + q_{ij0}^c + q_{i0j}^c$
and $q_{ijk} = q_{ijk}^c$ for $i,j,k \in \{1,\ldots,N\}$.

%=======================================================================
\subsection{Mean-field modeling}
\label{Sec:MeanField}

Mean-field modeling allows a dramatic simplification 
of a general Galerkin system \eqref{Eqn:DynamicalSystem}
close to bifurcations.
In this section, we derive constitutive equations
with a small number of more general assumptions.

In the spirit of the Reynolds decomposition,
the velocity field is decomposed into a slowly varying distorted mean flow $\bm{u}^D$
and fluctuation $\bm{u}^{\prime}$ with first-order (relaxational) or second-order (oscillatory) dynamics:
\begin{equation}
\label{Eqn:MFAnsatz}
\bm{u} (\bm{x}, t)
= \bm{u}^D (\bm{x},t) 
+ \bm{u}^{\prime} ( \bm{x}, t),
\quad\quad
 \bm{u}^D (\bm{x},t) =  \bm{u}_s (\bm{x})  + \bm{u}_{\Delta} (\bm{x},t)
\end{equation}
Here, the \emph{mean-field deformation} $\bm{u}_{\Delta}$ 
is the difference between the distorted mean flow 
and the steady solution. 
For the oscillatory dynamics considered, 
the distorted mean flow can be defined as an average over one local fluctuation period $T$ denoted by  $\langle \cdot \rangle$.
Thus, 
\begin{equation}
\label{Eqn:PhaseAverage}
\bm{u}^D  (\bm{x},t) 
= \langle \bm{u} ( \bm{x}, t ) \rangle  
:= \frac{1}{T} \int\limits_{t-T/2}^{t+T/2} \!\!\! d\tau  \> \bm{u} ( \bm{x}, \tau ).
\end{equation}
After the pitchfork bifurcation
into two mirror-conjugated flows $\bm{u}^+$, $\bm{u}^-$,
a symmetric distorted mean flow is enforced via
\begin{equation}
\label{Eqn:PhaseAveragePF}
\bm{u}^D  (\bm{x},t) 
= \frac{1}{2}(\langle \bm{u^+} ( \bm{x}, t ) \rangle  + \langle \bm{u^-} ( \bm{x}, t ) \rangle).
\end{equation}
We note that $\bm{u}^D (\bm{x},t)$ is not the mean flow, 
which is defined by the post-transient limit
\begin{equation}
\label{Eqn:TimeAverage}
\bm{\bar{u}} (\bm{x})
= \lim_{T\rightarrow \infty }\frac{1}{T}\int\limits_{0}^{T} \bm{u} (\bm{x}, \tau) d\tau.
\end{equation}
The distorted mean flow coincides with mean flow 
for the post-transient phase before the pitchfork bifurcation.  
Technically, 
a harmonic fluctuation is assumed 
and this one period average is computed as an average of all phases in $[0,2\pi]$.
The somewhat loaded term ``distorted mean flow'' is directly adopted 
from the original publications of mean-field theory \citep{Stuart1958jfm}.
J.T.~Stuart considers this flow as ``distorted'' 
from the steady solution by the Reynolds stress associated with the instability mode(s). 

For a nominally symmetric cylindrical obstacle,
the distorted mean flow can be expected to be symmetric 
while the dominant fluctuation is antisymmetric.
This leads to a symmetry-based decomposition
of the flow into a symmetric contribution $\bm{u}^s =\left( u^s, v^s \right)\in \mathcal{U}^s$ with
\begin{equation}
\label{Eqn:Symmetry}
u^s(x,-y) = u^s(x,y), \quad 
v^s(x,-y) = -v^s(x,y)
\end{equation}
and an antisymmetric component $\bm{u}^a=\left( u^a, v^a \right) \in \mathcal{U}^a$  satisfying 
\begin{equation}
\label{Eqn:Antisymmetry}
u^a(x,-y) = -u^a(x,y), \quad 
v^a(x,-y) =  v^a(x,y).
\end{equation}
Here, $\mathcal{U}^s$ and $\mathcal{U}^a$ denote the set of symmetric and antisymmetric vector fields, respectively.  
The resulting decomposition reads
\begin{equation}
\bm{u} ( \bm{x}, t) = \bm{u}^s  ( \bm{x}, t) +  \bm{u}^a  ( \bm{x}, t) .
\end{equation}
In the sequel, we will identify the distorted mean flow 
with the symmetric component and the fluctuation with the antisymmetric one:
\begin{equation}
\label{Eqn:SymmetryAssumption}
\mathbf{u}^D (\mathbf{x},t)  = \mathbf{u}^s ( \mathbf{x} , t), \quad
\bm{u}^{\prime} (\bm{x},t)  = \bm{u}^a ( \bm{x} , t).
\end{equation}
This identification is justified for symmetry-breaking bifurcations
with first- or second-order dynamics with neglected higher harmonics.
For brevity, $\mathcal{U}^s$ and $\mathcal{U}^a$ are introduced
as symmetric and antisymmetric subsets of $\mathcal{L}^2 ( \Omega )$.

The convective term is easily shown to have the following symmetry properties:
\begin{subequations}
\label{Eqn:SymmetriesConvectiveTerm}
\begin{eqnarray} 
   \nabla \cdot \bm{u}^s \otimes \bm{u}^s & \in & \mathcal{U}^s,
\\ \nabla \cdot \bm{u}^a \otimes \bm{u}^a & \in & \mathcal{U}^s,
\\ \nabla \cdot \bm{u}^s \otimes \bm{u}^a & \in & \mathcal{U}^a,
\\ \nabla \cdot \bm{u}^a \otimes \bm{u}^s & \in & \mathcal{U}^a.
\end{eqnarray}
\end{subequations}

The antisymmetric component 
is derived starting with \eqref{Eqn:NSE}, 
subtracting the steady version of \eqref{Eqn:NSE:Steady}
and 
exploiting the symmetry of $\bm{u}^D$, 
the antisymmetry of $\bm{u}^{\prime}$ as well as the symmetry relations \eqref{Eqn:SymmetriesConvectiveTerm}. 
The fluctuation dynamics reads:
\begin{equation}
\label{Eqn:NSE:Antisymmetric}
\partial_t \bm{u}^{\prime} + \nabla \cdot  \left[ \bm{u}^D \otimes \bm{u}^{\prime} 
                                                 + \bm{u}^{\prime} \otimes \bm{u}^D  \right] 
= \nu \triangle \bm{u}^{\prime} - \nabla p^{\prime}.
\end{equation}
Analogously, the symmetric part describes the distorted mean flow dynamics:
\begin{equation}
\label{Eqn:NSE:Symmetric}
\partial_t \bm{u}_{\Delta} + \nabla \cdot  \left[ \bm{u}_s        \otimes \bm{u}_{\Delta} 
                                                 + \bm{u}_{\Delta} \otimes \bm{u}_s 
                                                 + \bm{u}_{\Delta} \otimes \bm{u}_{\Delta}
                                                 + \bm{u}^{\prime} \otimes \bm{u}^{\prime}  \right]
= \nu \triangle \bm{u}_{\Delta} - \nabla p_{\Delta}
\end{equation}
Note that this symmetric component of the Navier-Stokes equations has not yet been averaged and the sum of Eqs.~\eqref{Eqn:NSE:Steady}, \eqref{Eqn:NSE:Symmetric} and \eqref{Eqn:NSE:Antisymmetric} 
leads to the Navier-Stokes equations \eqref{Eqn:NSE}.
To this point, all equations are strict identities for the symmetric and antisymmetric part of the Navier-Stokes dynamics.

Next, we follow mean-field arguments and consider $\bm{u}^{\prime}$ and $\bm{u}_{\Delta}$ 
as small perturbations around the fixed point $\bm{u}_s$.
Let $\bm{u}^{\prime} \in O(\varepsilon)$ and $\bm{u}_{\Delta} \in O(\delta)$
where $\varepsilon$ and $\delta$ are smallness parameters.
Hence, $\bm{u}_{\Delta} \otimes \bm{u}_{\Delta} \in O(\delta^2)$ can be neglected in comparison to the $O(\delta)$ terms $\bm{u}_s \otimes \bm{u}_{\Delta}$, $\bm{u}_{\Delta} \otimes \bm{u}_s$.
We follow Stuart's original idea
to separate between the fluctuation $\bm{u}^{\prime}$ driven by the instability
and the resulting mean-field deformation $\bm{u}_{\Delta}$ 
and arrive at the unsteady linearized Reynolds equation,
\begin{equation}
\label{Eqn:RE:Linearized}
\partial_t \bm{u}_{\Delta} + \nabla \cdot  \left[ \bm{u}_s        \otimes \bm{u}_{\Delta} 
                                                 + \bm{u}_{\Delta} \otimes \bm{u}_s 
                                                 + \left\langle \bm{u}^{\prime} \otimes \bm{u}^{\prime} \right\rangle  \right]
= \nu \triangle \bm{u}_{\Delta} - \nabla p_{\Delta}.
\end{equation}
The mean-field deformation $\bm{u}_{\Delta}$ characterized by the scale $\delta$ 
is seen to respond linearly to the Reynolds stress force 
$ -\nabla \cdot   \left\langle \bm{u}^{\prime} \otimes \bm{u}^{\prime} \right\rangle $ scaling with $\epsilon^2$.
Hence,  $\delta \sim \epsilon^2$.

Summarizing, 
Eqs.~\eqref{Eqn:NSE:Antisymmetric} and \eqref{Eqn:RE:Linearized} are the constitutive equations of mean-field theory
exploiting only symmetry and smallness of the mean-field deformation.

Close to the critical Reynolds number  $Re_c$,
the temporal growth rate can be Taylor expanded to $\sigma = \alpha (Re-Re_c)$ 
and can be assumed to be small.
In this case, 
$\partial_t \bm{u}_{\Delta} \in O(\sigma \delta)$, 
i.e. the time derivative of \eqref{Eqn:RE:Linearized}
can be neglected with respect to the other terms $\in O(\delta) = O(\varepsilon^2)$.
%A deeper analysis reveals $\partial_t \bm{u}_{\Delta} \in O(\delta^2)$.
This leads to the steady linearized Reynolds equation:
\begin{equation}
\nabla \cdot  \left[ \bm{u}_s  \otimes \bm{u}_{\Delta} 
                  + \bm{u}_{\Delta} \otimes \bm{u}_s 
                  + \langle {\bm{u}^{\prime} \otimes \bm{u}^{\prime}}  \rangle \right]
= \nu \triangle \bm{u}_{\Delta} - \nabla p_{\Delta}.
\label{Eqn:RE:LinSteady}
\end{equation}
This equation is also true for the post-transient solution,
e.g.\ the limit cycle of a Hopf bifurcation or the asymmetric state of a pitchfork bifurcation.
Often, the distorted mean flow $\bm{u}^D$ quickly responds to the Reynolds stress even far away from the bifurcation.

%=======================================================================
\subsection{Supercritical Hopf bifurcation}
\label{Sec:Hopf}

At low Reynolds numbers, a symmetric stable steady solution 
$\bm{u}_s \in \mathcal{U}^s$ is observed.
Periodic vortex shedding sets in with the occurrence 
of an unstable oscillatory antisymmetric  eigenmode at $Re \ge Re_{1}$.
The Reynolds-number dependent initial growth rate 
and frequency are denoted by $\sigma_1$ and $\omega_1$, respectively.
The real and imaginary parts of this eigenmode
are $\bm{u}_1$ and $\bm{u}_2$, respectively, both antisymmetric modes.
In the following, these modes are assumed to be orthonormalized.

This oscillation generates a Reynolds stress, which changes the mean flow via \eqref{Eqn:RE:Linearized}.
The mean flow deformation is described by the symmetric shift mode $\bm{u}_3$ with unit norm.
By symmetry, the first two modes are orthogonal with respect to the shift mode.
Thus, the modes form an orthonormal basis.
The resulting Galerkin expansion reads:
\begin{equation}
\label{Eqn:MFAnsatz:Hopf}
\bm{u} ( \bm{x},t ) = \bm{u}_s ( \bm{x} ) 
                      + \underbrace{a_1(t) \bm{u}_1(\bm{x}) + a_2(t) \bm{u}_2(\bm{x})}_{ \bm{u}^{\prime}}
                      + \underbrace{a_3(t) \bm{u}_3(\bm{x})}_{\bm{u}_{\Delta}}.
\end{equation}
Moreover, polar coordinates are introduced 
$a_1 (t) = r(t) \cos \theta (t) $, $a_2(t) = r(t) \sin \theta (t) $,  $d\theta/dt = \omega(t) $
where $r$ and $\omega$ are assumed to be slowly varying functions of time.

Substituting \eqref{Eqn:MFAnsatz:Hopf} in \eqref{Eqn:NSE:Antisymmetric},
projecting on $\bm{u}_i$, $i=1,2$ and applying the Krylov-Bogoliubov \citep{jordan1999nonlinear} 
averaging method yields:
\begin{subequations}
\label{Eqn:MFSystem:Hopf:Antisymmetric}
\begin{eqnarray}
   da_1/dt &= \sigma a_1 - \omega a_2, \quad \quad &\sigma = \sigma_1 - \beta a_3,
\\ da_2/dt &= \sigma a_2 + \omega a_1, \quad\quad &\omega = \omega_1 + \gamma a_3.
\end{eqnarray}
\end{subequations}
Here, $\sigma_1, \omega_1, \beta > 0$ for a supercritical Hopf bifurcation.
We refer to \citet{Noack2003jfm} for details.

%Krylov-Bogoliubov averaging implies the employed harmonic balancing based on a slowly varying amplitude and frequency.
Krylov-Bogoliubov averaging implies a harmonic balancing on the slowly varying amplitude and frequency of oscillatory $a_{1,2}$ and the slowly varying $a_3$ dynamics.
The corresponding original theorem 
includes a convergence proof of this approximation
for a second-order ordinary differential equation
for oscillations in the limit of small nonlinearity.
We cannot perform this limit 
but justify the operation
on the a priori observation 
that quadratic Galerkin system terms $q_{ijk}$
are typically two orders of magnitude smaller 
than the dominant linear coefficients,
i.e.\ describe a small nonlinearity.
A posteriori the operation is justified by the results,
i.e.\ by obtaining amplitudes and frequencies with up to a few percent error.

Substituting \eqref{Eqn:MFAnsatz:Hopf} in \eqref{Eqn:RE:LinSteady} 
replaces \eqref{Eqn:MFSystem:Hopf:Symmetric} by the mean-field manifold:
\begin{equation}
\label{Eqn:MFManifold:Hopf:Symmetric}
   a_3 = \kappa  \> \left( a_1^2 + a_2^2 \right)
\end{equation}
with derivable proportionality constant $\kappa$.

Alternatively, the mean-field manifold may be obtained from the Galerkin system.
Substituting \eqref{Eqn:MFAnsatz:Hopf} in \eqref{Eqn:RE:Linearized}
and projecting on $\bm{u}_3$  yields:
\begin{equation}
\label{Eqn:MFSystem:Hopf:Symmetric}
   da_3/dt = \sigma_3 a_3 + \beta_3 \left( a_1^2 + a_2^2 \right),
\end{equation}
where $\sigma_3 < 0$ and $\beta_3>0$ are necessary 
for a globally stable limit cycle.
Note that \eqref{Eqn:MFSystem:Hopf:Symmetric}  can be rewritten as:
\begin{equation}
\label{Eqn:MFSystem:Hopf:Symmetric:2}
   da_3/dt = \sigma_3 \left [ a_3  - \kappa \left( a_1^2 + a_2^2 \right) \right].
\end{equation}
Now, the slaving process which leads to the mean-field manifold of \eqref{Eqn:MFManifold:Hopf:Symmetric} 
can be appreciated from the mean-field Galerkin system.
If $\vert \sigma_3 \vert \gg \sigma_1$, 
the timescale of slaving $a_3$ 
to the fluctuation level $a_1^2+a_2^2$ is much smaller
than the timescale of the transient
and $da_3/dt$ can be set to zero.

Eqs.~\eqref{Eqn:MFSystem:Hopf:Antisymmetric} and 
\eqref{Eqn:MFManifold:Hopf:Symmetric} 
yield the famous Landau equations with the cubic damping term:
\begin{equation}
\label{Eqn:Landau}
dr/dt = \sigma_1 r - \beta \kappa r^3,
\quad d\theta/dt = \omega_1 + \gamma \kappa r^2.
\end{equation}
The Landau oscillator leads to a stable limit cycle with
$r^{\circ} = \sqrt{\sigma_1/\beta \kappa}$, 
frequency $\omega^{\circ} = \omega_1 +  \sigma_1 \gamma / \beta$
and shift-mode amplitude $a_3^{\circ} = \sigma_1/\beta$.
The three nonlinearity parameters $\beta$, $\gamma$ and $\kappa$
can be uniquely derived from the limit cycle parameters
$r^{\circ}$, $\omega^{\circ}$, and $a_3^{\circ}$.
The growth rate $\sigma_3$ needs to be chosen sufficiently large,
e.g.\ $\sigma_3= - 10 \sigma_1$ to ensure slaving on the manifold.

Eqs.~\eqref{Eqn:MFSystem:Hopf:Antisymmetric}, \eqref{Eqn:MFSystem:Hopf:Symmetric} are the mean-field Galerkin system,
while Eqs.~\eqref{Eqn:MFSystem:Hopf:Antisymmetric}, \eqref{Eqn:MFManifold:Hopf:Symmetric} characterize the original mean-field model,
\textit{i.e.} the slaved Galerkin system.
Near the Hopf bifurcation, when $\sigma_1(Re)$ crosses the zero line at $Re_{1}$,
the growth rate is approximated by 
$\sigma_1 = \alpha \left( Re -Re_{1} \right)$ 
implying the square-root law 
$r^{\circ} = \sqrt{\alpha / \beta \kappa}  \sqrt{Re - Re_{1}}$.

The Landau equation has been proposed by Landau \citep[see, \textit{e.g.}][]{Landau1987book},
derived from the Navier-Stokes equation by \citet{Stuart1958jfm},
generalized for Galerkin systems by \citet{Noack2003jfm},
and validated in numerous simulations 
and experiments for cylinder wakes \citep{Schumm1994jfm}
and other soft onsets of oscillatory flows.
We note that the proposed derivation from symmetry considerations 
constrains the model to symmetric obstacles
but liberates the mean-field Galerkin model from typical assumptions,
likes closeness to the Hopf bifurcation 
or the need for frequency filtered Navier-Stokes equations.

%=======================================================================
\subsection{Supercritical pitchfork bifurcation}
\label{Sec:Pitchfork}

\begin{table}
 \label{Tab:EqClassification}
 \begin{center}
 \includegraphics[width=.9\linewidth]{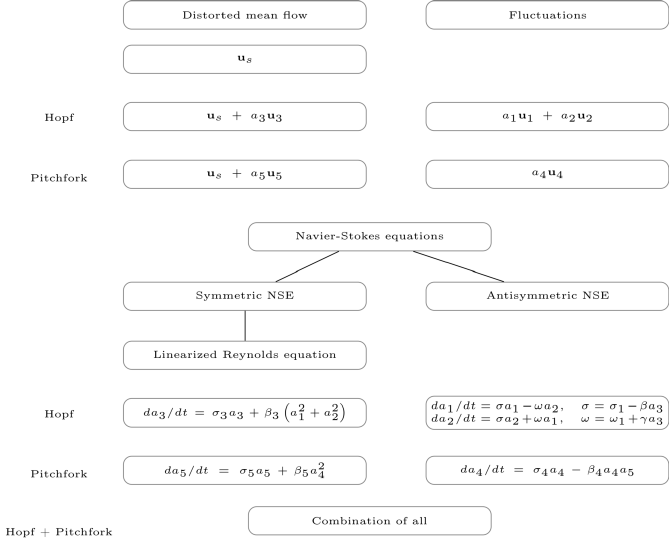}
 \end{center}
 \caption{Symmetries and hierarchy of equations. }
\end{table}

Next, the symmetry-breaking pitchfork bifurcation of a steady symmetric Navier-Stokes equation is considered.
Now, mode $\bm{u}_4$ describes the antisymmetric instability with positive growth rate $\sigma_4$.
The shift mode  $\bm{u}_5$ prevents unbounded exponential growth.
The corresponding Galerkin expansion reads
\begin{equation}
\label{Eqn:MFAnsatz:Pitchfork}
\bm{u} ( \bm{x},t ) = \bm{u}_s ( \bm{x} ) 
                      + \underbrace{a_4(t) \bm{u}_4(\bm{x})}_{ \bm{u}^{\prime}}
                      + \underbrace{a_5(t) \bm{u}_5(\bm{x})}_{\bm{u}_{\Delta}}.
\end{equation}

Substituting \eqref{Eqn:MFAnsatz:Pitchfork} 
in Eqs.~\eqref{Eqn:NSE:Antisymmetric} and \eqref{Eqn:RE:Linearized},
and exploiting the symmetry of the modes,
yields:
\begin{subequations}
\label{Eqn:MFSystem:Pitchfork}
\begin{eqnarray}
   da_4/dt &=& \sigma_4 a_4 - \beta_4 a_4 a_5,
\label{Eqn:MFSystem:Pitchfork:Antisymmetric}
\\ da_5/dt &=& \sigma_5 a_5 + \beta_5 a_4^2.
\label{Eqn:MFSystem:Pitchfork:Symmetric}
\end{eqnarray}
\end{subequations}
Note that a linear $a_5$ term and 
quadratic $a_4 a_4$ and $a_5 a_5$ terms 
in \eqref{Eqn:MFSystem:Pitchfork:Antisymmetric}
are ruled out by symmetry.
Similarly, a linear $a_4$ term or a mixed quadratic term $a_4 a_5$ 
in \eqref{Eqn:MFSystem:Pitchfork:Symmetric} are prohibited by symmetry.
The quadratic  $a_5 a_5$ term is not consistent with the linearized Reynolds equation \eqref{Eqn:RE:Linearized}.
The pitchfork bifurcation can be considered as a Hopf bifurcation with $\omega=0$ and a single mode.
Replacing $a_1$ by $a_4$, $a_3$ by $a_5$ and setting $a_2=0$ 
yields \eqref{Eqn:MFSystem:Pitchfork:Antisymmetric} 
from \eqref{Eqn:MFSystem:Hopf:Antisymmetric} and
\eqref{Eqn:MFSystem:Pitchfork:Symmetric} 
from \eqref{Eqn:MFSystem:Hopf:Symmetric}---modulo names of the coefficients.

Eqs. \eqref{Eqn:MFSystem:Pitchfork:Antisymmetric}, 
\eqref{Eqn:MFSystem:Pitchfork:Symmetric} 
are the mean-field Galerkin system.
Substituting \eqref{Eqn:MFAnsatz:Pitchfork} 
in \eqref{Eqn:RE:LinSteady} yields the manifold:
\begin{equation}
\label{Eqn:MFManifold:Pitchfork:Symmetric}
   a_5 = \kappa_5  a_4 ^2,
\end{equation}
with $\kappa_5=-\beta_5/\sigma_5$. 
The asymmetric steady solutions read 
$a_4^{\pm} = \pm \sqrt{\sigma_4 /\beta_4 \kappa_5}$,
$a_5= \sigma_4/\beta_4$. 
The two nonlinearity parameters $\kappa_4$, $\beta_4$ are readily determined
from the two asymptotic values $a_4$ and $a_5$.
The growth rate can be set in analogy to the previous model
to $\sigma_5= -10 \sigma_4$ to ensure slaving on the manifold.

From \eqref{Eqn:MFSystem:Pitchfork:Antisymmetric} and \eqref{Eqn:MFManifold:Pitchfork:Symmetric},
the famous unstable dynamics with cubic damping term is obtained:
 $$ da_4/dt = \sigma_4 a_4 - \kappa_5 \beta_4 a_4^3 ,$$
where $\kappa_5\beta_4>0$ for a supercritical bifurcation. 
Eqs. \eqref{Eqn:MFSystem:Pitchfork:Antisymmetric}, \eqref{Eqn:MFSystem:Pitchfork:Symmetric} 
are the mean-field Galerkin system.

Near the secondary pitchfork bifurcation,
$\sigma_4 = \alpha_2 \left( Re -Re_{2} \right)$
and $a_4 \propto \sqrt{Re-Re_{2}}$.
The parameters of the pitchfork Galerkin system
can be derived from the eigenmode and the asymptotic state
in complete analogy to \S~\ref{Sec:Hopf}.
The growth rate $\sigma_5 = - 10 \sigma_4$ will ensure 
the slaving of \eqref{Eqn:MFManifold:Pitchfork:Symmetric}.
We emphasize that this pitchfork model is derived primarily from symmetry considerations
and does not require closeness to the critical parameter.

%=======================================================================
\subsection{Pitchfork bifurcation of periodic solution}
\label{Sec:MultipleBifurcation}

In the final modeling effort, a low-dimensional model
from a primary supercritical Hopf bifurcation at $Re=Re_{1}$
and a secondary supercritical pitchfork bifurcation at $Re=Re_{2}> Re_{1}$ is derived following the numerical observations of the fluidic pinball in \S~\ref{Sec:FlowConfig}.
For simplicity,
 closeness to the secondary bifurcation is assumed.
For the same reason, 
the mean-field Galerkin system shall still describe the periodic solution. 
In this case, the generalized 5-mode mean-field expansion:
\begin{equation}
\label{Eqn:MFAnsatz:Hopf+Pitchfork}
\bm{u} ( \bm{x},t ) = \bm{u}_s ( \bm{x} )  + \sum\limits_{i=1}^5 a_i (t) \bm{u}_i (\bm{x} )
\end{equation}
describes the flow where $a_1,a_2,a_3 \in O(1)$ and $a_4 \in O(\varepsilon)$ and $a_5 \in O(\delta)$,
$\varepsilon$, $\delta$ being smallness parameters associated with the pitchfork bifurcation.
We project Eqs.~\eqref{Eqn:NSE:Antisymmetric} and \eqref{Eqn:RE:Linearized}
on  \eqref{Eqn:MFAnsatz:Hopf+Pitchfork}.
The $O(1)$ terms encapsulate the original Hopf model
while the low-pass filtered $O(\varepsilon,\delta)$  terms yield the original pitchfork system.
This yields the following generalized mean-field system:
\begin{subequations}
\label{Eqn:MFSystem:Hopf+Pitchfork}
\begin{eqnarray}
   da_1/dt &=& \sigma a_1 - \omega a_2, \quad \quad \sigma = \sigma_1 - \beta a_3
\\ da_2/dt &=& \sigma a_2 + \omega a_1, \quad\quad \omega = \omega_1 + \gamma a_3
\\ da_3/dt &=& \sigma_3 a_3 + \beta_3 \left( a_1^2 + a_2^2 \right)
\\ da_4/dt &=& \sigma_4 a_4 - \beta_4 a_4 a_5 
\\ da_5/dt &=& \sigma_5 a_5 + \beta_5 a_4^2 
\end{eqnarray}
\end{subequations}
The linear instability parameters $\sigma_1$, $\omega_1$, $\sigma_4$
are obtained from the corresponding global stability analysis.
Slaving is ensured with $\sigma_3=-10 \sigma_1$ and $\sigma_5=-10 \sigma_4$.
The nonlinearity parameters $\beta$, $\gamma$, $\beta_3$, $\beta_4$ and $\beta_5$
are determined from the limit cycle parameters $r^{\circ}$, $\omega^{\circ}$ and $a_3^{\circ}$
and pitchfork parameters $a_4^{\pm}$ and $a_5^{\pm}$ in the asymptotic regime.

As the amplitude of the pitchfork bifurcation grows,
the smallness argument does not hold and we get cross-terms,
like  $\sigma = \sigma_1 - \beta a_3 - \beta_{15} a_5$.
We shall not pause to elaborate on the possible generalizations now,
but will return to the topic in the result section.

%=======================================================================
\subsection{Sparse Galerkin model from mean-field considerations}
\label{Sec:Galerkin+Symmetries}

The mean-field Galerkin system \eqref{Eqn:MFSystem:Hopf+Pitchfork} 
with decoupled Hopf and pitchfork dynamics can, by construction,
only be expected to hold near the pitchfork bifurcation $Re \approx Re_{2}$.
At higher Reynolds numbers $Re > Re_{2}$, 
cross-terms will appear,
e.g., the growth rate $\sigma$ may also depend on the pitchfork-related shift mode amplitude $a_5$.
The most general Galerkin system \eqref{Eqn:DynamicalSystem}
contains $5 \times 5=25$ linear terms and $5 \times 5 \times 6/2=75$ quadratic terms.

%Six terms vanish by neglecting the quadratic mean-field deformation terms in \eqref{Eqn:NSE:Symmetric}, i.e., $q_{i33} a_3 a_3$, $q_{i35} a_3 a_5$, $q_{i55} a_5 a_5$ for $i=3,5$. The corresponding coefficients are set to zero.

The assumed symmetry of the modes excludes roughly half of these 100 coefficients.
Let $\chi_i = 0$ for symmetric mean flow modes $\bm{u}_i$, $i=3,5$ 
and $\chi_i=1$ for the antisymmetric fluctuation modes $\bm{u}_i$, $i=1,2,4$.
The linear coefficients 
 $$
 l_{ij} = - \nu  \left ( \bm{u}_i, \triangle \bm{u}_i \right)_{\Omega}
        + \left ( \bm{u}_i, \nabla \cdot \bm{u}_s \otimes \bm{u}_j \right)_{\Omega}
        + \left ( \bm{u}_i, \nabla \cdot \bm{u}_j \otimes \bm{u}_s \right)_{\Omega}
 $$
 can be shown to vanish if $\mod (\chi_i + \chi_j, 2) =1 $.
In other words, the coefficients $l_{ij}$ vanish 
if the modes $\bm{u}_i$ and $\bm{u}_j$ have opposite symmetries.
This excludes 12 of the 25 linear coefficients.
Analogously, 
the quadratic coefficient $q_{ijk}$ can be shown to vanish if $\chi_i \not = \mod ( \chi_j + \chi_k, 2 ) $.
In other words, the $q_{ijk}$  vanishes 
if the symmetry of $\bm{u}_i$
does not coincide with the symmetry of quadratic term $\bm{u}_j \otimes \bm{u}_k$.
The quadratic term is symmetric if the modes $\bm{u}_j$ and $\bm{u}_k$ 
are both symmetric or both antisymmetric
and is antisymmetric if both modes have opposite symmetries.
In summary, $q_{ijk}$ vanishes if one or three modes are antisymmetric.

An additional sparsity of the coefficients arises from the temporal dynamics.
Modes $\bm{u}_i$, $i=1,2$ have oscillatory behaviour with angular frequency $\omega$,
while the other modes show first-order dynamics, i.e., relaxation to asymptotic values.
We apply the Krylov-Bogoliubov approximation with oscillatory $a_1,a_2$ and slow $a_3,a_4,a_5$ dynamics.
Thus, for instance, $l_{41} a_1$ vanishes on a one-period average and should not contribute to $da_4/dt$.
The linear coefficient $l_{41}$ can hence be set to zero.
Taking the quadratic terms for example, $a_1 a_3$ generates a first harmonic.
Hence, $q_{413} a_1 a_3$ cannot contribute to $da_4/dt$ but $q_{113} a_1 a_3$ can contribute to the oscillatory behaviour of  $da_1/dt$. Similarly, $a_1 a_2$ generates a second harmonic,  $a_1 a_2 = r \cos \omega t \times r \sin \omega t =  (\nicefrac{1}{2}) \>  r^2 \> \sin 2 \omega t$ does not have a steady contribution, so $q_{312}$ can be set to zero.

From symmetry and Krylov-Bogoliubov considerations,
only 9 coefficients contribute to the linear term:
$l_{11}$, $l_{12}$, $l_{21}$, $l_{22}$, $l_{33}$, $l_{35}$, $l_{44}$, $l_{53}$, $l_{55}$. 
Note that the oscillator equations contain the $2 \times 2 $ block,
while the shift-mode equations $i=3,5$ have cross-terms
and the pitchfork amplitude dynamics $i=4$ has no cross-terms.
Similarly, only 16 quadratic coefficients survive.
The first 8 coefficients
$q_{113}$, 
$q_{115}$, 
$q_{123}$, 
$q_{125}$, 
$q_{213}$, 
$q_{215}$, 
$q_{223}$, 
$q_{225}$
are consistent with the Landau oscillator but with cross-terms to the pitchfork-related shift-mode amplitude,
i.e.\ $\sigma = \sigma_1 - \beta a_3 - \beta_{15} a_5$ and
$\omega = \omega_1 + \gamma a_3 + \gamma_{15} a_5$,
introducing $\beta_{15}$ and $\gamma_{15}$ as new coefficients.
The first and second shift-mode equations $i=3,5$ may contain six quadratic terms
$q_{311}$, 
$q_{322}$, 
$q_{344}$,
$q_{511}$, 
$q_{522}$, 
$q_{544}$
from the Reynolds stresses.
The amplification of the pitchfork dynamics
is affected by the shift modes via 
$q_{443}$, 
$q_{445}$.

%=======================================================================

\section{Primary flow regime}
\label{Sec:Primary}

The primary flow regime covers the range of Reynolds numbers $Re_{1}<Re<Re_{2}$. 
We consider the flow and reduction of the dynamics at $Re=30$, as a representative case. In \S~\ref{Sec:SteadySolution30}, a linear stability analysis is done on the steady solution and the three degrees of freedom of the flow dynamics are identified. In \S~\ref{Sec:ROM30}, we propose a least-order model of the flow dynamics at $Re=30$ and compare its performance with respect to the full flow dynamics. 

\subsection{Eigenspectra of the steady solution}
\label{Sec:SteadySolution30}

\begin{figure}
\begin{center}
\begin{tabular}{c }
(a) \\
 \includegraphics[width=0.75\textwidth]{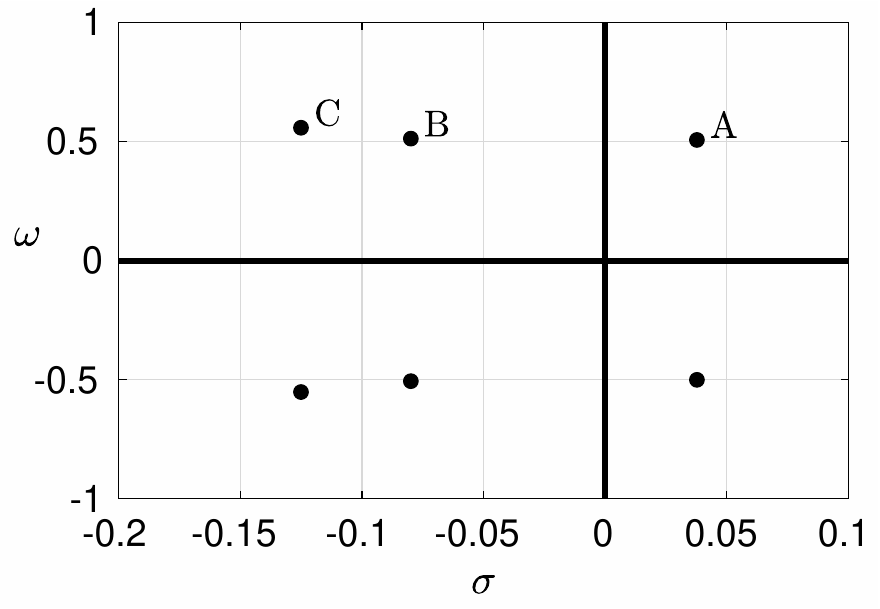} \\

\end{tabular}

\begin{tabular}{c  c }
& (b) \\
{A} & \raisebox{-0.5\height}{ \includegraphics[width=0.45\textwidth]{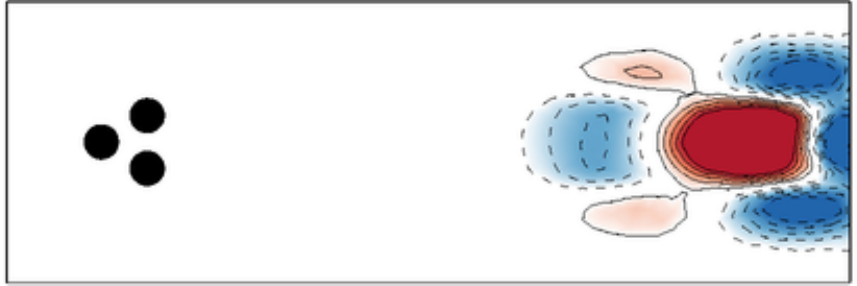}} \\
{B} & \raisebox{-0.5\height}{ \includegraphics[width=0.45\textwidth]{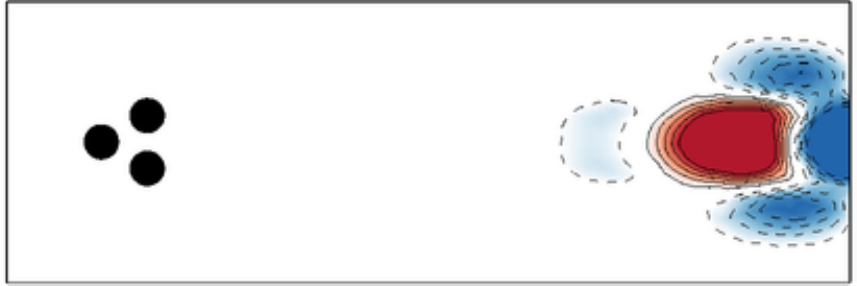}}\\
{C} & \raisebox{-0.5\height}{ \includegraphics[width=0.45\textwidth]{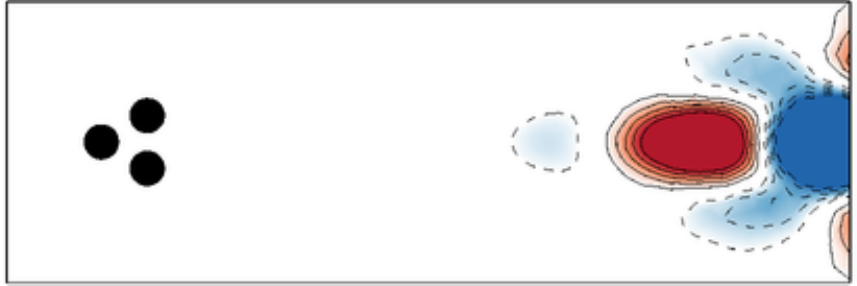}}\\
%\includegraphics[width=0.75\textwidth]{./Figures/Deng_JFM_spectrum30.png}   
%& \includegraphics[width=0.45\textwidth]{}  \\ 
\end{tabular}
\caption{(a) Eigenspectrum resulting from the linear stability analysis of the steady solution $\bm{u}_s$, together with (b) the first three leading eigenmodes, at $Re=30$. Only the real part of the complex eigenmodes is shown. Red color and solid contours are positive values of the vorticity, blue color and dashed contours are negative values. }
\label{Fig:EigenModes30}
\end{center}
\end{figure}

\begin{figure}
\begin{center}
\begin{tabular}{c}
 (a) \\
\includegraphics[width=0.5\textwidth]{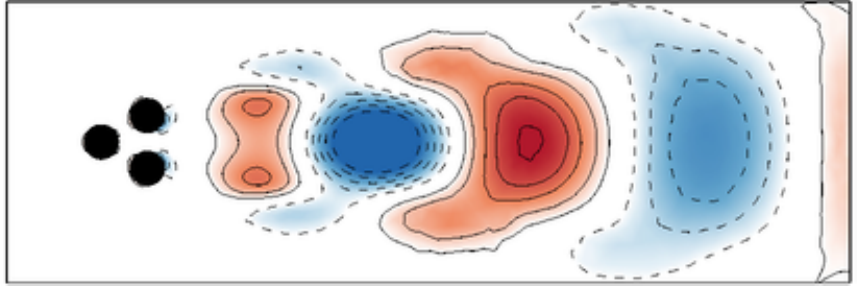}   \\
 (b) \\
\includegraphics[width=0.5\textwidth]{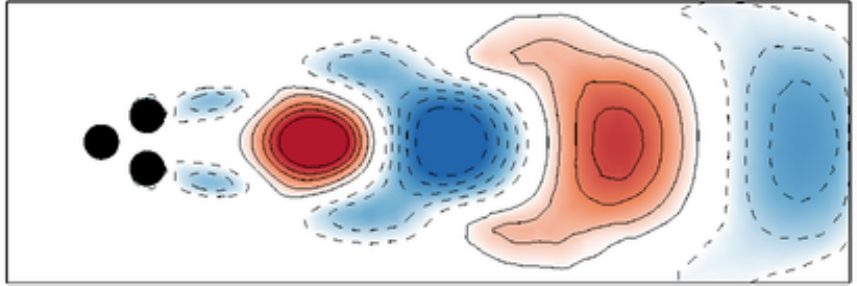}  \\
 (c) \\
\includegraphics[width=0.5\textwidth]{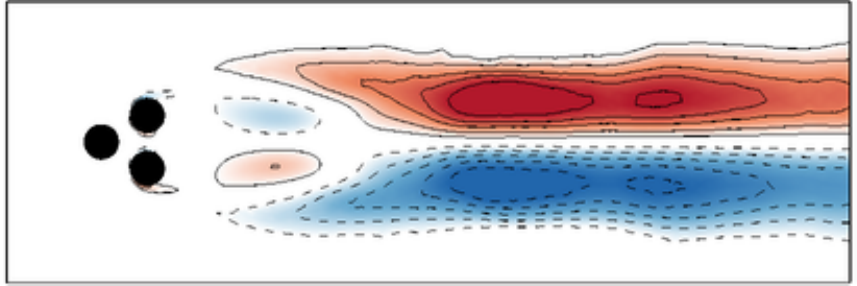}  \\
\end{tabular}
\caption{First two leading POD modes $\bm{u}_{1,2}$ at $Re=30$ (a) \& (b) and shift mode $\bm{u}_3$ (c). Red color and solid contours are positive values of the vorticity, blue color and dashed contours are negative values.}
\label{Fig:POD30}
\end{center}
\end{figure}

The steady solution becomes unstable beyond $Re=Re_{1}$, as reported in \S~\ref{Sec:FlowConfig}. A linear stability analysis indicates that one pair of complex-conjugated eigenmodes have a positive growth rate on the range $Re_{1}<Re<Re_{2}$, as shown in figure~\ref{Fig:EigenModes30} for $Re=30$. These two leading eigenmodes are associated with vortical structures shed downstream in the wake, at the angular frequency 1/2. As the instability grows, the distorted mean flow $\bm{u}^D=\bm{u}_s+\bm{u}_\Delta$ changes, as expected by Eq.~\eqref{Eqn:NSE:Steady} and \eqref{Eqn:NSE:Symmetric}. The shift mode $\bm{u}_3$, involved in $\bm{u}_\Delta $ at $Re=30$, is shown in figure~\ref{Fig:POD30}(c). In the permanent (time-periodic) flow regime, the distorted mean flow $\bm{u}^D$ eventually matches the asymptotic mean flow field $\overline{\bm{u}}$, and vortex shedding is well established with frequency $8.7\times 10^{-2}$. %In this regime, the first two modes $\bm{u}_{1,2}$ arise from a Proper Orthogonal Decomposition (POD) of the limit cycle data, contribute to almost 95\% of the total fluctuating kinetic energy at $Re=30$ and are clearly associated with the von K\'arm\'an street of shed vortices, as shown in figure~\ref{Fig:POD30} (a)\&(b). For the construction of the ROM, POD modes $\bm{u}_{1,2}$ are preferred to the two leading eigenmodes, because we demand an accurate representation of the asymptotic periodic dynamics.

\subsection{Reduced-order model (ROM) of the primary flow regime}
\label{Sec:ROM30}

\begin{table}
 \begin{center}
 \begin{tabular}{cccccc}
 $\sigma_1$ & $\omega _1$ & $\sigma_3$ & $\beta$ & $\gamma$ & $\kappa$ \\
 \hline \hline
 $3.80\times 10^{-2}$\quad  & \quad $5.00\times 10^{-1}$\quad  & \quad $-10\sigma_1$ \quad & \quad $1.40\times 10^{-2}$\quad  & \quad $1.70\times 10^{-2}$\quad  & \quad $2.10\times 10^{-1}$ \\
 \end{tabular}
 \end{center}
 \caption{Coefficients of the reduced-order model (ROM) at $Re=30$. See text for details. }
 \label{Tab:Hopf}
\end{table}

As introduced in \S~\ref{Sec:Methodology}, the Galerkin ansatz for the Hopf bifurcation reads:
\begin{equation}
	\bm{u}(\bm{x},t) \approx \bm{u}_s(\bm{x}) + a_1(t)\bm{u}_1(\bm{x}) + a_2(t)\bm{u}_2(\bm{x}) + a_3(t)\bm{u}_3(\bm{x}).
\label{Eq:HopfAnsatz}
\end{equation}
The von K\'arm\'an modes $\bm{u}_{1,2}$ could be chosen as the real and imaginary part of the first eigenmode, respectively. This choice would make sense to describe the transient dynamics close to the steady solution. A better choice for describing the dynamics on the asymptotic limit cycle is to choose $\bm{u}_{1,2}$ as the first two modes of a Proper Orthogonal Decomposition (POD) of the limit cycle data. The first two POD modes actually contribute to almost 95\% of the total fluctuating kinetic energy at $Re=30$ and are clearly associated with the von K\'arm\'an street of shed vortices, as shown in  figure~\ref{Fig:POD30} (a)\&(b). For the construction of the ROM, POD modes $\bm{u}_{1,2}$ are preferred to the two leading eigenmodes, because we demand an accurate representation of the asymptotic periodic dynamics.
Following Eq.~\eqref{Eqn:MFSystem:Hopf:Antisymmetric}-\eqref{Eqn:MFManifold:Hopf:Symmetric}, the dynamical system resulting from the Galerkin projection of ansatz \eqref{Eq:HopfAnsatz} on the Navier-Stokes equations, after Krylov-Bogoliubov simplifications, reads
\begin{eqnarray}
	da_1/dt & = & \sigma a_1 - \omega a_2 \label{Eq:Hopf1}\\
	da_2/dt & = & \sigma a_2 + \omega a_1 \label{Eq:Hopf2}\\
	da_3/dt & = & \sigma_3\left(a_3 - \kappa (a_1^2 + a_2^2) \right), \label{Eq:Hopf3}
\end{eqnarray}
with $\sigma = \sigma _1 - \beta a_3$ and $\omega  = \omega _1 + \gamma a_3$. The value of the coefficients at $Re=30$ for the resulting ROM are summarized in table~\ref{Tab:Hopf}. Note that all coefficients but $\gamma$ and $\sigma_3$ are fixed by either the linear stability analysis or the asymptotic dynamics, see \S~\ref{Sec:Methodology}. The coefficient $\sigma_3$ can be chosen arbitrarily large as $a_3$ is slaved to $a_1,a_2$ (here we chose $\sigma_3=-10\sigma_1$), while $\gamma $ had to be calibrated in order to better match the asymptotic angular frequency. 

\begin{figure}
\begin{center}
\begin{tabular}{c}
 \includegraphics[width=0.75\textwidth]{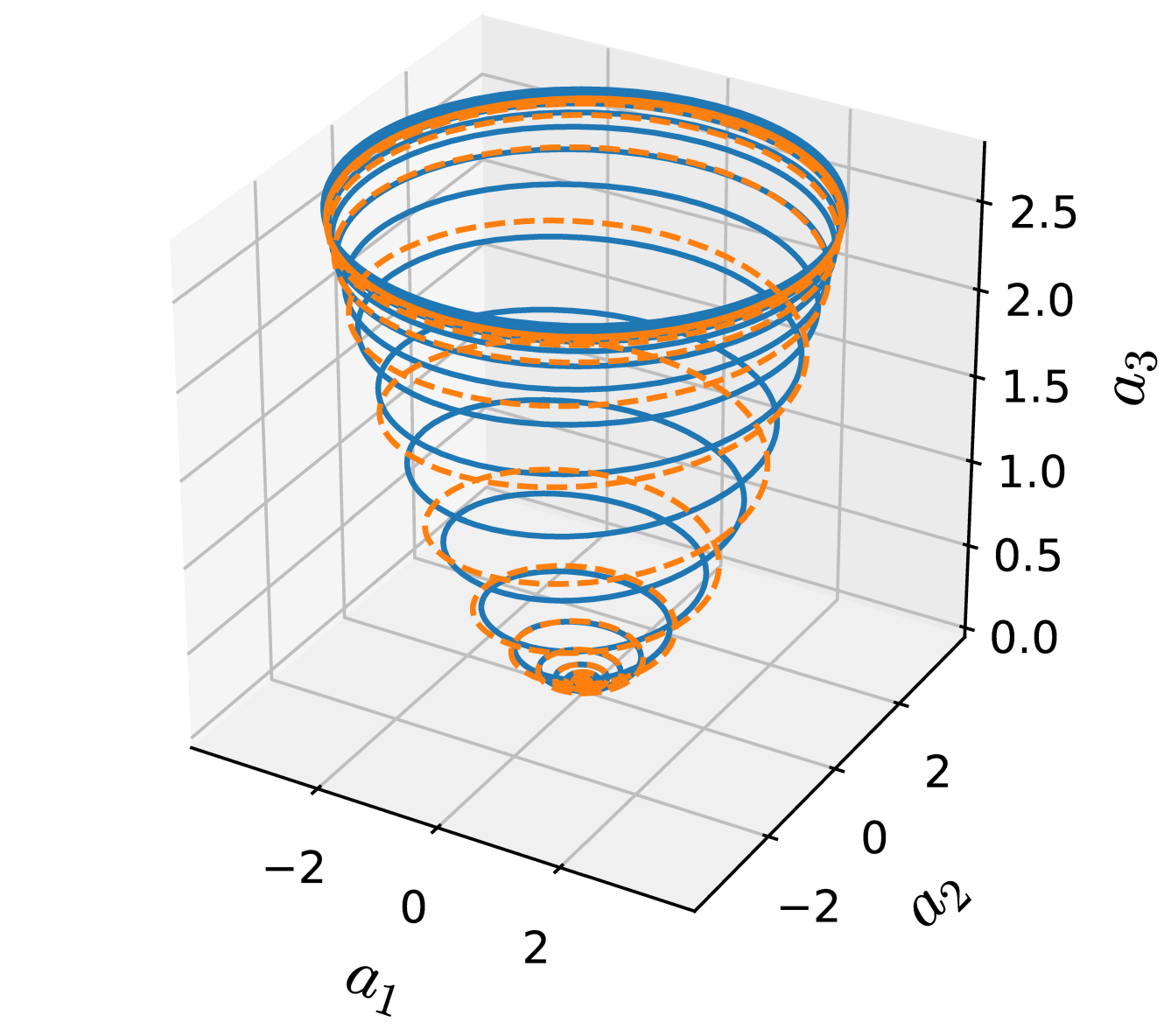} \\
 \includegraphics[width=0.9\textwidth]{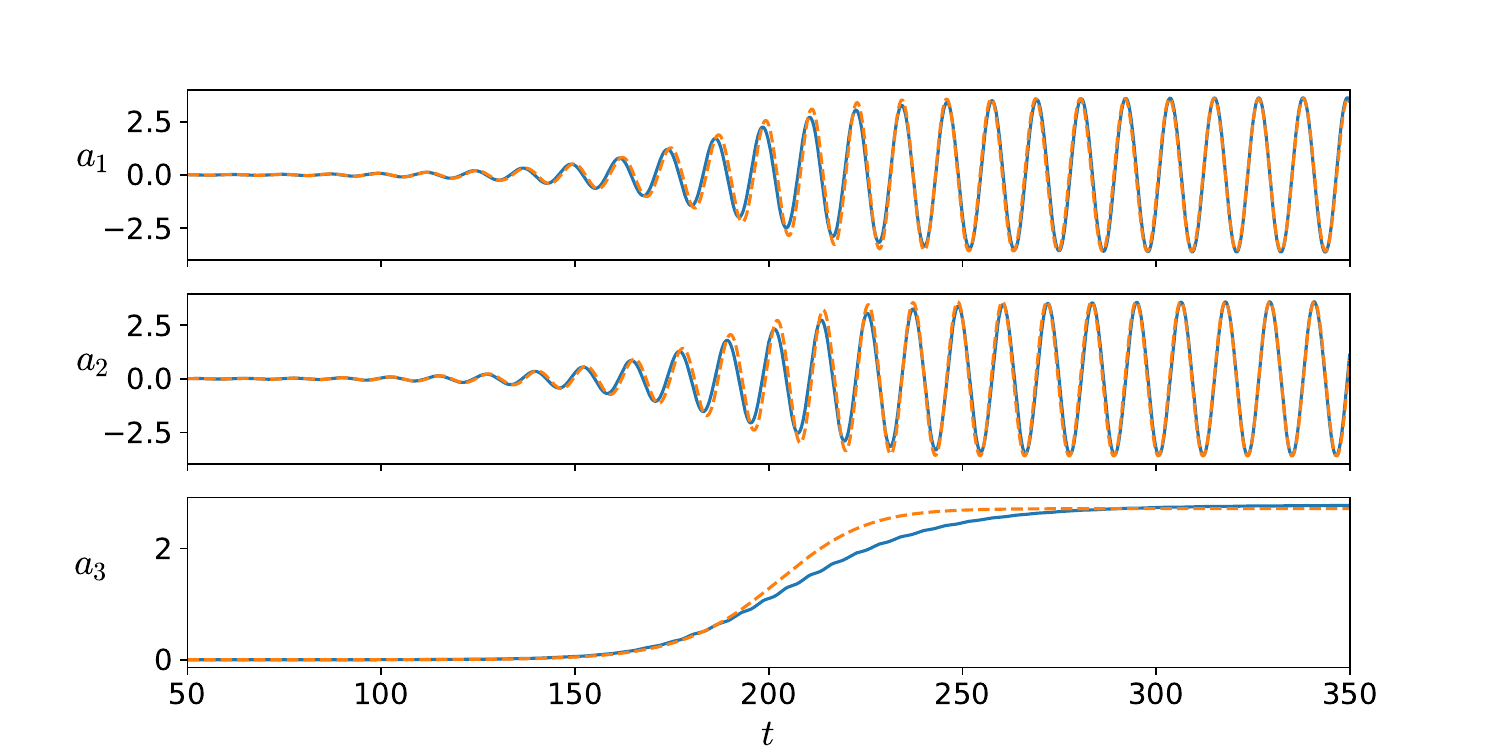} \\
\end{tabular}
\caption{Top figure: three-dimensional state space spanned by $a_1$, $a_2$, $a_3$,  at $Re=30$. 
From direct numerical simulations of the fluidic pinball (solid blue line) and from the mean-field reduced-order model (dashed red line). The initial condition, identical in both systems, starts close to the fixed point (steady solution) before evolving on the parabolic manifold $a_3 = \kappa (a_1^2+a_2^2)$ toward the asymptotic limit cycle.  Bottom figure: corresponding time series for $a_1$ to $a_3$. }
\label{Fig:Portrait30}
\end{center}
\end{figure}

The dynamics of both the fluidic pinball (solid blue curve) and the ROM (dashed red curve) are compared in the three-dimensional subspace spanned by $a_1$, $a_2$, $a_3$, see the top of figure~\ref{Fig:Portrait30}. In figure~\ref{Fig:Portrait30} are also shown the individual time series of $a_1$ to $a_3$ for both the fluidic pinball and the ROM (same representation). As expected from the POD modes $\bm{u}_{1,2}$, the dynamics on the asymptotic (permanent) limit cycle is well described in amplitude $r$ and angular frequency $\omega$ by the ROM. Moreover, the ROM also captures the transient dynamics on the parabolic manifold $a_3\equiv \kappa (a_1^2+a_2^2)$. 
Henceforth,  although all coefficients but one are fixed, the Galerkin system \eqref{Eq:HopfAnsatz} is able to reproduce the most salient dynamical features of the flow in both the transient and the permanent regimes. 

\section{Secondary flow regime}
\label{Sec:Secondary}

The secondary flow regime ranges over $Re_{2}<Re<Re_{3}$. For illustration, we focus on the flow at $Re=80$, \textit{i.e.} at a finite distance from the secondary bifurcation. In \S~\ref{Sec:SteadySolutions80} a linear stability analysis of the resulting three steady solutions is performed. 
A least-order model is proposed and discussed in \S~\ref{Sec:ROM80}.

\subsection{Eigenspectra of the steady solutions}
\label{Sec:SteadySolutions80}

As a result of the pitchfork bifurcation, there exist three steady solutions beyond $Re_{2}$: 
the symmetric steady solution $\bm{u}_s$, unstable to the periodic vortex shedding beyond $Re_{1}$, 
and two mirror-conjugated asymmetric steady solutions $\bm{u}_s^\pm$, also unstable to vortex shedding but only existing beyond $Re_{2}$, see figure~\ref{Fig:FlowStates}. 

\begin{figure}
\begin{tabular}{c  c }
(a) & (b) \\
\includegraphics[width=0.5\textwidth]{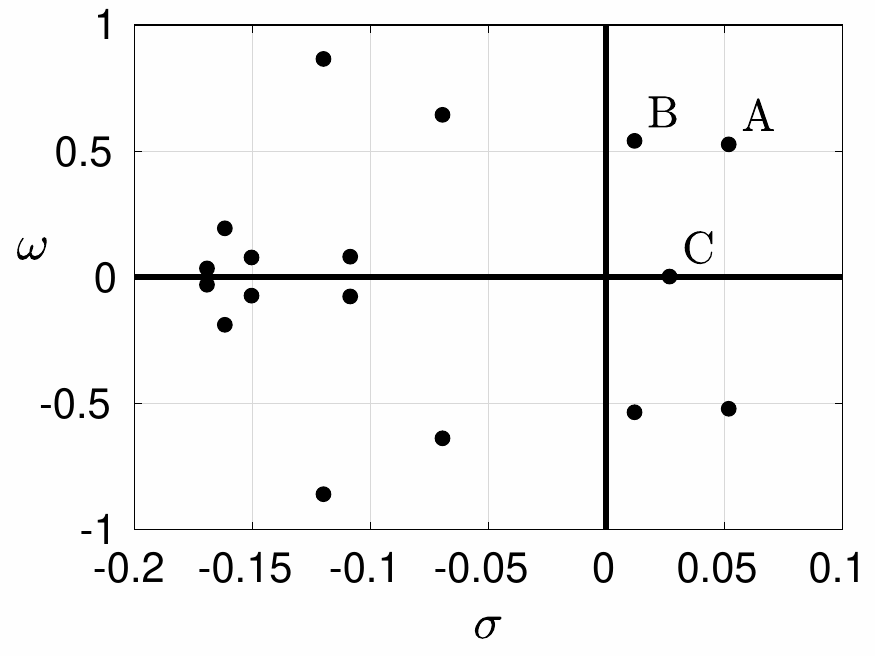} & \includegraphics[width=0.5\textwidth]{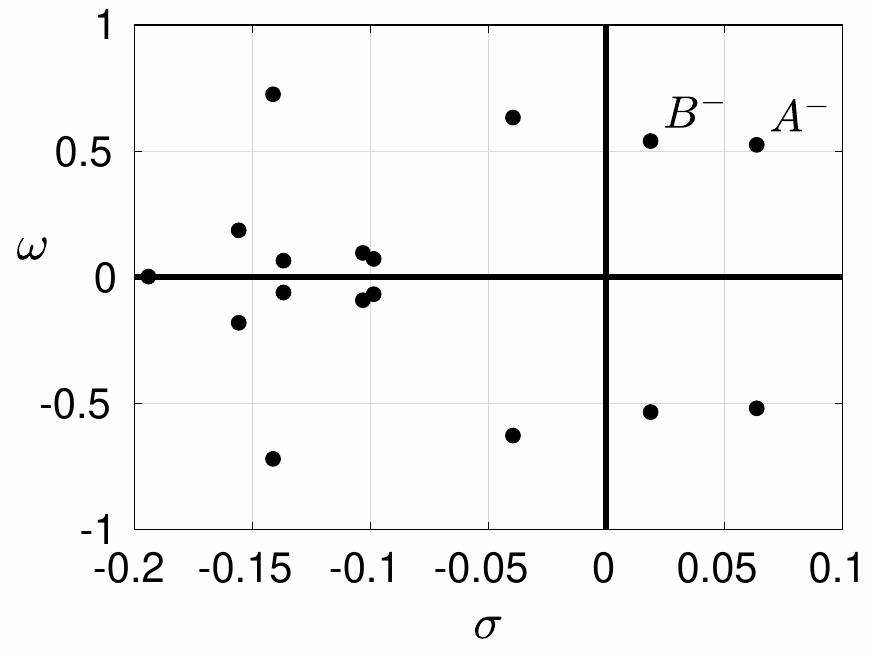} \\
%\end{tabular}
\begin{minipage}[t]{0.45\textwidth}
\begin{tabular}{c  c }
{A} & \raisebox{-0.5\height}{\includegraphics[width=0.9\textwidth]{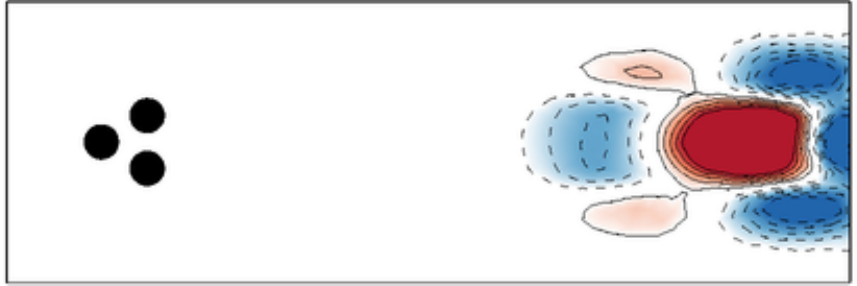}} \\
{B} & \raisebox{-0.5\height}{\includegraphics[width=0.9\textwidth]{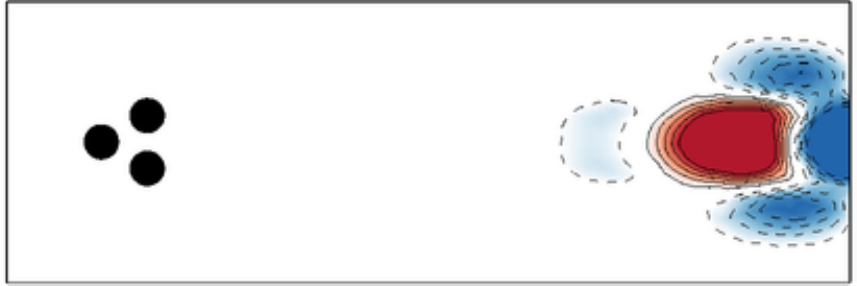}} \\
{C} & \raisebox{-0.5\height}{\includegraphics[width=0.9\textwidth]{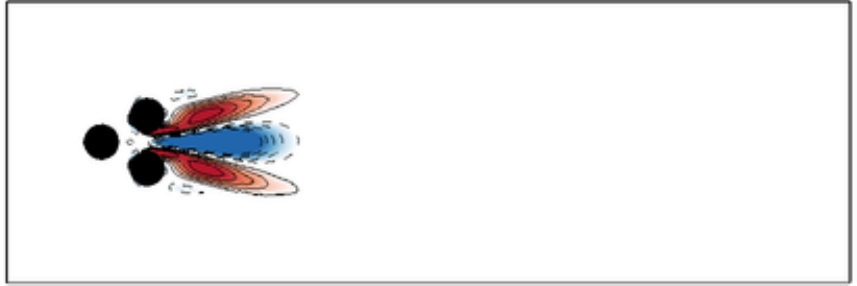}} \\
\end{tabular} 
\end{minipage} &
\begin{minipage}[t]{0.45\textwidth}
\raisebox{0.5\height}{
\begin{tabular}{c  c }
{A$^-$} & \raisebox{-0.5\height}{\includegraphics[width=0.9\textwidth]{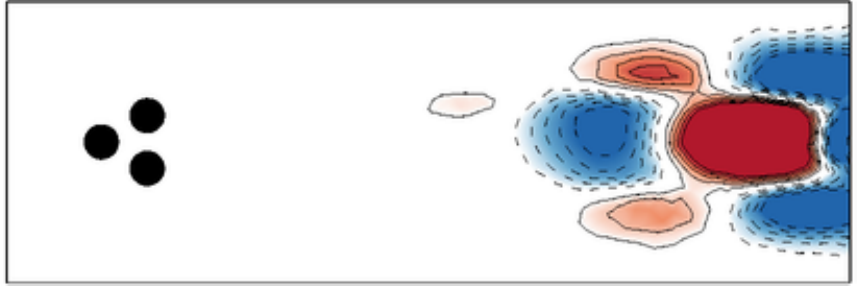}} \\
{B$^-$} & \raisebox{-0.5\height}{\includegraphics[width=0.9\textwidth]{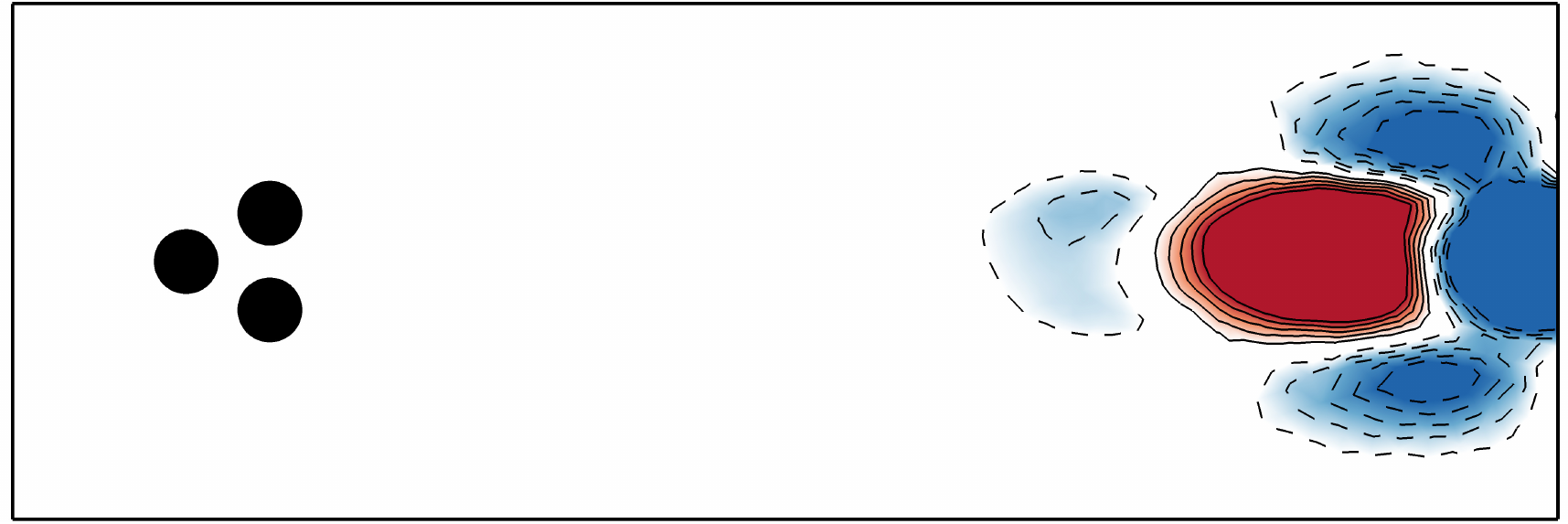}} \\
\end{tabular}
}
\end{minipage}
\end{tabular}
\caption{Eigenspectrum (top) and real part of the eigenvectors (bottom) of the symmetry-preserving steady solution $\bm{u}_s$ (left), 
of the symmetry-breaking steady solution $\bm{u}_s^-$ (right), both at $Re=80$. The red color and solid contours in the eigenvectors are positive values of the vorticity, blue color and dashed contours are negative values.}
\label{Fig:EigenModes80} 
\end{figure}

The linear stability analysis on $\bm{u}_s$ reveals two pairs of complex-conjugated eigenmodes with positive growth rate, and one eigenmode of zero frequency, see figure~\ref{Fig:EigenModes80}(a). The steady eigenmode is antisymmetric and reflects the symmetry broken by the pitchfork bifurcation. It is clearly associated with the base-bleeding jet, with all its energy concentrated in the near-field.  The two pairs of complex-conjugated eigenmodes are each associated with von K\'arm\'an streets of shed vortices. Both pairs of complex eigenmodes are antisymmetric and have quite similar angular frequencies. A closer view of the second pair of complex eigenmodes indicates that its growth rate cancels when the real eigenmode crosses the zero axis. This indicates that the new oscillatory mode is intimately connected to the symmetry breaking occurring at $Re_{2}$. At $Re>Re_{2}$, this gives rise to the only stable limit cycle for the flow dynamics, while the limit cycle associated with the leading pair of complex eigenmode has become unstable and can only be visited transiently in time. 

The linear stability analysis on $\bm{u}_s^-$ (resp. $\bm{u}_s^+$) reveals two pairs of complex-conjugated eigenmodes with positive growth rate, centered on an asymmetric mean flow, see figure~\ref{Fig:EigenModes80}(b). All eigenmodes are asymmetric, a property inherited from the steady solution. 

In the permanent regime, the mean flow field will inherit the symmetry of one of the three (unstable) steady solutions, depending on the details of the initial perturbation. 

\subsection{Reduced-order model in the secondary flow regime}
\label{Sec:ROM80}

\begin{figure}
\begin{center}
\begin{tabular}{c}
 (a) \\
 \includegraphics[width=0.5\linewidth]{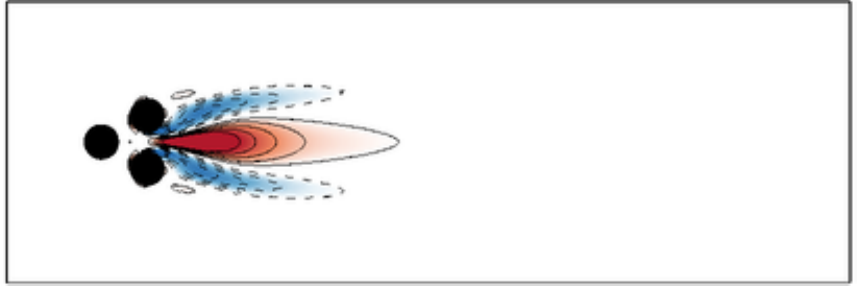}   \\
 (b) \\
 \includegraphics[width=0.5\linewidth]{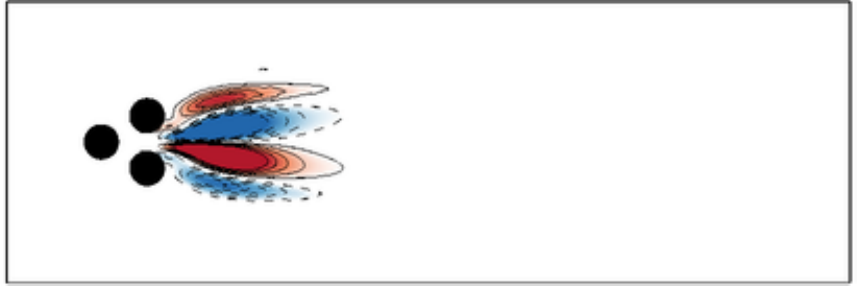}  \\ 
\end{tabular}
\caption{Additional modes arising from the pitchfork bifurcation, at $Re=80$, (a) mode $\bm{u}_4$, (b) mode $\bm{u}_5$. 
Red color and solid contours are positive values of the vorticity, blue color and dashed contours are negative values. 
See text for details about the computation of these two modes.}
\label{Fig:POD80}
\end{center}
\end{figure}

As discussed in \S~\ref{Sec:Methodology}, an ansatz of the flow state can now be written as:
\begin{equation}
 \bm{u}(\bm{x},t) \approx \underbrace{\bm{u}_s(\bm{x})} + \underbrace{a_1(t)\bm{u}_1(\bm{x}) + a_2(t)\bm{u}_2(\bm{x})}_{\textrm{leading POD modes at $Re=80$}} + \underbrace{a_3(t)\bm{u}_3(\bm{x}) }_{\textrm{shift mode}} + \underbrace{a_4(t)\bm{u}_4(\bm{x}) + a_5(t)\bm{u}_5(\bm{x}) }_{\textrm{pitchfork degrees of freedom}} 
 \label{eq:pitchforkreduction}
\end{equation}
It is worthwhile noticing that, in the frame of this ansatz, the two asymmetric steady solutions $\bm{u}_s^\pm$ are related to the symmetric steady solution $\bm{u}_s$ \textit{via} the additional antisymmetric mode $\bm{u}_4$:
\begin{equation}
 \label{Eq:AsymSteady}
 \bm{u}_s^\pm = \bm{u}_s \pm \overline{a}_4\bm{u}_4 + \overline{a}_5\bm{u}_5, 
\end{equation}
where $\overline{a}_4$ and $\overline{a}_5$ are the time-averaged coefficients in the permanent regime. 
Consequently, $\bm{u}_4$ can be easily computed as: 
\begin{equation}
 \label{Eq:Mode4}
 \bm{u}_4 \propto (\bm{u}_s^+ - \bm{u}_s^-),
\end{equation}
and further orthonormalized to $u_1$, $u_2$, $u_3$ by a Gram-Schmidt procedure. 
The resulting mode $\bm{u}_4$ is shown in figure~\ref{Fig:POD80}(a). A comparison with the eigenmode associated with the real eigenvalue, in figure~\ref{Fig:EigenModes80}(a), shows that the shift mode $\bm{u}_4$ is just the real eigenmode against which the symmetric steady solution is unstable at $Re=80$, as expected by the definition of mode $\bm{u}_4$.

In a similar way, the additional mode $\bm{u}_5$ can be constructed as: 
\begin{equation}
 \label{Eq:Mode5}
 \bm{u}_5 \propto (\bm{u}^+_s + \bm{u}^-_s)/2-\bm{u}_s.
\end{equation}
Mode $\bm{u}_5$ is shown in figure~\ref{Fig:POD80}(b) after orthonormalization. 

\begin{table}
 \begin{center}
 \begin{tabular}{lrlr}
 $\sigma_1$ & $5.22 \times 10^{-2}$ \quad & \quad $\beta$ & $1.31 \times 10^{-2}$ \\
 $\omega_1$ & $5.24 \times 10^{-1}$ \quad & \quad $\gamma$ & $2.95 \times 10^{-2}$ \\
 $\sigma_3$ & $-5.22 \times 10^{-1}$ \quad & \quad $\beta_{3}$ & $1.53 \times 10^{-1}$ \\
 $\sigma_4$ & $2.72 \times 10^{-2}$ \quad & \quad $\beta_{4}$ & $2.45 \times 10^{-1}$ \\
 $\sigma_5$ & $-2.72 \times 10^{-1}$ \quad & \quad $\beta_{5}$ & $2.14 \times 10^{-1}$ \\
 \end{tabular}
 \end{center}
 \caption{Coefficients of the least reduced-order model  \eqref{Eqn:MFSystem:Hopf+Pitchfork} at $Re=80$. }
 \label{Tab:leastROM80}
\end{table}

\begin{figure}
\begin{center}
 \includegraphics[width=.9\textwidth]{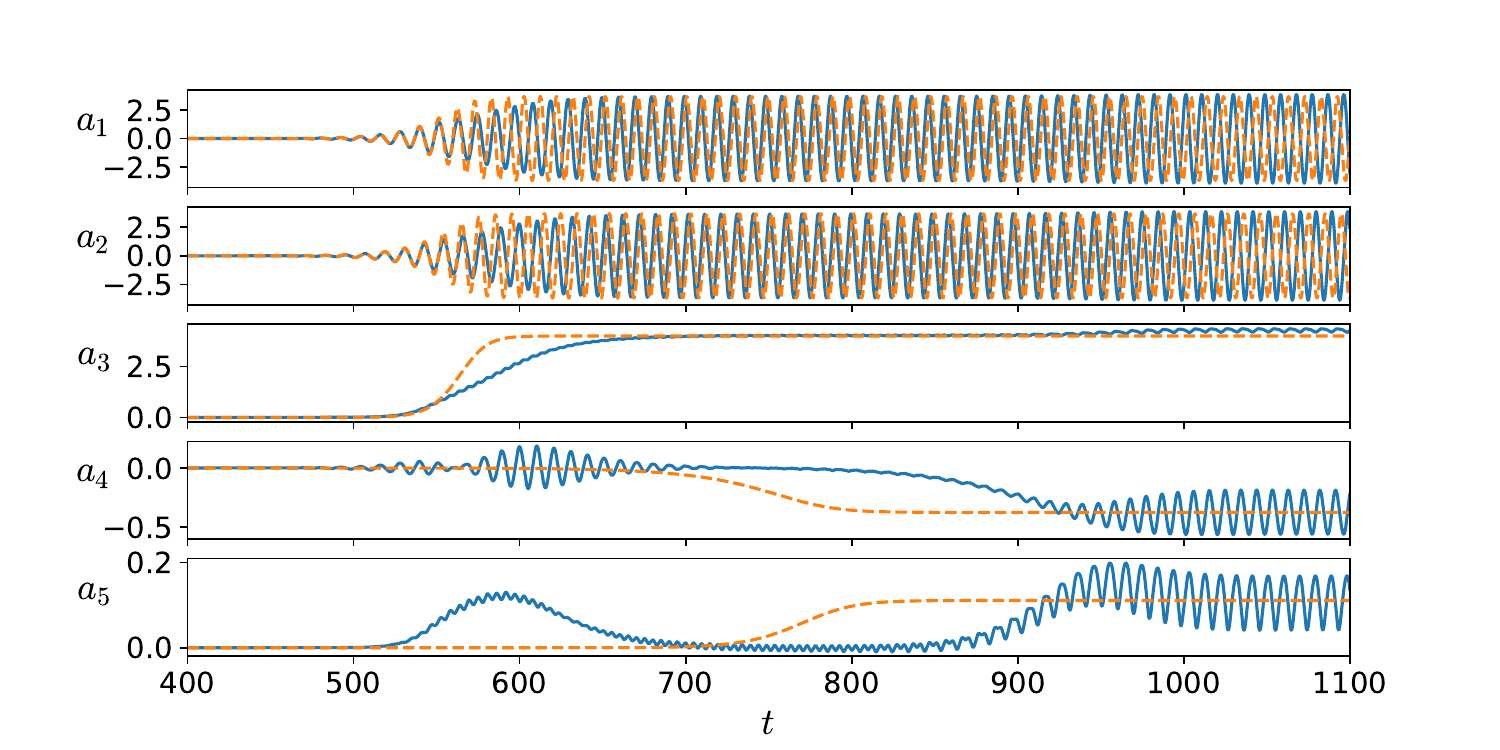}  
\caption{Time evolution of coefficients $a_1$ to $a_5$ in the full flow dynamics (solid blue line) and the ROM (dashed red line) without cross-terms and coefficients fixed by the linear stability analysis and the asymptotic dynamics. The initial condition for both systems is the same.  }
\label{Fig:leastROM80} 
\end{center}
\end{figure}

Close to the pitchfork bifurcation, the resulting dynamical system is described by Eqs.~\eqref{Eqn:MFSystem:Hopf+Pitchfork}. 
At the threshold, the degrees of freedom $a_4$, $a_5$ associated with the pitchfork bifurcation are expected to be fully uncoupled to the degrees of freedom $a_1$, $a_2$, $a_3$ associated with the Hopf bifurcation, and reciprocally, see \S~\ref{Sec:Hopf}. 
In this case, an accurate linear and nonlinear dynamics from a Galerkin projection relies on deformable modes from eigenmodes near the fixed point to POD modes near the limit cycle \citep{Loiseau2018jfm}. 
We avoid this complication by model identification from simulation data. 
The coefficients of the mean-field system \eqref{Eqn:MFSystem:Hopf+Pitchfork} 
reported in table~\ref{Tab:leastROM80} were identified 
from the linear stability analysis of the symmetric steady solution 
and the asymptotic dynamics on the unstable symmetric preserving limit cycle 
and the stable symmetric-breaking limit cycle
(see \S~\ref{Sec:Hopf} to \ref{Sec:MultipleBifurcation}). 
The resulting ROM dynamics is compared to the  flow dynamics in figure~\ref{Fig:leastROM80}.
Inspection of figure~\ref{Fig:leastROM80} shows that such a model, reduced to only five degrees of freedom, is able to reproduce many features of the original dynamics: both the early transient and asymptotic dynamics of $a_1$ to $a_3$ are well reproduced, as well as the large timescale evolution of $a_4,a_5$. However, the growth of coefficients $a_1$ to $a_3$ appears to be faster for the ROM, and $a_4$, $a_5$ also reach their asymptotic value significantly sooner than their values for the full simulations of the fluidic pinball. In addition, the transient kick in $a_4,a_5$ is absent from the mean-field system~\eqref{Eqn:MFSystem:Hopf+Pitchfork}, as well as the transient and asymptotic oscillations of $a_4$ and $a_5$ visible in figure~\ref{Fig:leastROM80}, which would require coupling to $a_1$ or $a_2$, or both. All these features indicate that at Reynolds number $Re=80$, the Krylov-Bogoliubov assumption of pure harmonic behaviour with slowly varying amplitude and frequency no longer holds. 

As a consequence, cross-terms must be included in the ROM. 
The assumption 
of non-oscillatory dynamics of the shift mode amplitude $a_3$ 
and of the two pitchfork modes  $a_4$, $a_5$ is relaxed 
to reproduce the oscillatory behaviour evidenced in figure~\ref{Fig:leastROM80}. 
Following \S~\ref{Sec:Galerkin+Symmetries}, 
the model identification process reads:
\begin{enumerate}[Step 1: ]
\item Keep the five-dimensional linear-quadratic form of the dynamical system from the Galerkin projection
with 25 linear and 75 quadratic terms.
\item Remove vanishing terms arising from the symmetry of the modes. 
Thus, only 13 linear and 36 quadratic terms are left to be determined.
\item Enforce the linear dynamics of the unstable Hopf and pitchfork eigenmodes from stability analysis
in the Galerkin system.
This implies that the growth rate $\sigma_1$ and frequency $\omega_1$ characterize
the initial growth and angular frequency of  $a_1$, $a_2$
and $\sigma_4$ represents the linear growth rate for $a_4$. 
\item The slaving of the Reynolds-stress-induced  modes $\bm{u}_3$ and $\bm{u}_5$
to the fluctuation level is imposed by setting the damping rate 
10 times larger than the growth rate of the corresponding fluctuation:
$\sigma_3 = -10 \sigma_1$, $\sigma_5 = -10 \sigma_4$.
This strong damping rate quickly forces the trajectory onto the mean-field manifold.
\item Enforce  phase invariance for $a_1$, $a_2$ in the first two equations.
This is implied by the mean-field theory and is found to be a good approximation from numerical inspection. 
Thus, the oscillatory dynamics of $a_1$, $a_2$ 
are governed by $\sigma_1$, $\omega_1$, $\beta$, $\beta_{15}$, $\gamma$, $\gamma_{15}$, and $\beta_3$.
\item Impose the asymptotic dynamics of the unstable symmetric limit cycle by fixing $\beta$, $\gamma$, and $\beta_3$.
\item Apply the SINDy algorithm \citep{brunton2016pnas} to the remaining unknown terms,
i.e.\ 6 linear terms and 28 quadratic terms.
This step alone typically fails 
to yield a physics-based globally stable Galerkin system.
This is not surprising in view of 
the necessary and sufficient conditions for global boundedness of Galerkin systems 
by \citet{schlegel2015jfm}. 
A  Galerkin system identification 
with $\ell1$-norm penalization of the coefficients
is found to have a performance similar to SINDy
for the chosen constraints.
\item Two additional simplifying physics-based constraints 
are found to make the dynamic system identified by SINDy robust for a large range of initial conditions. Enforcing $q_{145}=q_{245}=0$ avoids the initial oscillatory dynamics being influenced by the initial kick of $a_{5}$, and setting $q_{344} = q_{434} =0$ leads to the right asymptotic transition of the pitchfork bifurcation.
\end{enumerate}

\begin{figure}
\begin{center}
 \includegraphics[width=.9\textwidth]{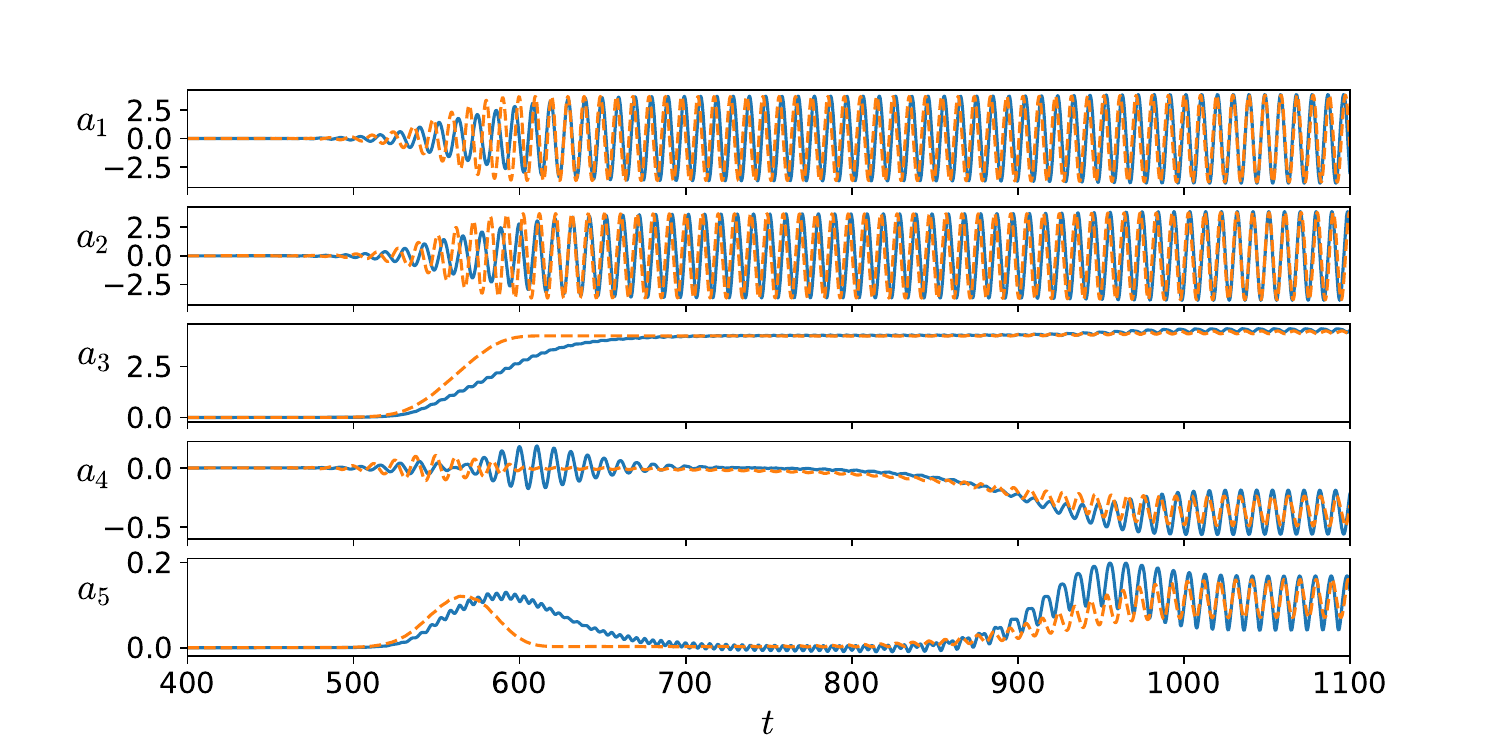}  
\caption{Performance of the ROM with cross-terms. Time evolution of coefficients $a_1$ to $a_5$ in the full flow dynamics (solid blue line) and for the ROM (red dashed line). The initial condition is the same for the ROM and the full flow dynamics. }
\label{Fig:TS-Re80} 
\end{center}
\end{figure}

The resulting ROM reads:
\begin{subequations}
\label{Eqn:Hopf+Pitchfork}
\begin{eqnarray}
	da_1/dt & = & a_1(\sigma_1 - \beta\ a_3 - \beta_{15}\ a_5) - a_2(\omega_1 + \gamma\ a_3 + \gamma_{15}\ a_5) + l_{14}\ a_4 + q_{134}\ a_3a_4,\\
	da_2/dt & = & a_2(\sigma_1 - \beta\ a_3  - \beta_{15}\ a_5) + a_1(\omega_1 + \gamma\ a_3 + \gamma_{15}\ a_5) + l_{24}\ a_4 + q_{234}\ a_3a_4,\\
	da_3/dt & = & \sigma_3\ a_3 +\beta_3\ r^2 + l_{35}\ a_5 + q_{314}\ a_1a_4 + q_{335}\ a_3a_5 + q_{355}\  a_5^2, \\
	da_4/dt & = & \sigma_4\ a_4 - \beta_4\ a_4a_5 + a_1(l_{41} + q_{413}\ a_3 + q_{415}\ a_5) + a_2(l_{42} + q_{423}\ a_3 + q_{425}\ a_5), \qquad \qquad   \\
	da_5/dt & = & \sigma_5\ a_5 + \beta_{5}\ a_4^2 + l_{53}\ a_3 + q_{514}\ a_1a_4 + q_{533}\ a_3^2 + q_{535}\ a_3a_5.
\end{eqnarray}
\end{subequations}
where $r^2=a_1^2+a_2^2$. The coefficients are summarized in table~\ref{Tab:Pitchfork}. The dynamics of the system~\eqref{Eqn:Hopf+Pitchfork} (dashed red line) is compared to the full flow dynamics of the fluidic pinball (solid blue line) in figure~\ref{Fig:TS-Re80}. Now the initial stage of the dynamics is much better reproduced, as well as the asymptotic oscillations of $a_3$, $a_4$ and $a_5$. Interestingly, the faster growth in $a_1$ to $a_3$, on the time-range around 600, could not be completely corrected. It is worthwhile noticing that this range of time also corresponds to oscillations in $a_4$, which could not be reproduced by any cross-terms compatible with the symmetries of the system. Noticing that $a_4\neq 0$ on this range of time may also question our choice for $\sigma_1=5.22\times 10^{-2}$, which is the linear growth rate of the leading eigenmode around the symmetric steady solution $\bm{u}_s$. Indeed, although the initial condition is close to this point, the large amplitude oscillations of $a_4$ on the time range around 600 mean that the trajectory transiently escapes the symmetric subspace not only along $a_3$, but also along $a_4$ and $a_5$, before coming back close to the $a_3$ axis, on the time range from 700 to about 800.
Therefore, it would be reasonable here to keep all the coefficients of the model unconstrained, but the number of free parameters is now too large for the identification process to be computationally tractable --- for instance, it provides positive $\sigma_3$ when it must necessarily be strongly negative. 

Model identification can be very challenging for a number of reasons.
First, the conditions for global boundedness of the attractor 
for the linear-quadratic Galerkin system
are rather restrictive~\citep{schlegel2015jfm}. 
Second, the Galerkin method assumes fixed expansion modes.
Yet, least-order models often have deformable modes changing with the fluctuation level
\citep{Tadmor2011ptrsa}. 
The von K\'arm\'an vortex shedding metamorphosis 
from stability modes to POD modes may serve as an example.
These deformations may also affect the structure of the dynamical system.
Third, the Navier-Stokes dynamics may live on a strongly attracting manifold.
This restriction of the state space may make certain Galerkin system coefficients numerically unobservable,
like $\sigma_3$ and $\sigma_5$ in our case. 
Despite these challenges,
the model identified in table~\ref{Tab:Pitchfork}, as exemplified by figure~\ref{Fig:TS-Re80}, constitutes a faithful least-order model for the fluidic pinball at $Re=80$. 
Although the model is only five-dimensional, 
it can reproduce most of the key features, timescales, transient and asymptotic behaviour of the full dynamics. 
%Of course, this is not coincidental for the reasons developed in \S~\ref{Sec:Methodology}. 
\begin{table}
 \begin{center}
 \begin{tabular}{lrlrlrlr}
  $\sigma_1$ & $5.22 \times 10^{-2}$ \quad & \quad $\beta$ & $1.31 \times 10^{-2}$ \quad & \quad $l_{14}$ & $2.93 \times 10^{-1}$ \quad & \quad $l_{24}$ & $-4.87 \times 10^{-1}$ \\
 $\omega _1$ & $5.24 \times 10^{-1}$ \quad & \quad $\gamma$ & $2.95 \times 10^{-2}$ \quad & \quad $q_{134}$ & $-5.87 \times 10^{-2}$ \quad & \quad $q_{234}$ & $1.18 \times 10^{-1}$ \\
 $\sigma_3$ & $-5.22 \times 10^{-1}$ \quad & \quad $\beta_{3}$ & $1.53 \times 10^{-1}$ \quad & \quad $l_{41}$ & $3.14 \times 10^{-2}$ \quad & \quad $l_{42}$ & $-5.14 \times 10^{-2}$ \\
 $\sigma_4$ & $2.72 \times 10^{-2}$ \quad & \quad $\beta_{4}$ & $5.78 \times 10^{-2}$ \quad & \quad $q_{413}$ & $-7.56 \times 10^{-3}$ \quad & \quad $q_{423}$ & $1.28 \times 10^{-2}$ \\
 $\sigma_5$ & $-2.72 \times 10^{-1}$ \quad & \quad $\beta_{5}$ & $1.91 \times 10^{-1}$ \quad & \quad $q_{415}$ & $2.99 \times 10^{-2}$ \quad & \quad $q_{425}$ & $1.71 \times 10^{-1}$ \\
 & & \quad $\beta_{15}$ & $-2.42 \times 10^{-2}$ \quad & \quad $l_{35}$ & \quad $4.28$ \quad & \quad $l_{53}$ & $2.89 \times 10^{-2}$ \\
 & & \quad $\gamma_{15}$ & $1.70 \times 10^{-2}$ \quad & \quad $q_{335}$ & $-1.11$ \quad & \quad $q_{533}$ & $-7.22 \times 10^{-3}$\\
 & & & & \quad $q_{355}$ & $-5.13 \times 10^{-1}$ \quad & \quad $q_{535}$ & $1.48 \times 10^{-2}$ \\
 & & & & \quad $q_{314}$ & $1.57 \times 10^{-2}$ \quad & \quad $q_{514}$ & $-9.44 \times 10^{-3}$ \\ 
 \end{tabular}
 \end{center}
 \caption{Coefficients of the reduced-order model at $Re=80$. See text for details.}
 \label{Tab:Pitchfork}
\end{table}

\section{Conclusions and outlooks}
\label{Sec:Conclusions}

Reduced-order models (ROM) serve a number of purposes.
For instance, ROMs facilitate a deeper understanding of the physical mechanisms at play in a flow configuration, by extracting the low-dimensional manifold on which evolve the dynamics \citep{Manneville2010book}. 
In that respect, 
the linear stability analysis of the steady solution shows the nonlinear amplitude saturation mechanism through the distorted mean flow.
The difference between steady solution and the mean flow
is caused by the Reynolds stress, captured by the shift mode $\bm{u}_\Delta$, 
and affects the stability properties 
\citep{Noack2003jfm,barkley2006europhys,Sipp2007jfm,turton2015pre}. 
Deeper investigations would certainly deserve to be carried out on the relation between the stability analysis of fixed points, Floquet analysis of limit cycles and Lyapunov exponents of chaotic flow regimes. 
In addition, because Hopf and pitchfork bifurcations are generic bifurcations in fluid flows, the nonlinear dynamics identified in this study are expected to be extended 
and generalizable to other flows exhibiting similar bifurcations. 
Last but not least, a ROM provides fast estimators for predicting the forward evolution of the system. Such estimators could be used for control purposes \citep{Brunton2015amr,rowley2017arfm}.  All these considerations motivated the present study, 
whose main results are summarized in \S~\ref{Sec:Discussion}. 
Outlooks of this work are listed in \S~\ref{Sec:Outlook}.

\subsection{Concluding remarks and discussion}
\label{Sec:Discussion}

Flow configurations undergoing successive Hopf and pitchfork bifurcations are common in fluid mechanics. This is, for instance, the case of three-dimensional wake flows such as spheres \citep{mittal_AIAA1999,gumowski_PRE2008,szaltys_JFS2012,Grandemange_EXIF2014} or bluff body wake flows \citep{grandemange_PRE2012,Grandemange2013jfm,cadot_PRE2015,bonnavion_JFM2018,rigas_JFM2014}. 
The drag crisis and stalled flows are also characterized by the pitchfork bifurcation of a primarily Hopf-bifurcated flow, but the secondary transition is subcritical in this case. 

In this study, we have considered the fluidic pinball on its way to chaos 
and have identified least-order models of the flow dynamics in the primary Hopf-bifurcated and secondary pitchfork-bifurcated flow regimes. 
Reduced-order modeling of Hopf bifurcations was already addressed in \citet{Noack2003jfm} for the cylinder wake flow, while \citet{Meliga2009jfm} derived the amplitude equation for a codimension two bifurcation (pitchfork and Hopf) based on a weakly nonlinear analysis in the wake of a disk, 
and \citet{Fabre2008pof} derived the same equation solely  based on symmetry arguments, in the wake of axisymmetric bodies. In the present contribution, we could demonstrate that the dynamics resulting from the successive Hopf and pitchfork bifurcation could be well-captured by a five-dimensional model whose degrees of freedom couple through quadratic non-linearities, as imposed by the Navier-Stokes equations.

For the fluidic pinball, the route to chaos is characterized by a primary supercritical Hopf bifurcation at $Re\approx 18$, followed by a secondary supercritical pitchfork bifurcation at $Re\approx  68$. 
The Hopf bifurcation corresponds to the destabilization of the steady solution 
with respect to vortex shedding, while the pitchfork bifurcation occurs when the mean flow breaks the symmetry with respect to the mirror-plane. 
The fluctuation amplitude of the von K\'arm\'an street is reduced, over a finite range of the Reynolds number around $Re_{2}$, when the base-bleeding jet is rising. 
This means that energy is withdrawn from the fluctuations 
to feed the mean flow transformation. 
Before the next transition occurs, the fluctuation amplitude starts to grow again, up to the largest value of the Reynolds number considered in this work. 

\begin{figure}
\begin{center}
\includegraphics[height=.95\linewidth,angle=90]{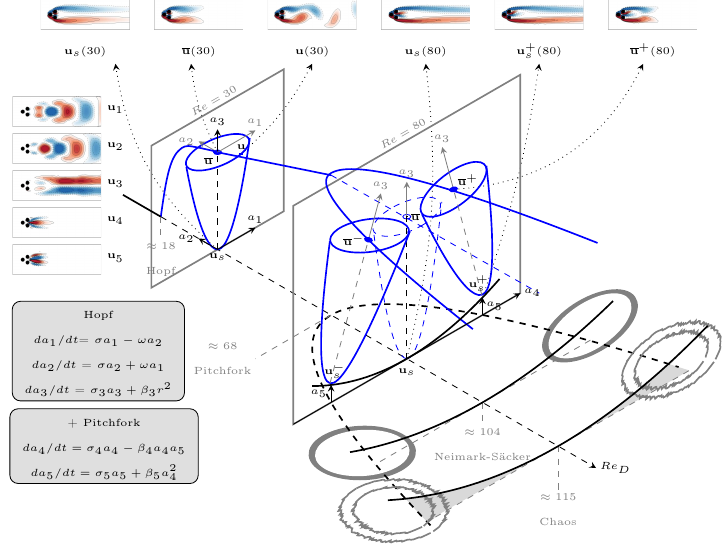} 
\caption{Phenomenogram. The route to chaos develops along the $Re_D$-axis. 
On this route are highlighted the least-order models identified at $Re=30$ and $Re=80$,
where the manifolds on which the dynamics take place are schematically drawn. The branches of the steady solution and periodic solution are presented by the black and blue curves, with the solid/dashed curves for the stable/unstable states.
The degrees of freedom $a_1$,$a_2$ span the limit cycle subspace,
$a_3$ is the axis of the parabolically-shaped manifold, while $a_4$ is transversally associated with the pitchfork bifurcation, together with $a_5$ which slightly bend the surface to which the steady solutions $\bm{u}_s^\pm$ and $\bm{u}_s$ belong. The gray-shaded shadows of the quasi-periodic and chaotic regimes are represented for the sake of illustration. Gray-shaded inserts provide the least-reduced order models at the threshold for both the Hopf and pitchfork bifurcations under the constraint of the Navier-Stokes equations. 
Also shown at the top are the figure snapshots of the steady solutions, mean flow fields, and instantaneous flow field, at either $Re=30$ or $Re=80$. 
The dotted arrows connect these snapshots to the corresponding points in the phenomenogram. 
In addition, individual degrees of freedom $\bm{u}_1$ to $\bm{u}_5$ are shown at the left.}
\label{Fig:Phenomenogram}
\end{center}
\end{figure}

There is strong evidence that the next transition is a Neimark-S\"acker bifurcation. 
The resulting flow regime is most likely quasi-periodic over the range $[Re_{_3},Re_{4}]$. In this regime, a new oscillatory phenomenon takes place, 
characterized by slow oscillations of the base-bleeding jet. 
Three additional degrees of freedom might be necessary to deal with the newly arising oscillator.
The flow dynamics eventually bifurcates into a chaotic regime, characterized by the random switching of the base-bleeding jet between two symmetric deflected positions. 
The overall route to chaos is summarized in the phenomenogram of figure~\ref{Fig:Phenomenogram}. 

The reduced-order models derived by Galerkin projections of the Navier-Stokes equations, based on the symmetry of the individual degrees of freedom, under Krylov-Bogoliubov simplifications, faithfully extract the manifolds on which the flow dynamics sets in. 
The ROM for the primary flow regime is only three dimensional: 
two degrees of freedom are associated with the asymptotic stable limit cycle resulting from the vortex shedding. 
The third degree of freedom is a mode slaved to the two dominant modes 
and is mandatory for the description of the transient flow dynamics from the unstable steady solution to the post-transient mean flow, 
as already demonstrated in \citet{Noack2003jfm}. 
The least-order model in the secondary flow regime has only five degrees of freedom, 
three of which are associated with the Hopf bifurcation,
the two remaining degrees of freedom being associated with the pitchfork bifurcation. 
In the phenomenogram of figure~\ref{Fig:Phenomenogram} are reported the structure of both reduced-order models close to the Hopf and the pitchfork bifurcations. When the two sets of degrees of freedom are fully uncoupled, some features of the flow dynamics are well-reproduced (asymptotic mean behaviour, parabolic manifolds of the Hopf and pitchfork bifurcations), but many details are missing. 
To reproduce most of the transient and asymptotic flow features far from the bifurcation point, 
additional cross-terms have been included in the model, 
which relaxes the steadiness constraint usually assumed for the shift modes.  

\subsection{Outlook}
\label{Sec:Outlook}
The current generalized mean-field model captures the Hopf bifurcation 
and subsequent pitchfork bifurcation of the steady solution and limit cycles.
The following onset of a quasi-periodic regime with slow oscillations
of the deflected base-bleeding jet might presumably be incorporated by another Hopf bifurcation,
leading to a 8-dimensional mean-field Galerkin model.
The transition to chaos is accompanied by a return to a statistically symmetric flow,
i.e.\ the base-bleeding jet oscillates around one asymmetric state 
before it stochastically switches to the other mirror-symmetric one.
This behaviour is reminiscent of the transition to chaos of a harmonically forced Duffing oscillator.
In the case of the fluidic pinball, 
the vortex shedding would constitute a forcing.
Hence, one may speculate that the transition to chaos 
may already be resolved by the 8-dimensional Galerkin model
in which the effect of vortex shedding on the jet oscillation becomes stronger with an increasing Reynolds number.

An alternative direction is to increase the accuracy of the mean-field Galerkin model.
While the structure of the Galerkin system prevails for a large range of Reynolds numbers,
the modes and all Galerkin system coefficients change in a non-trivial manner,
e.g., the growth-rate formula should read $ \sigma = \sigma_1(Re) - \beta (Re) a_3$.
The transients can be expected to be much more accurately resolved by the mean flow dependent modes,
e.g.\ $\bm{u}_i(Re, a_3, a_5, \vec{x} )$, $i=1,2$ for the resolution of vortex shedding \citep{Loiseau2018jfm}.
More generally, the flow lives on a low-dimensional manifold which includes mode deformations \citep{Noack2016jfm2}.
Locally linear embedding (LLE) is a powerful technique for identifying the dimension and a parameterization
in an automatic manner \citep{Roweis2000s}.
The normal form of the bifurcations can be expected to coincide with the dynamics on the LLE feature coordinates.

A third direction follows an observation of \citet{Rempfer1994pf}
that Galerkin systems of many fluid flows can be considered as nonlinearly coupled oscillators.
For two incommensurable shedding frequencies, this observation has been formalized 
in a generalized mean-field model by \citet{Noack2008jnet,Luchtenburg2009jfm}.
Such multi-frequency models may be extended to resolve  broadband frequency dynamics 
taking, for instance, the most dominant DMD modes \citep{Rowley2009jfm,Schmid2010jfm}. Strengths and weaknesses of techniques currently used for model reduction are discussed in \citet{taira_AIAA2017}. 
While mean-field consideration expressly ignores non-trivial triadic interactions,
their quantitative effect on the frequency crosstalk may still be well approximated by the mean flow interaction terms.
Such multi-frequency mean-field models may eventually describe the effect of open-loop forcing on turbulence, see for instance the recent thorough review by \citet{jimenez2018jfm}  on  turbulent flow modeling. 

A fourth direction aligned with the large success of machine learning / artificial intelligence
is the automated learning of state spaces, modes, and dynamical systems.
For the latter, SINDy provides an established elegant framework \citep{brunton2016pnas}.
The choice of the state spaces might be facilitated by manifold learning from many solution snapshots \citep{Gorban2005book}.
The authors actively pursue all the mentioned directions.

\section*{Acknowledgements}

This work is supported
by a public grant overseen by the French National Research Agency (ANR)
as part of the ``Investissement d'Avenir'' program, through the  ``iCODE Institute project''
funded by the IDEX Paris-Saclay, ANR-11-IDEX-0003-02,
by the ANR grants `ACTIV\_ROAD' and `FlowCon' (ANR-17-ASTR-0022),
and by
Polish Ministry of Science and Higher Education
(MNiSW) under the Grant
No.: 05/54/DSPB/6492.

We appreciate valuable stimulating discussions with
Steven Brunton, Alessandro Bucci, Nathan Kutz, Onofrio Semeraro, Yohann Duguet, Laurette Tuckerman and 
the French-German-Canadian-American pinball team:
Fran\c{c}ois Lusseyran, Guy Cornejo-Maceda, Jean-Christophe Loiseau, Robert Martinuzzi, Cedric Raibaudo, Richard Semaan, and Arthur Ehlert.

\begin{appendix}
\section{Asymmetric steady solutions}
\label{Sec:AsymSS}
%As mentioned in \S~\ref{Sec:DNS}, the steady solution is computed by a Newton-Raphson method of the discretized steady Naiver-Stokes equations. This steady solution solver is integrated into the Navier-Stokes solver as an optional function called ``steady computation'' \citep{Noack2017put}.
For a Reynolds number larger than the critical value of the pitchfork bifurcation $Re_2$, we can obtain two additional asymmetric steady solutions, one associated with the base-bleeding jet deflected upward, the other downward. These two asymmetric steady solutions are obtained by the steady Navier-Stokes solver initialized with a flow field (snapshot) with the same state of the base-bleeding jet.
From the two asymmetric vortex shedding, their corresponding asymmetric steady solutions can be obtained by the following steps:
\begin{enumerate}[Step 1: ]
\item Run the unsteady Navier-Stokes solver with a time scale larger than the vortex shedding period, initialized from a snapshot of the asymmetric vortex shedding. The vortex shedding will quickly vanish for this artificially large time scale and approach the corresponding approximation of the steady state.
\item Run the steady Navier-Stokes solver, restarted from this vortex shedding vanished solution to further refine the steady state.
\end{enumerate}

\begin{figure}
\centering
\includegraphics[width=0.45\textwidth]{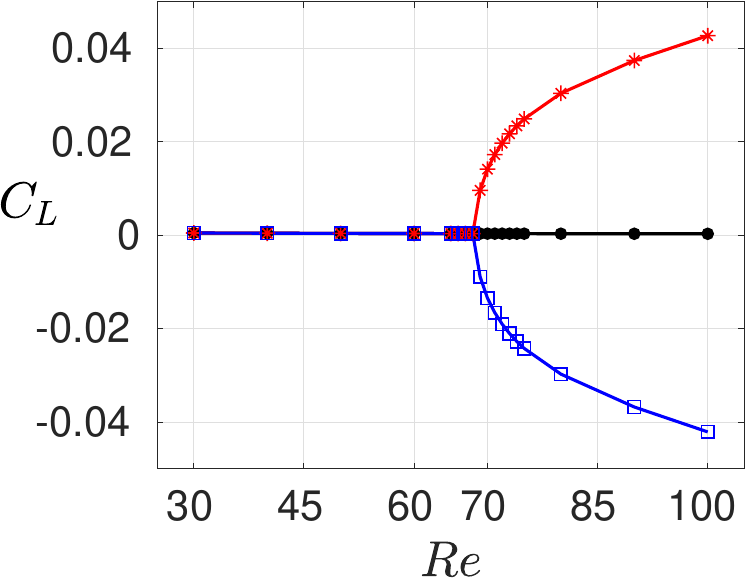}
\caption{Pressure lift coefficient of steady solutions at different values of the Reynolds number resulting from the steady Navier-Stokes solver starting with the steady solutions: $\bm{u}_s(\bm{x})$ (point + black curve),  $\bm{u}_s^+(\bm{x})$ (star + red curve), and $\bm{u}_s^-(\bm{x})$ (square + blue line), at $Re=80$. Three curves overlap on the $C_L=0$ level as $Re\leq68$.}
\label{Fig:SS_Bif_Lift}
\end{figure}

Initialized with the three steady solutions with different states of the base-bleeding jet at $Re=80$, the steady solutions for other Reynolds numbers can be obtained by the steady Navier-Stokes solver.
%Initialized with an asymmetric steady solution for a given Reynolds number are prepared, we can easily compute the steady solution with the same base-bleeding jet state for the other Reynolds numbers. Starting with the three steady solutions with different states of the base-bleeding jet at $Re=80$, we do the steady computation at the other values of Reynolds number respectively. 
At $Re=68$, all three steady solutions converge to a unique solution, which indicates that the critical value of the pitchfork bifurcation of the steady solution is between $68$ to $69$, as shown in figure~\ref{Fig:SS_Bif_Lift} based on the pressure lift coefficient of the steady solutions.

\section{Linear stability analysis}
\label{Sec:LSA}
The linear stability problem for a base flow $\left ( \bm{U}(\bm{x}), P(\bm{x}) \right )$ with small perturbations $\left ( \bm{u'}(\bm{x},t), p'(\bm{x},t) \right )$ is governed by the linearized Navier-Stokes equations, which read:
\begin{equation}
\label{Eqn:LNSE}
\partial_t \bm{u'} + \left (  \bm{U} \cdot \nabla\right ) \bm{u'} +\left (  \bm{u'} \cdot \nabla\right ) \bm{U} = \nu \triangle \bm{u'} - \nabla p' ,\quad 
\nabla \cdot  \bm{u'} =0 .
\end{equation}
The assumption of small perturbation allows to linearize the equations, and we can separate the time and space dependence as:
\begin{equation}
\label{Eqn:SmallPert}
\bm{u'}(\bm{x},t)=\bm{\hat{u}}(\bm{x})e^{(\sigma+i\omega) t} ,\quad 
p'(\bm{x},t)=\hat{p}(\bm{x})e^{(\sigma+i\omega)  t} .
\end{equation}
By introducing the linear operator $\bm{L}(\bm{U})$, all the terms except the time derivative term and the continuity equation can be cast as $\bm{L}(\bm{U}) \bm{u'}$, we can rewrite (\ref{Eqn:LNSE}) as
\begin{equation}
\label{Eqn:LNSE2}
\partial_t \bm{u'} = \bm{L}(\bm{U}) \bm{u'} .
\end{equation}
Introducing (\ref{Eqn:SmallPert}) into (\ref{Eqn:LNSE2}), the equations can be written as:
\begin{equation}
\label{Eqn:LPDE}
%\lambda\bm{\hat{u}} + \left (  \bm{U} \cdot \nabla\right ) \bm{\hat{u}} +\left (  \bm{\hat{u}} \cdot \nabla\right ) \bm{U} - \nu \triangle\bm{\hat{u}} + \nabla \hat{p} =0 ,\quad 
%\nabla \cdot  \bm{\hat{u}} =0
(\sigma+i\omega) \bm{\hat{u}} = \bm{L}(\bm{U}) \bm{\hat{u}} .
\end{equation}
We use subspace iteration to solve this eigenvalue problem. A detailed review can be found in \citet{MORZYNSKI1999161}.
%We use subspace iteration to solve this eigenvalue problem of the form:
%\begin{equation}
%\label{Eqn:eigenAB}
%A x = (\sigma+i\omega) x
%\end{equation}
%A detailed review can be found in \citet{MORZYNSKI1999161}.

\begin{figure}
\begin{center}
\begin{tabular}{cccc}
& $Re=18$ & $Re=19$ & $Re=20$\\
\raisebox{7.5\height}{(a)}&
\includegraphics[width=0.3\textwidth]{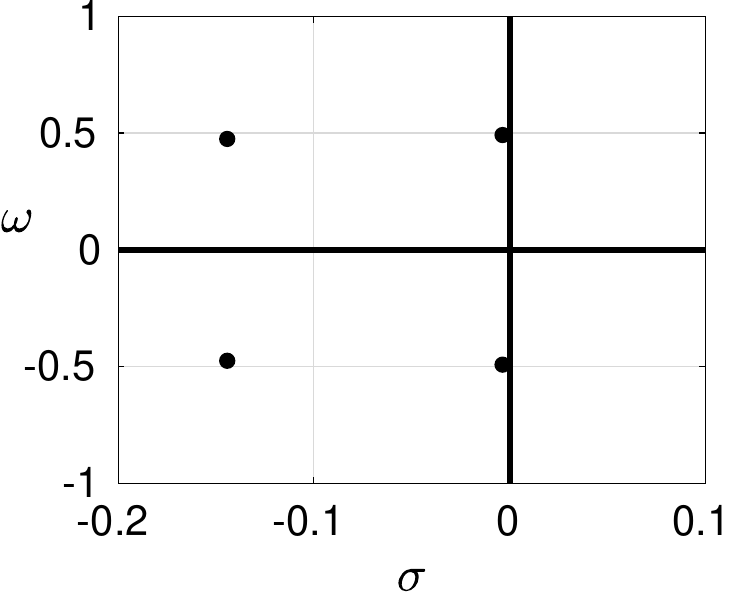}  &
\includegraphics[width=0.3\textwidth]{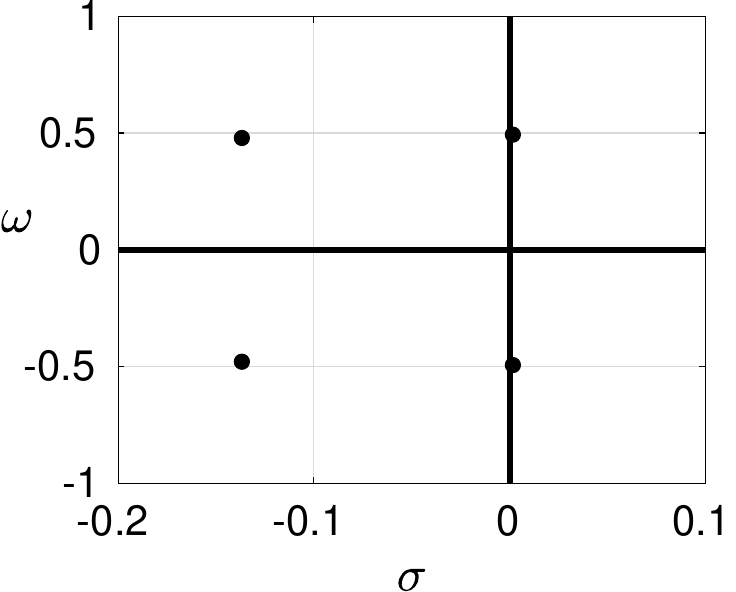}  &
\includegraphics[width=0.3\textwidth]{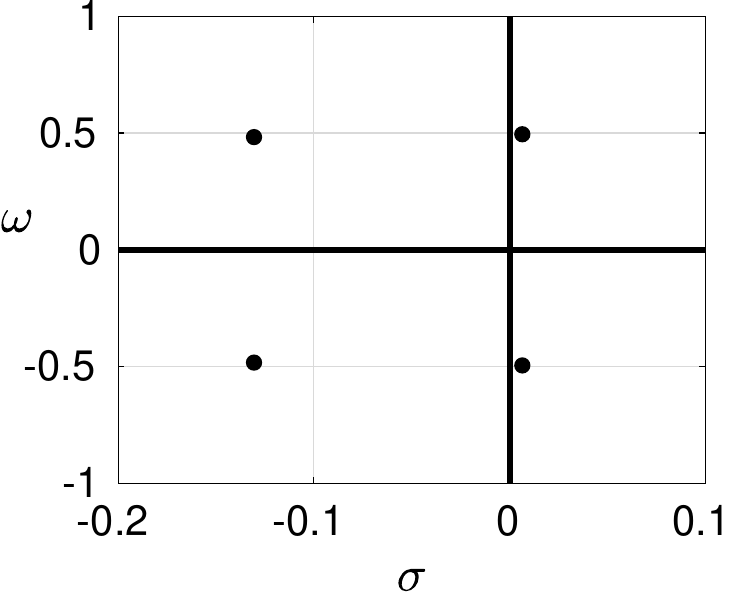}  \\
& $Re=68$ & $Re=69$ & $Re=70$\\
\raisebox{7.5\height}{(b)}&
\includegraphics[width=0.3\textwidth]{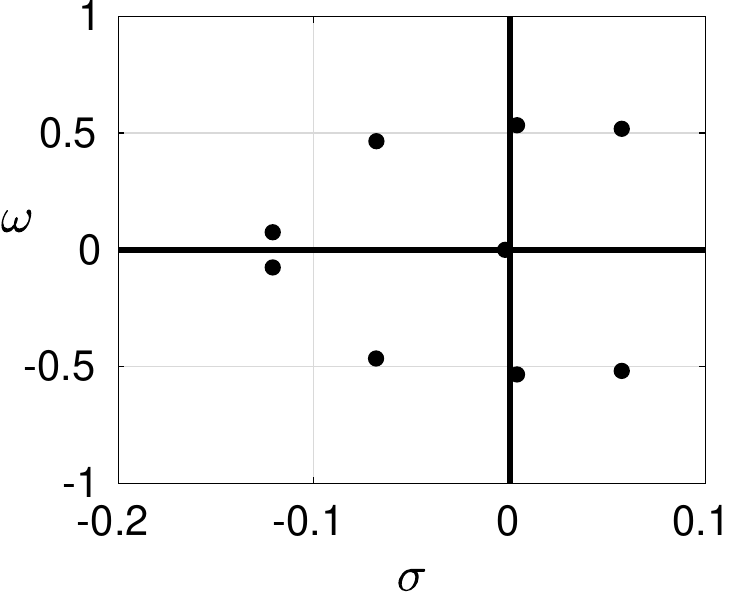}  &
\includegraphics[width=0.3\textwidth]{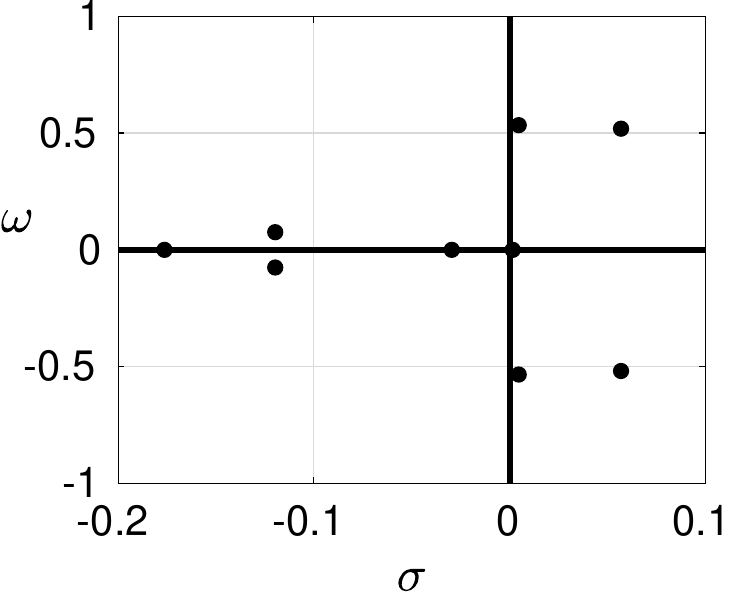}  &
\includegraphics[width=0.3\textwidth]{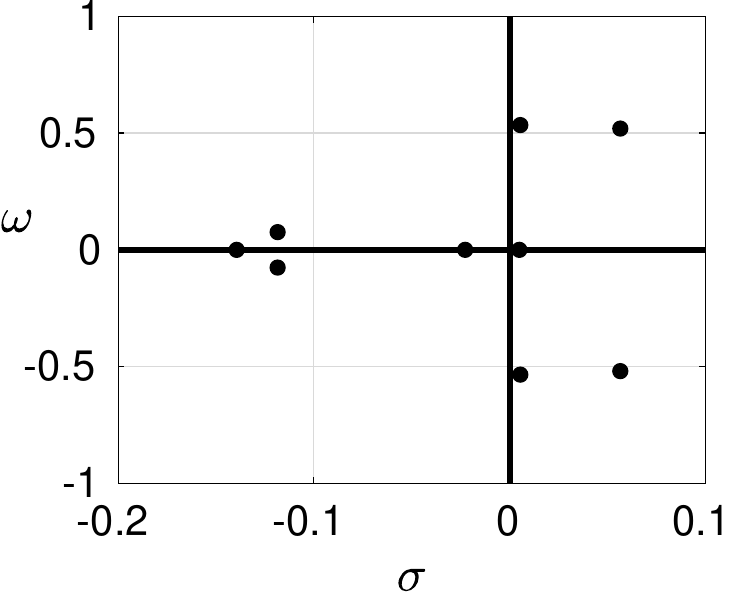}  \\
\end{tabular}
\caption{Eigenspectrum resulting from the linear stability analysis of the symmetric steady solution $\bm{u}_s$. With increasing $Re$, (a) a complex-conjugated eigenvalue pair crosses the imaginary axis at Re changing from 18 to 19, the critical value of the Hopf bifurcation $Re_1 \approx 18$, (b) a real eigenvalue crosses the imaginary axis at Re changing from 68 to 69, the critical value of the pitchfork bifurcation $Re_2 \approx 68$. }
\label{Fig:LSA_SS}
\end{center}
\end{figure}

The global stability analysis of the steady solutions at different Reynolds numbers has been performed on a Krylov subspace of dimension 9-20. This converges after 50-100 iterations. The linear stability analysis on the symmetric steady solution $\bm{u}_s$ reveals a pair of conjugated eigenvalues with positive real part first appearing as the Reynolds number is changing from 18 to 19, see figure~\ref{Fig:LSA_SS}(a). A real eigenvalue becomes positive between $Re=68$ and $69$, see figure~\ref{Fig:LSA_SS}(b). This confirms that a Hopf bifurcation occurs on the symmetric steady solution at $Re_1\approx18$, and a pitchfork bifurcation at $Re_2\approx68$. 

\section{Floquet stability analysis}
\label{Sec:FSA}
Similar to the linear stability framework, the Floquet stability problem works with a T-periodic base flow $\left ( \bm{U}(\bm{x},t), P(\bm{x},t) \right )$. The linear operator now reads:
\begin{equation}
\label{Eqn:LNSE_Floquet}
\partial_t \bm{u'} = \bm{L}(\bm{U}(\bm{x},t)) \bm{u'} .
\end{equation}

The linear operator $\bm{L}(\bm{U}(\bm{x},t))$ is T-periodic because of the base flow $\bm{U}(\bm{x},t)$. The solutions to (\ref{Eqn:LNSE_Floquet}) are seeked as:
\begin{equation}
\label{Eqn:SmallPert_Floquet}
\bm{u'}(\bm{x},t)=\bm{\hat{u}}(\bm{x},t)e^{(\sigma+i\omega) t} ,
\end{equation}
with the T-periodic Floquet modes $\bm{\hat{u}}(\bm{x},t)$ and the corresponding Floquet exponents $\sigma+i\omega$. We define the Floquet operator as the time-integrated $\bm{L}(\bm{U}(\bm{x},t))$ with the pre-stored periodic solutions over one period \citep{barkley_henderson_1996,schatz1995instability}, which reads: 
\begin{equation}
\label{Eqn:eigenFloquet}
A_F=\exp\left ( \int_{0}^{T}\bm{L}(\bm{U}(\bm{x},t))dt \right ) .
\end{equation}
The method used to solve the eigenproblem is the same as the linear stability analysis. The Floquet multipliers of $A_F$ can be written as $\lambda_F = e^{(\sigma+i\omega)T}$. In our case, we consider the multipliers $\lambda = e^{(\sigma+i\omega)}$.

\begin{figure}
\begin{center}
\begin{tabular}{ccc}
\includegraphics[width=0.3\textwidth]{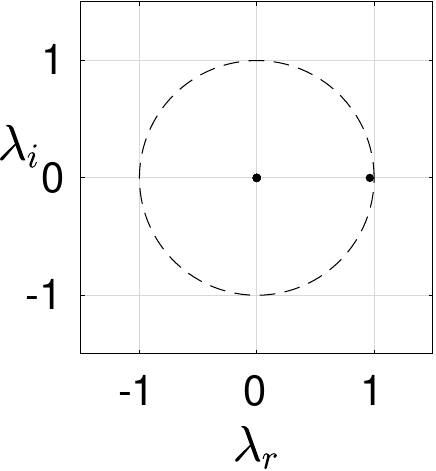}  &
\includegraphics[width=0.3\textwidth]{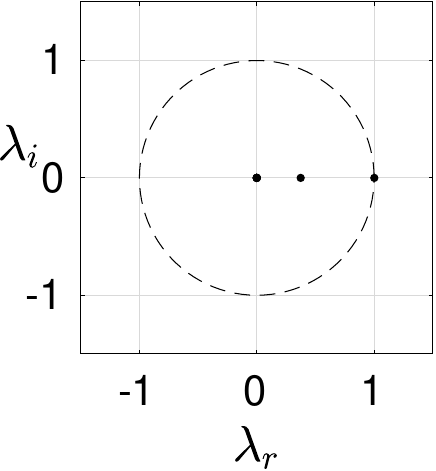}  &
\includegraphics[width=0.3\textwidth]{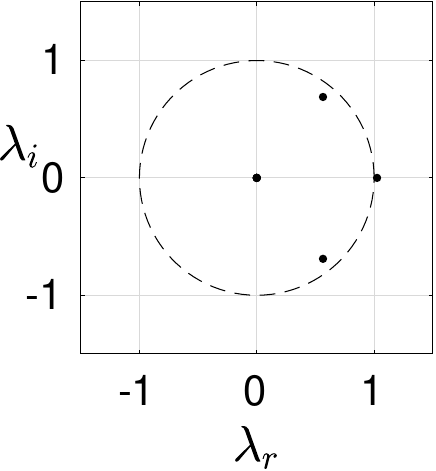}  \\
 (a) & (b) & (c)\\
\end{tabular}
\caption{Multipliers resulting from the Floquet analysis of the symmetry-preserving periodic solution for (a) $Re = 60$(a), (b) $70$, and (c) $80$.}
\label{Fig:FloquetSpect}
\end{center}
\end{figure}

\begin{figure}
\begin{tabular}{c  c }
(a) & (b) \\
 \includegraphics[width=0.3\textwidth]{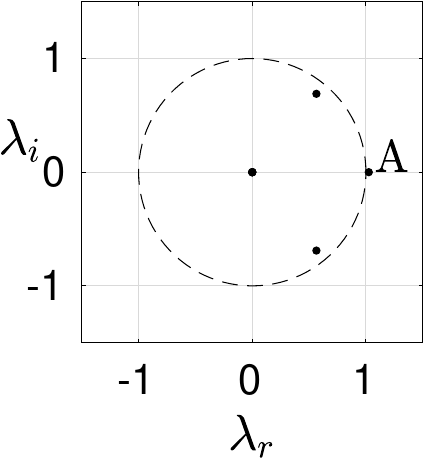} & \includegraphics[width=0.3\textwidth]{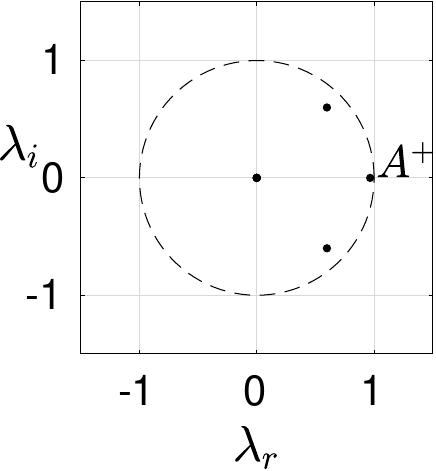} \\
\begin{minipage}[t]{0.45\textwidth}
\begin{tabular}{c  c }
{A} & \raisebox{-0.5\height}{\includegraphics[width=0.9\textwidth]{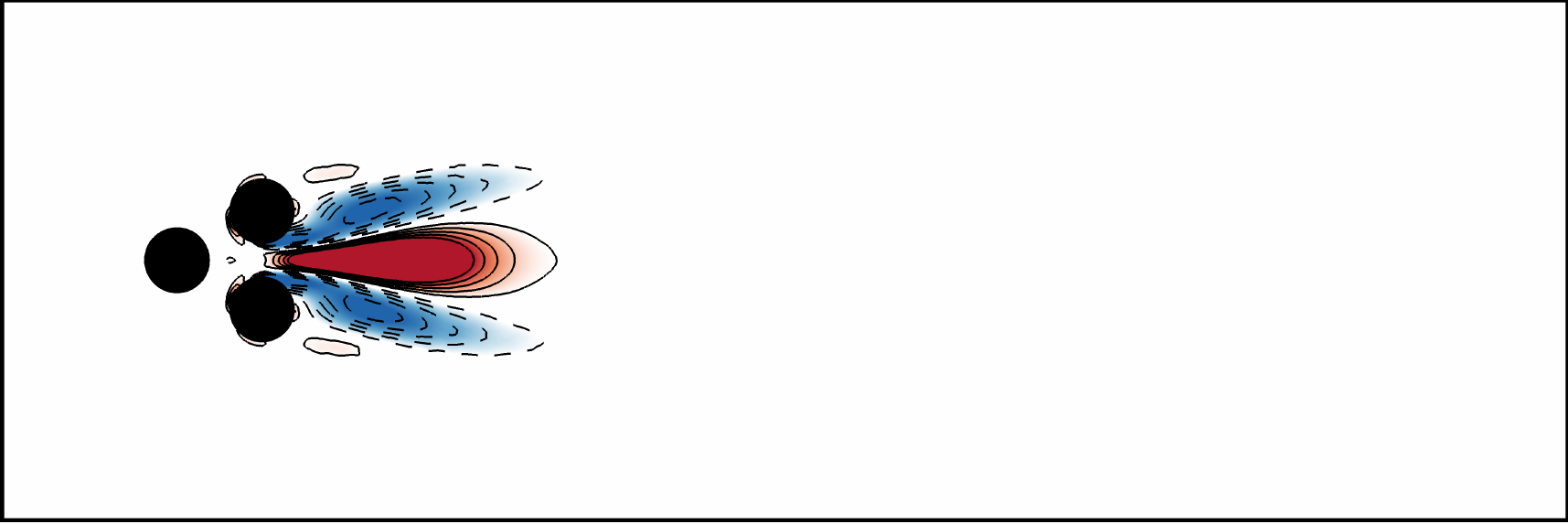}}  \\
\end{tabular} 
\end{minipage} &
\begin{minipage}[t]{0.45\textwidth}
\begin{tabular}{c  c }
{A$^+$} & \raisebox{-0.5\height}{\includegraphics[width=0.9\textwidth]{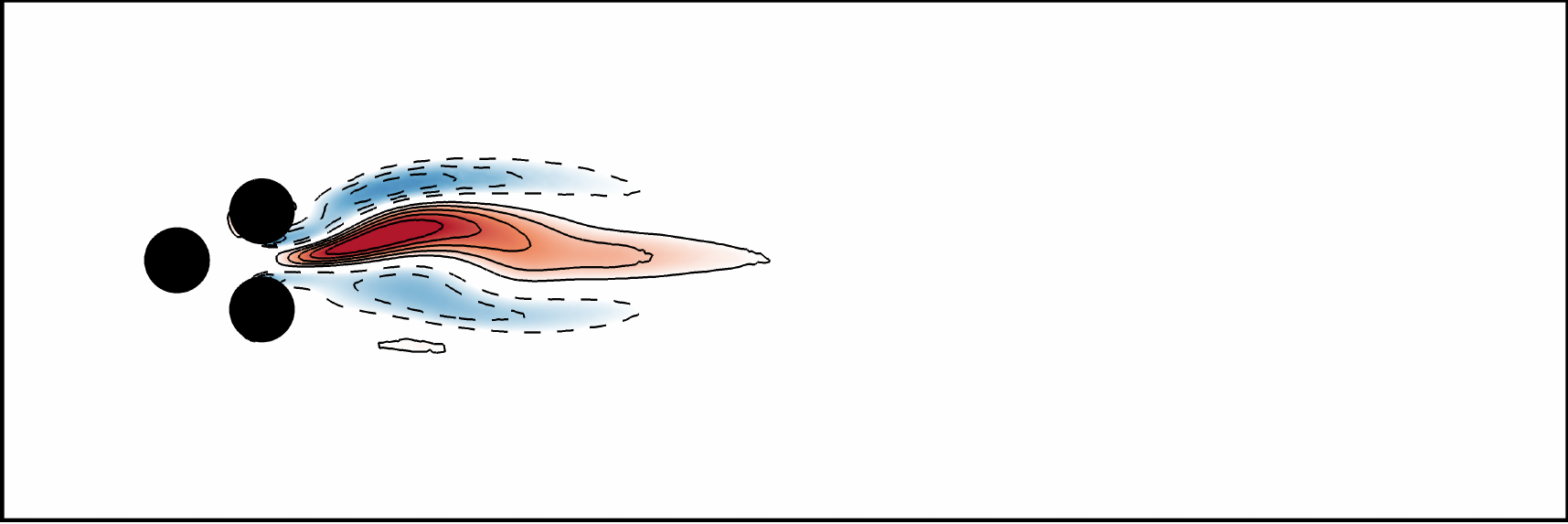}} \\
\end{tabular}
\end{minipage}
\end{tabular}
\caption{Multipliers resulting from the Floquet analysis (top) and the leading modes (bottom) of (a) the symmetry-preserving periodic solution, (b) the asymmetry solution, both at $Re=80$. Only the real part of the complex eigenmodes is shown. Red color and solid contours are positive values of the vorticity, blue color and dashed contours are negative values.}
\label{Fig:FloquetModes80}
\end{figure}

We performed the iteration on a Krylov subspace of dimension 9, using a symmetry-constrained T-periodic base flow. Below the critical Reynolds number, the base-bleeding jet is approximately steady and symmetric. This symmetry is also enforced at higher Reynolds numbers to compute the unstable periodic solution. The constraint is imposed on the central line as:
\begin{equation}
\label{Eqn:SymmetryCons}
v(0 \leq x \leq 1, \left |y  \right | \leq 5 \times 10^{-4})  = 0 .
\end{equation}
As (\ref{Eqn:SymmetryCons}) restricts the vertical velocity of the nodes, the symmetry-constrained periodic solution is very close to the symmetry-preserving periodic solution.  Normally, 10-20 iterations are enough to get a converged leading eigenvalue, initialized with a random vector or a Ritz vector computed at nearby Reynolds number with the periodic solution of the same symmetry. We do not attempt to calculate a complete, converged spectrum of all the eigenvalues, as the leading eigenvalue is associated with the instability of interest. The multipliers from the Floquet analysis around the critical value of the pitchfork bifurcation $Re_2$ are shown in figure~\ref{Fig:FloquetSpect}. When increasing $Re$, the leading real eigenvalue crosses the unit cycle at $(+1, 0)$ as $Re$ changes from 69 to 70. The critical value of the pitchfork bifurcation is, therefore, $Re_2 \approx 69$, identical to the critical value for the steady solution at the precision of the numerics. Both of them have the same eigenmode. At $Re=80$, the Floquet modes of both the unstable symmetric periodic solution and the stable asymmetric periodic solutions are shown in figure~\ref{Fig:FloquetModes80}. 

\begin{figure}
\begin{center}
\includegraphics[height=.9\linewidth]{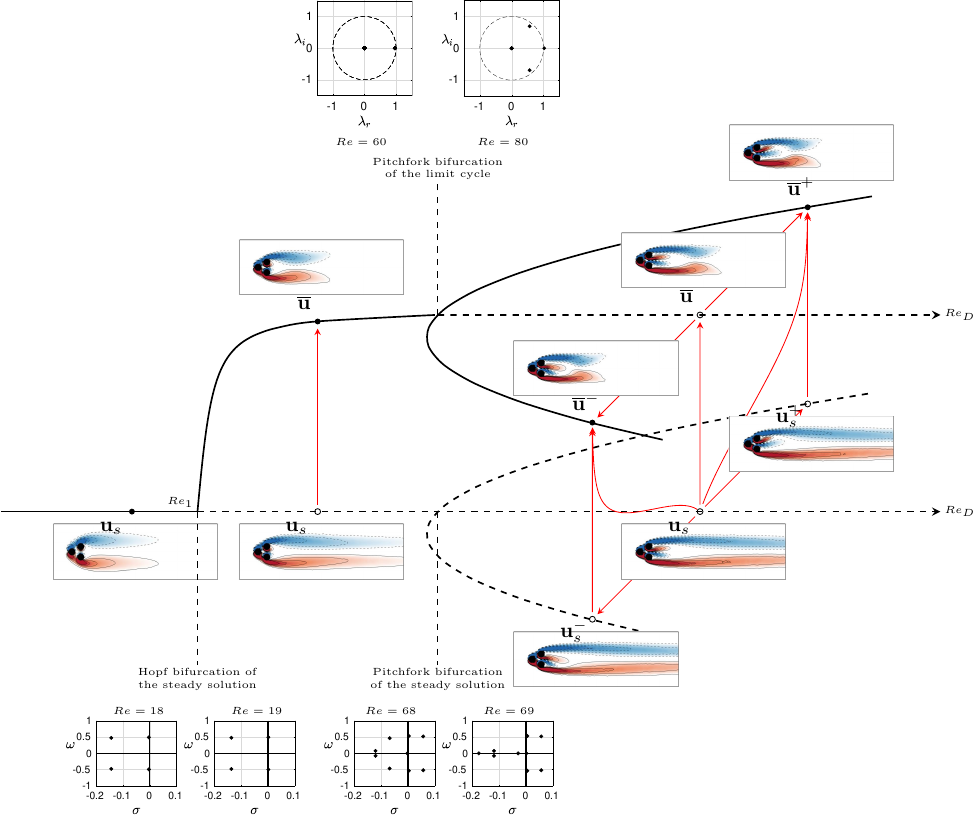} 
\caption{Bifurcation scenario along the $Re_D$-axis for the first two successive instabilities. The black curve indicates the stable branch, and dashed black curve for the unstable branch, combined with the linear/Floquet stability analysis results around the critical Reynolds number. The stability of the steady solutions $\bm{u}_s$ and $\bm{u}_s^\pm$, and the periodic solutions presented by the mean flow field $\overline{\bm{u}}$ and $\overline{\bm{u}}^\pm$ are illustrated by a dot for the stable state, or a circle for the unstable state. The red arrows show the possible transitions between them. The flow states in three stages are represented by the flow fields at $Re=10, 30, 80$.}
\label{Fig:Bif_Scenario}
\end{center}
\end{figure}

Overall, combined with the result of the linear stability analysis of the steady solutions, the bifurcation scenario at low Reynolds numbers can be shown in figure~\ref{Fig:Bif_Scenario}. The linear stability analysis on the steady solution and the periodic solution show a highly consistent result: the same kind of bifurcation with nearly the same critical Reynolds number, and the same eigenmodes. Besides, at $Re=80$, the growth rates of the real eigenmode are very close: 0.0272 from the symmetric steady solution, 0.0247 from the symmetric periodic solution. This similarity is understood as the result of a transverse effect of the symmetric subspace.

\section{Transient dynamics from different steady solutions}
\label{Sec:TranDyn}
\begin{figure}
\centering
\begin{tabular}{c c}
(a) & 
\raisebox{-0.5\height}{\includegraphics[width=0.8\textwidth]{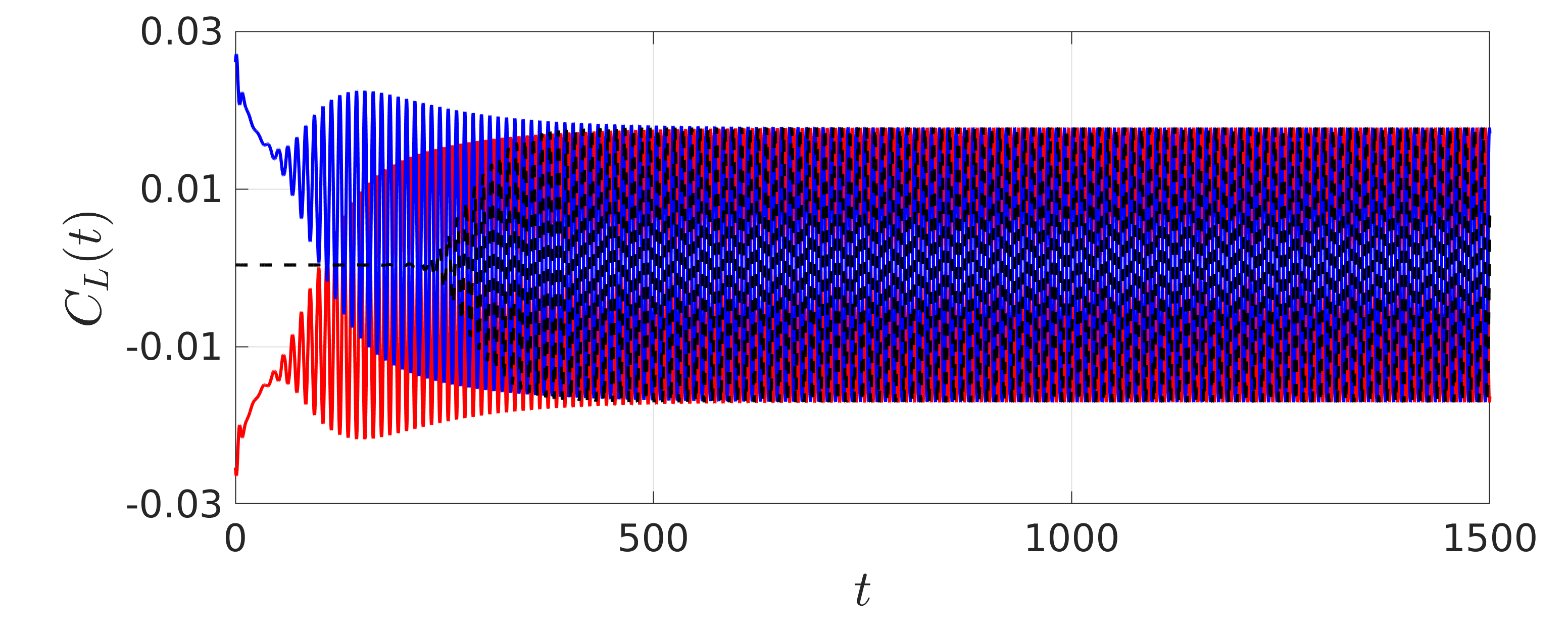}} \\
(b) & 
\raisebox{-0.5\height}{\includegraphics[width=0.8\textwidth]{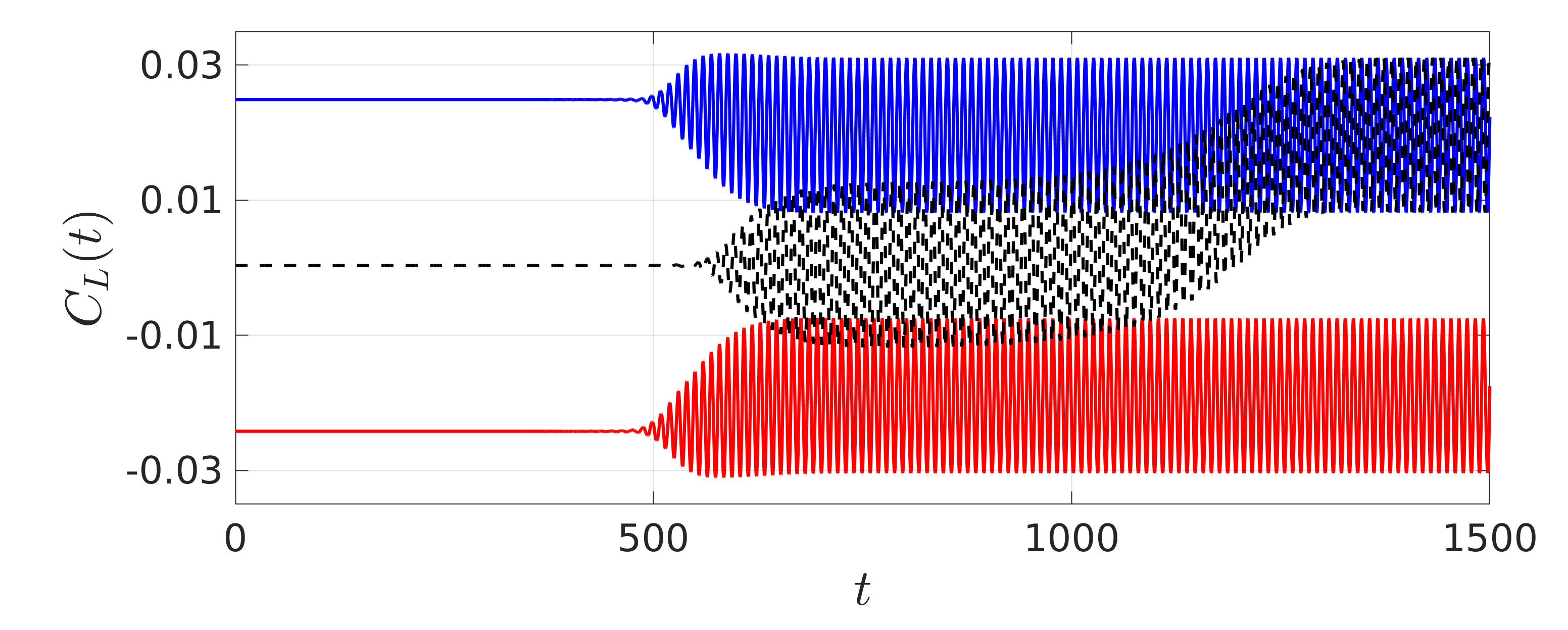}} \\
(c) & 
\raisebox{-0.5\height}{ \includegraphics[width=0.8\textwidth]{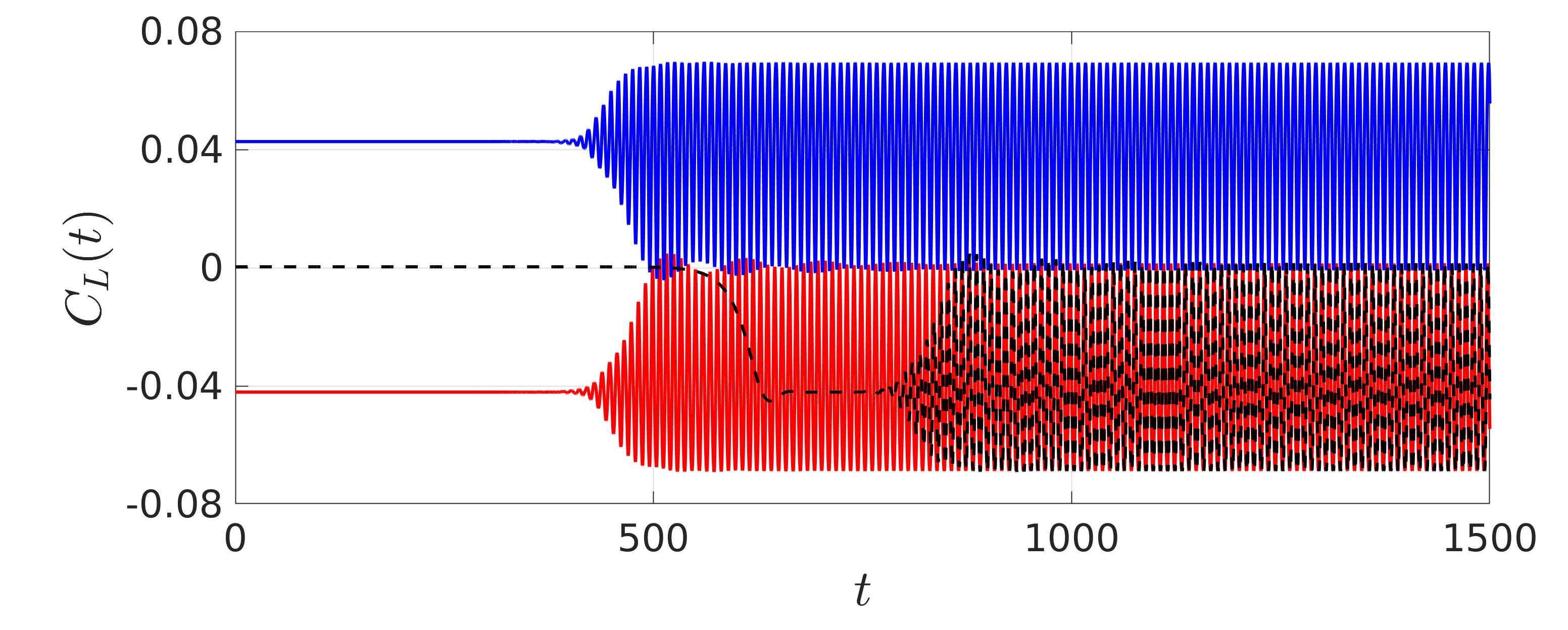}} \\
\end{tabular}
\caption{Transient dynamics based on the pressure lift coefficient $C_L$, resulting from the DNS starting with there steady solutions at different Reynolds numbers: (a) $Re=68$ starting with there steady solutions at $Re=75$, (b) $Re=75$ starting with there steady solutions at $Re=75$, and (c) $Re=100$ starting with there steady solutions at $Re=100$. The black dashed curve starts with the symmetric steady solution $\bm{u}_s$, the red curve starts with $\bm{u}_s^-$, and the blue curve starts with $\bm{u}_s^+$. }
\label{Fig:TranDynRe}
\end{figure}

\begin{figure}
\centering
\begin{tabular}{c c c c}
(a) & \raisebox{-0.5\height}{\includegraphics[width=0.4\textwidth]{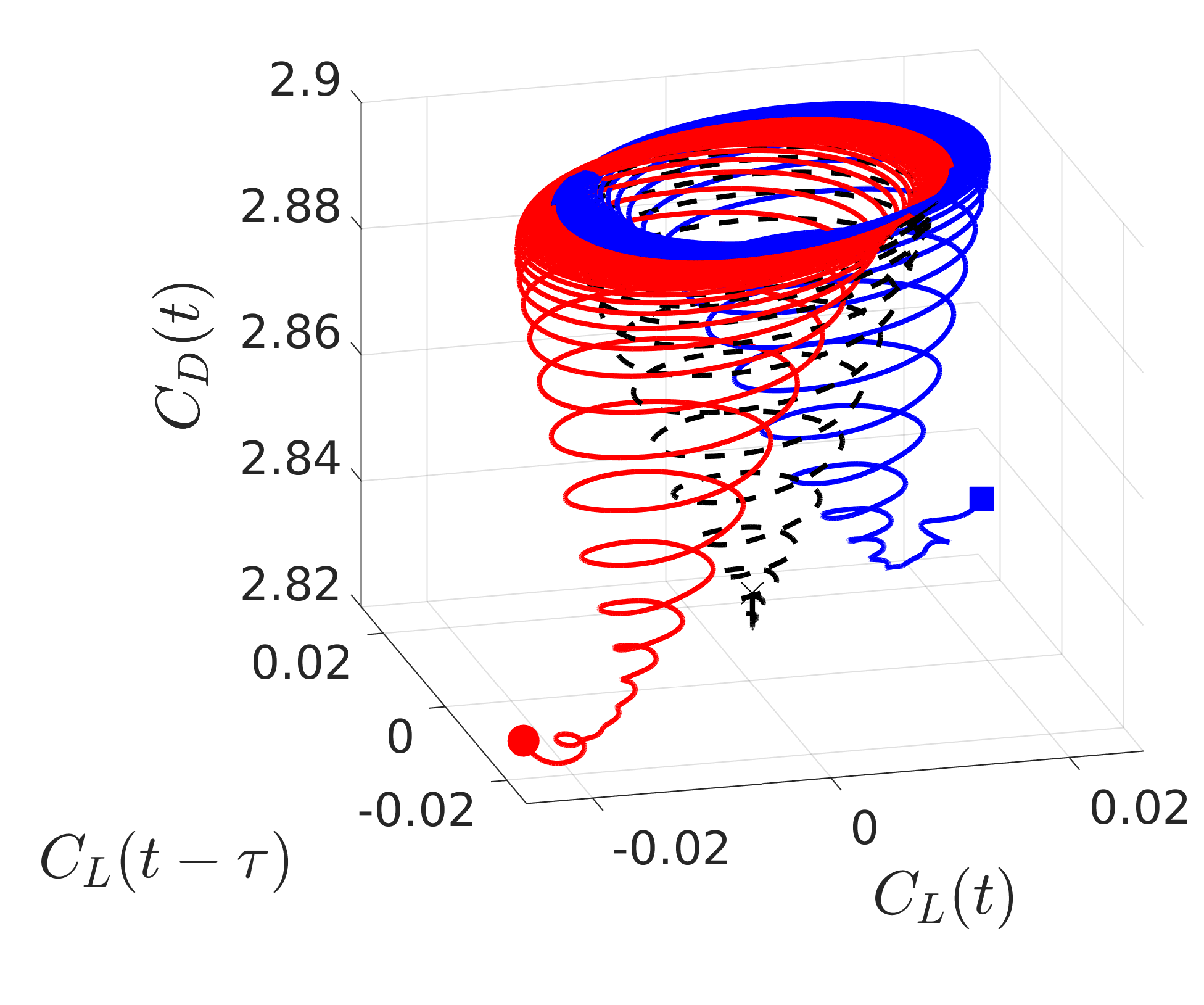}} &
(b) & \raisebox{-0.5\height}{\includegraphics[width=0.4\textwidth]{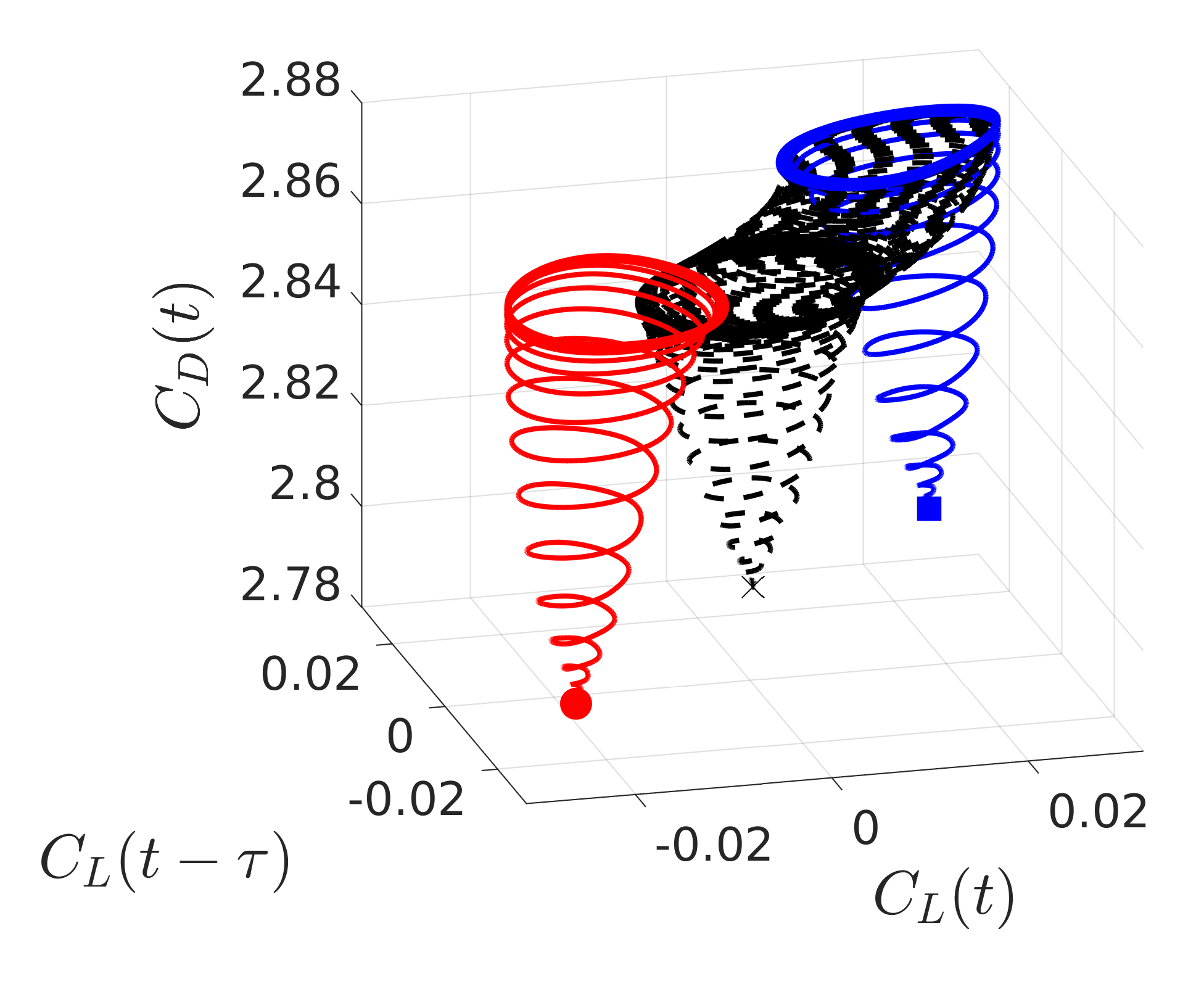}} \\
\end{tabular}
\begin{tabular}{c c}
(c) & \raisebox{-0.5\height}{\includegraphics[width=0.4\textwidth]{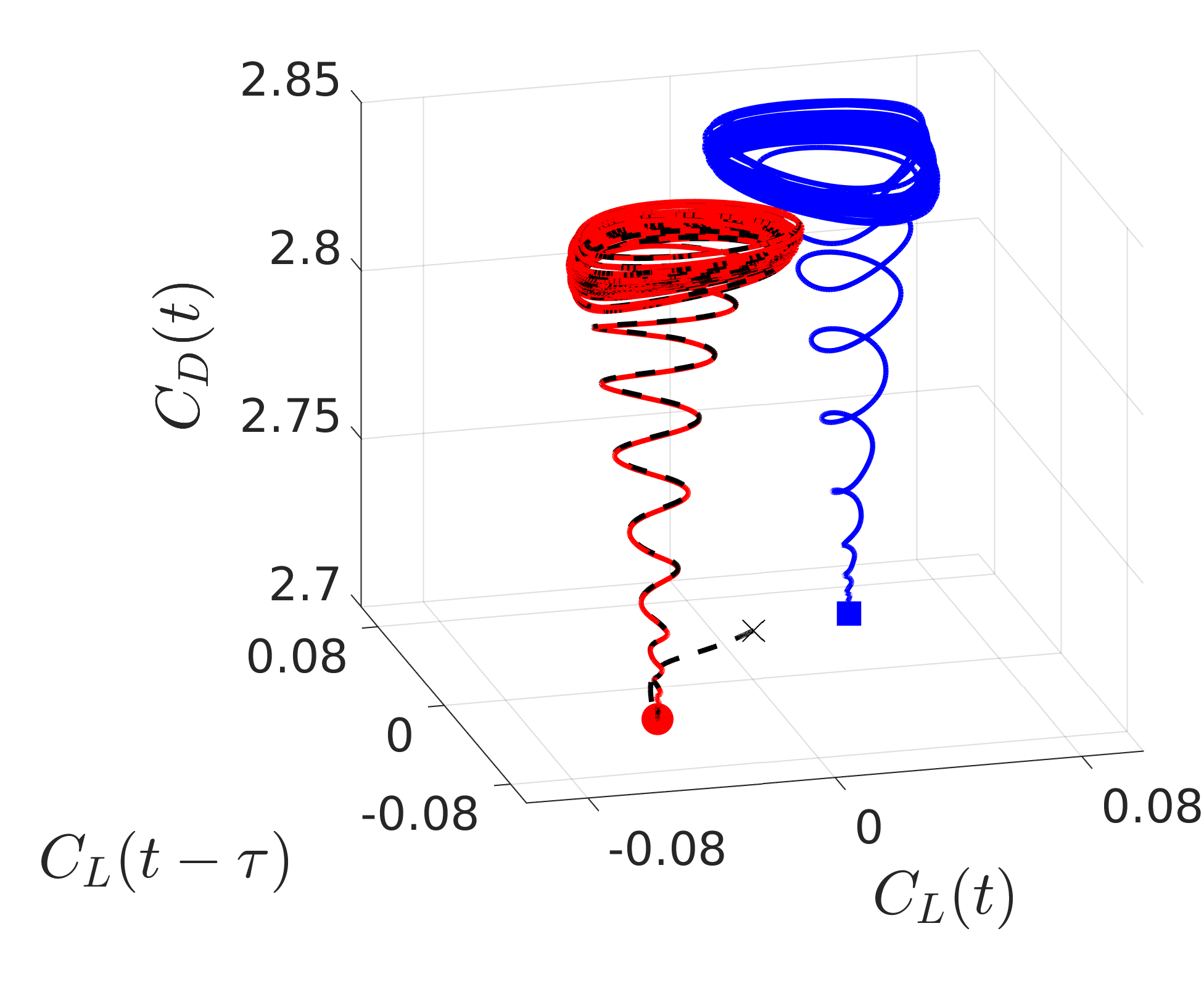}} 
\end{tabular}
\caption{Scenarios of transient dynamics, based on the pressure drag coefficient $C_D(t)$, pressure lift coefficient $C_L(t)$ and time-delayed pressure lift coefficient $C_L(t-\tau)$, from the three unstable symmetric/asymmetric steady solutions of (a)$Re=75$, (b)$Re=75$ and (c)$Re=100$, to their asymptotic stable limit cycles at (a)$Re=68$, (b)$Re=75$ and (c)$Re=100$: the black dashed curve from the symmetric steady solution $\bm{u}_s$({$\times$}), the red curve from the  asymmetric steady solution $\bm{u}_s^-$(\textcolor{red}{$\bullet$}), and the blue curve from the asymmetric steady solution $\bm{u}_s^+$(\textcolor{blue}{$\blacksquare$}).}
\label{Fig:PP_TranDynRe}
\end{figure}

In this section, we show in figure~\ref{Fig:TranDynRe} some typical transient dynamics starting with the unstable symmetric/asymmetric steady solutions at different Reynolds numbers, based on the pressure lift coefficient $C_L(t)$ from the resulting force on the three cylinders. Combining $C_L(t)$ with the pressure drag coefficient $C_D(t)$ and the time-delayed lift coefficient $C_L(t-\tau)$ in which $\tau$ is a quarter period, provides the phase portraits of figure~\ref{Fig:PP_TranDynRe}. Three comparative numerical simulations are shown, starting with the symmetric and the two mirror-conjugated asymmetric steady solutions at the same $Re$ respectively. The mirror-conjugated initial conditions provide mirror-conjugated transient dynamics.

Figure~\ref{Fig:TranDynRe}(a) shows the transient dynamics at the critical Reynolds number $Re_2$, initialized with the three steady solutions at $Re=75$. The lift coefficient starts oscillating quickly, and eventually reaches a unique oscillating state with zero mean value. This is consistent with the Floquet analysis at $Re=68$, where only one stable symmetry-centered limit cycle exists and any other state will eventually converge to this stable state.    

Figure~\ref{Fig:TranDynRe}(b) shows three different scenarios at $Re=75$ depending on the initial condition: from the symmetric steady solution $\bm{u}_s$ to the symmetry-centered limit cycle, from the symmetric centered limit cycle to the asymmetry-centered limit cycles, and from the asymmetric steady solutions $\bm{u}_s^\pm$ to the asymmetry-centered limit cycles. 
Starting with the symmetric steady solution, it will first reach the unstable symmetry-centered limit cycle, before asymptotically approaching one of two stable asymmetry-centered limit cycles. However, starting with the asymmetric steady solutions, it will directly reach the corresponding stable asymmetry-centered limit cycle. If the initial perturbation introduced to the symmetric steady solution has a certain bias of symmetry, a transition from the symmetric steady solution $\bm{u}_s$ to one of the two asymmetry-centered limit cycles will occur.

As we keep increasing the Reynolds number up to $Re=100$, there still exist six states, but the transient scenario from the symmetric steady solution $\bm{u}_s$ is different. It will first reach one of two unstable asymmetric steady solutions $\bm{u}_s^\pm$, before asymptotically approaching the corresponding asymmetry-centered limit cycle, as shown in figure~\ref{Fig:TranDynRe}(c).

All these transient dynamics mentioned above are highlighted with the red arrows in figure~\ref{Fig:Bif_Scenario}.

The phase portraits starting with different steady solutions, as shown in figure~\ref{Fig:PP_TranDynRe}, reveal the above-mentioned transient dynamics which is affected by the initial condition and the Reynolds number. At the same time, it also reflects the global effect of the pitchfork bifurcation at $Re_2$, splitting the state space (see figure~\ref{Fig:PP_TranDynRe}(a)) into a symmetric sub-space and two mirror-conjugated asymmetric sub-spaces (see figure~\ref{Fig:PP_TranDynRe}(b),(c)).

\section{On the simultaneous instability of the fixed point and the limit cycle}
\label{Sec:ODE}
In this section, we exemplify the transverse effect of the pitchfork bifurcation on a three-dimensional dynamical system equivalent to the system of Eq. \eqref{Eqn:MFSystem:Hopf+Pitchfork}. 

\subsection*{Dynamical system}          
\begin{figure}
\centering
\includegraphics[width=0.5\textwidth]{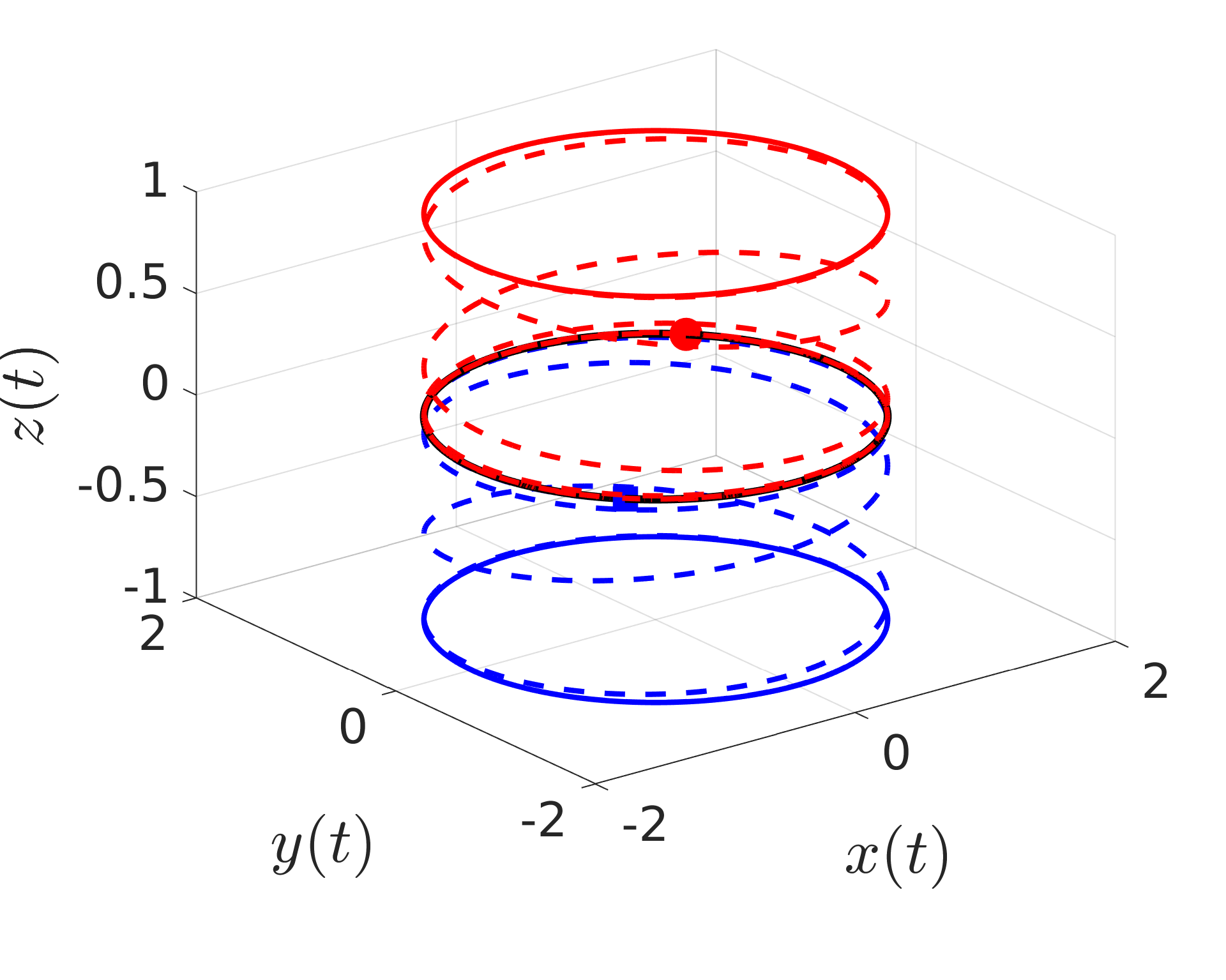} 
\caption{Three-dimensional schematic diagram of $\mu=3$, with the transition from the perturbed initial conditions on the unstable limit cycle $z=0$ (the black cycle) to the corresponding stable limit cycles $z=\pm\sqrt{\mu-\mu_2}$ (the blue and red cycles): the blue dashed curve from $(-1,-1,-0.0001)$(\textcolor{blue}{$\blacksquare$}), the red dashed curve from $(1,1,0.0001)$ (\textcolor{red}{$\bullet$}).}
\label{fig:ODE_A6}
\end{figure}

The dynamical system reads:
\begin{equation}
\left\{\begin{matrix}
\dot{x}&= &(\mu-\mu_1-(x^2+y^2))x+(\omega_0+(x^2+y^2))y\\ 
\dot{y}&= &(\mu-\mu_1-(x^2+y^2))y-(\omega_0+(x^2+y^2))x\\ 
\dot{z}&= &(\mu-\mu_2)z-z^3 , \\ 
\end{matrix}\right. 
\label{eq:ODE_A6}
\end{equation}
with $\mu_1=1, \mu_2=2, \omega_0=1$. This system undergoes a supercritical Hopf bifurcation in the $(x,y)$-plane at $\mu=\mu_1$ and a supercritical pitchfork bifurcation along the $z$-axis at $\mu=\mu_2$. For $\mu>\mu_1$, the stable fixed point at ${(0,0,0)}$ becomes unstable, and the limit cycle around $(0,0,0)$ with radius $r=\sqrt{\mu-\mu_1}$ and angular frequency $\omega=\omega_0+\mu-\mu_1$ is stable in the (x,y)-plane. Increasing $\mu$ until $\mu>\mu_2$, the fixed point undergoes a secondary instability, as well as the limit cycle. Three unstable fixed points ${(0,0,0)}$, ${(0,0,\pm\sqrt{\mu-\mu_2})}$, and three limit cycles around these fixed points with radius $r=\sqrt{\mu-\mu_1}$ and angular frequency $\omega=\omega_0+\mu-\mu_1$ in the (x,y)-plane are found, as shown in figure~\ref{fig:ODE_A6}.

\subsection*{Linear stability analysis} 
The ODE (\ref{eq:ODE_A6}) can be writen as:
\begin{equation}
\dot{\bm{q}}= \bm{F}(\bm{q}),\quad \bm{q}=(x,y,z)
\label{eq:A6LSA3}
\end{equation}
and we note $\bm{q}_s$ is the steady state,that is, $\bm{F}(\bm{q}_s)=0$. Consider a small perturbation $\bm{q'}$ around the steady state $\bm{q}_s$ by
\begin{equation}
 \bm{q}=\bm{q}_s+\bm{q'} .
\label{eq:A6LSA2}
\end{equation}

We derived the linearized evolution equation:
\begin{equation}
\dot{\bm{q}'}= \bm{DF}(\bm{q}_s)\bm{q'} ,
\end{equation}
where $\bm{DF(\bm{q}_s)}$ is the Jacobian matrix of the considered steady state $\bm{q}_s$.

\begin{figure}
\centering
\begin{tabular}{ccc}
\includegraphics[width=0.3\textwidth]{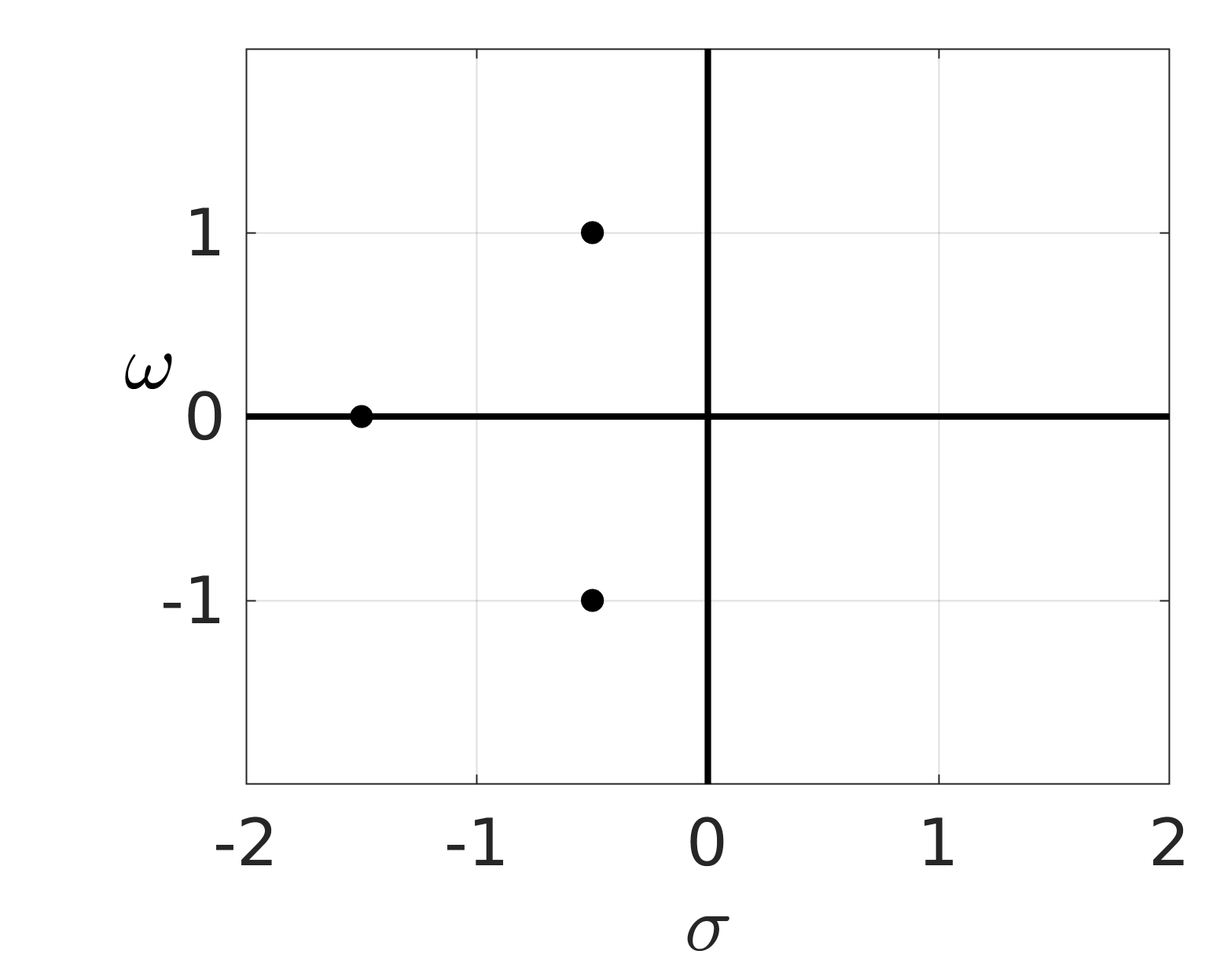}   &
\includegraphics[width=0.3\textwidth]{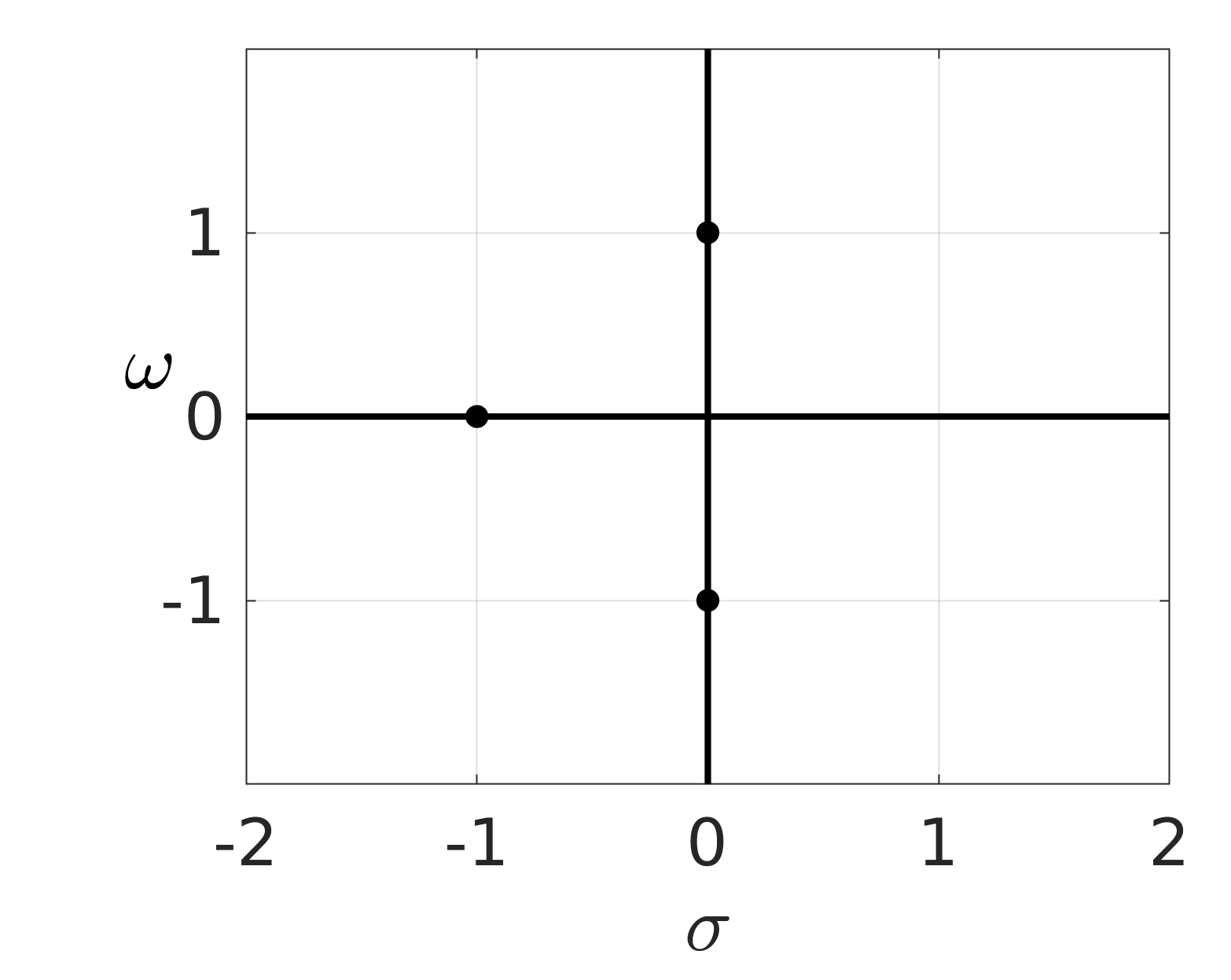}   &
\includegraphics[width=0.3\textwidth]{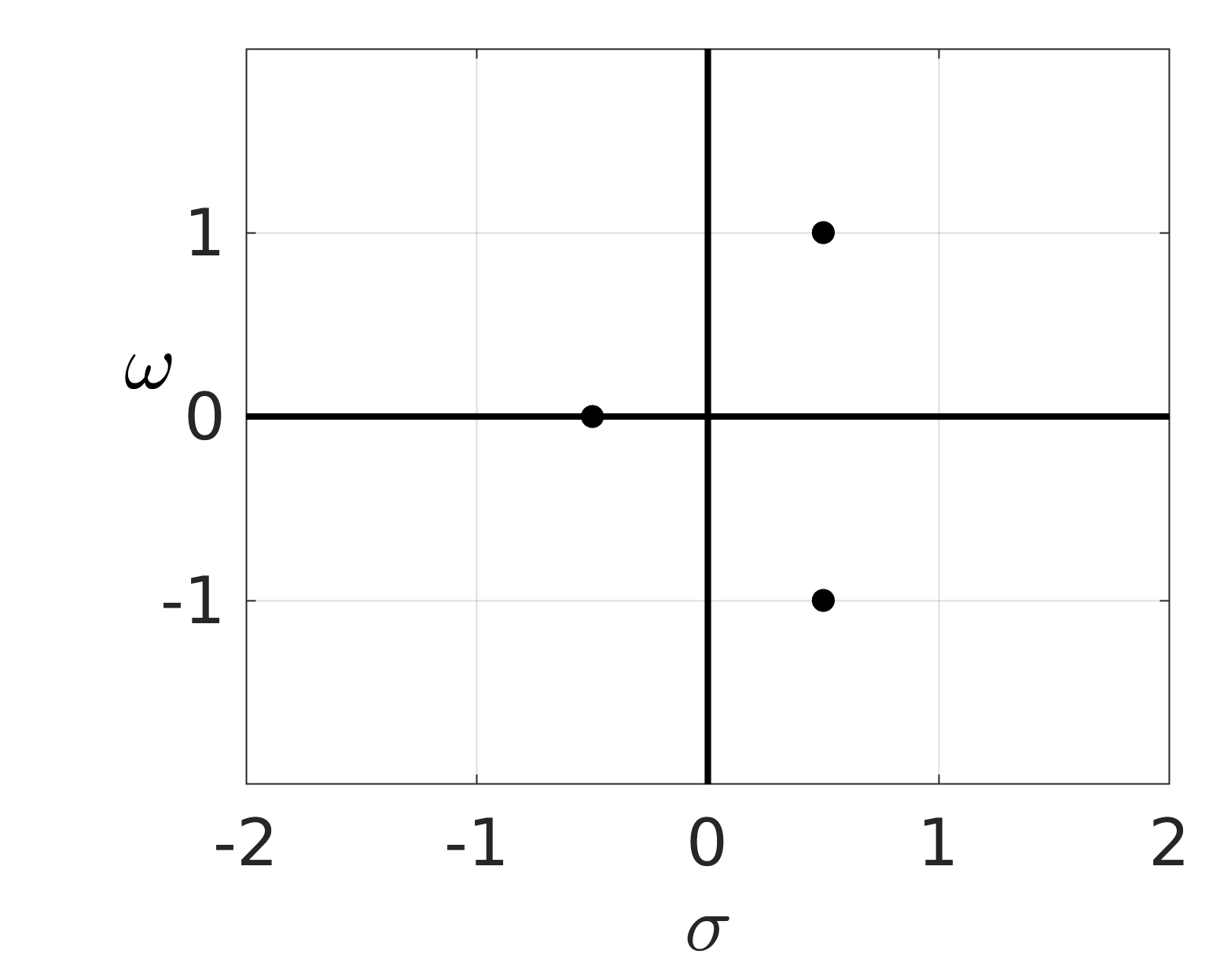}   \\
 (a) & (b) & (c) \\
\end{tabular}
\begin{tabular}{cc}
\includegraphics[width=0.3\textwidth]{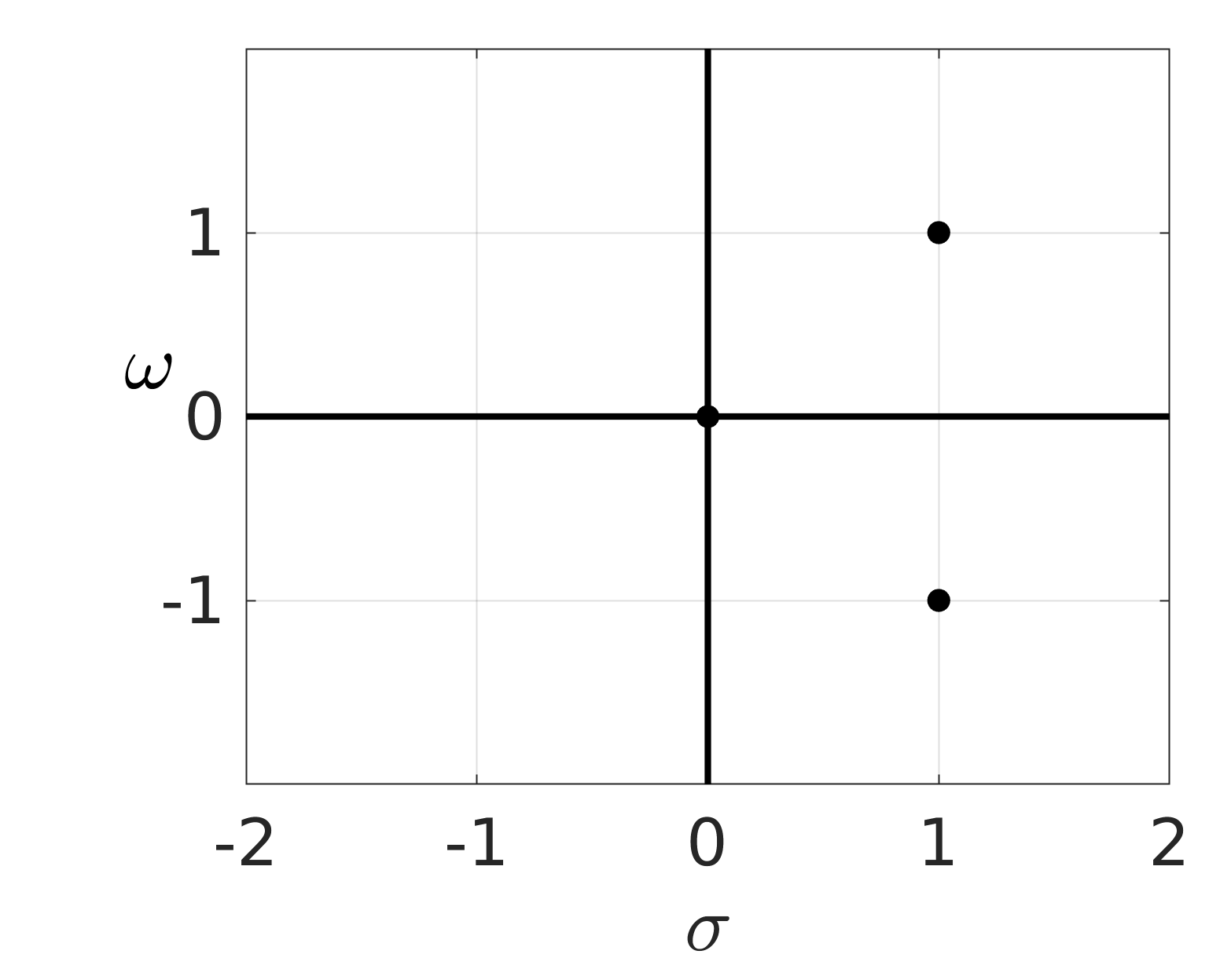}   &
\includegraphics[width=0.3\textwidth]{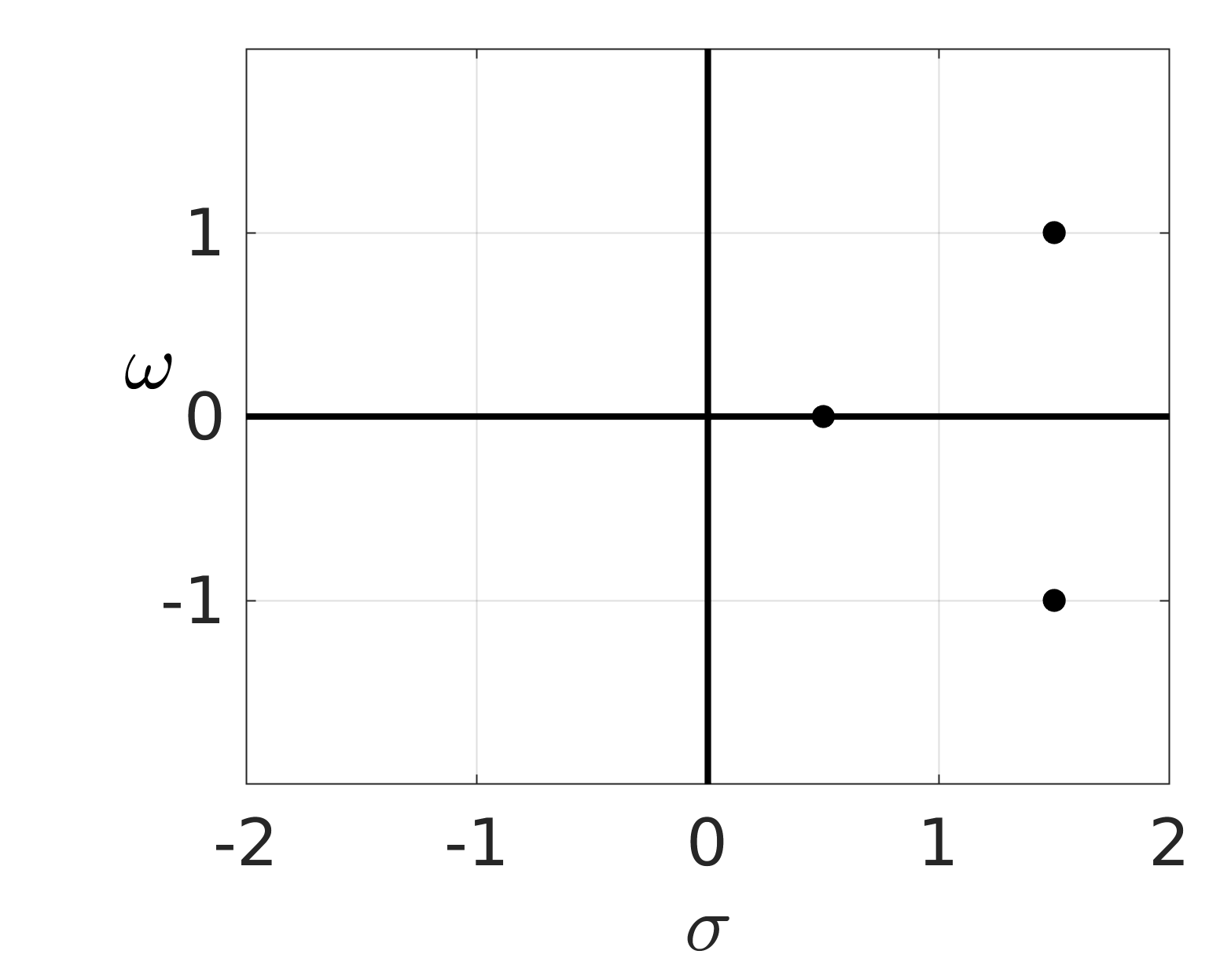} \\
 (d) & (e) \\
\end{tabular}
\caption{Eigenspectrum resulting from the linear stability analysis of the steady solution $\bm{q}_s^0$ at different $\mu$: (a)$\mu=0.5$, (b)$1.0$, (c)$1.5$, (d)$2.0$, (b)$2.5$.}
\label{fig:LSA_SS}
\end{figure}

The stability of this steady state is determined by the eigenvalues $\sigma+i\omega$ of the Jacobian matrix. The eigenspectrum of the fixed point $\bm{q}_s^0 = (0,0,0)$ is shown in figure~\ref{fig:LSA_SS}. The growth rate and angular frequency of the pair of conjugated eigenvalues are $\mu-\mu_1$ and $\omega_0$ respectively. The growth rate of the real eigenvalue is $\mu-\mu_2$.

The eigenspectrum of the steady solution $\bm{q}_s^\pm= (0,0,\sqrt{\mu-\mu_2})$ for $\mu>\mu_2$ is shown in figure~\ref{fig:LSA_SS_asym}(right). The growth rate is $\mu-\mu_2$ at the steady solution $\bm{q}_s^0$ and $-2(\mu-\mu_2)$ at the steady solution $\bm{q}_s^\pm$. 

\begin{figure}
\centering
\begin{tabular}{cc}
\includegraphics[width=0.3\textwidth]{./Figures/LSA_25.png}  &
\includegraphics[width=0.3\textwidth]{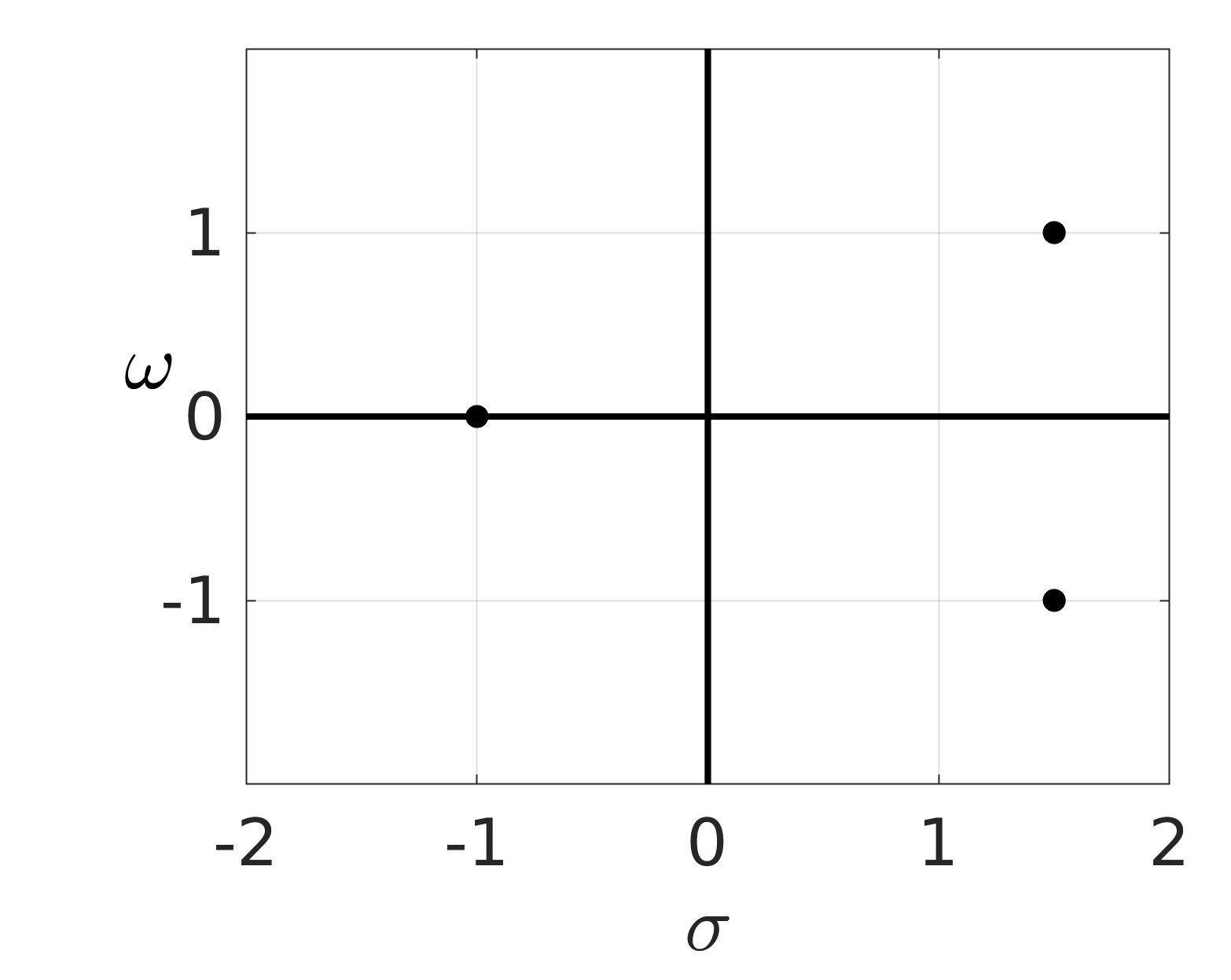}  \\
\end{tabular}
\caption{Eigenspectrum resulting from the linear stability analysis of the steady solutions $\bm{q}_s^0$(left) and $\bm{q}_s^\pm$(right), at $\mu=2.5$}
\label{fig:LSA_SS_asym}
\end{figure}

\subsection*{Floquet stability analysis} 
Now, we consider the periodic solution $\bm{q}_{p}(t)$ of the system of (\ref{eq:ODE_A6}), which can be writen as:
\begin{equation}
\bm{q}_{p}(t+T)=\bm{q}_{p}(t),\quad \hbox{with} \quad \dot{\bm{q}}_{p}(t)=F(\bm{q}_{p}(t)) .
\end{equation}

Consider a small perturbation $\bm{q'}$ around the periodic solution by
\begin{equation}
 \bm{q}(t)=\bm{q}_{p}(t)+\bm{q'}(t) .
\end{equation}

The first variational form reads:
\begin{equation}
\dot{\bm{q}'}(t)= \bm{DF}(\bm{q}_{p}(t))\bm{q'}(t) ,
\end{equation}
where $\bm{DF}(\bm{q}_{p}(t))$ is the Jacobian matrix of the considered periodic solution $\bm{q}_{p}(t)$, but now, the linear equation has periodic coefficients.

The monodromy matrix can be written as:
\begin{equation}
M_{\rm mono}=\exp\left ( \int^T_0\bm{DF}(\bm{q}_{p}(t))dt \right ) 	
\end{equation}
or equivalently:
\begin{equation}
M_{\rm mono}=\exp\left ( \frac{T}{2\pi}\int^{2\pi}_0\bm{DF}(\bm{q}_{p}(\theta))d\theta \right )  .
\end{equation}

\begin{figure}
\centering
\begin{tabular}{ccc}
\includegraphics[width=0.25\textwidth]{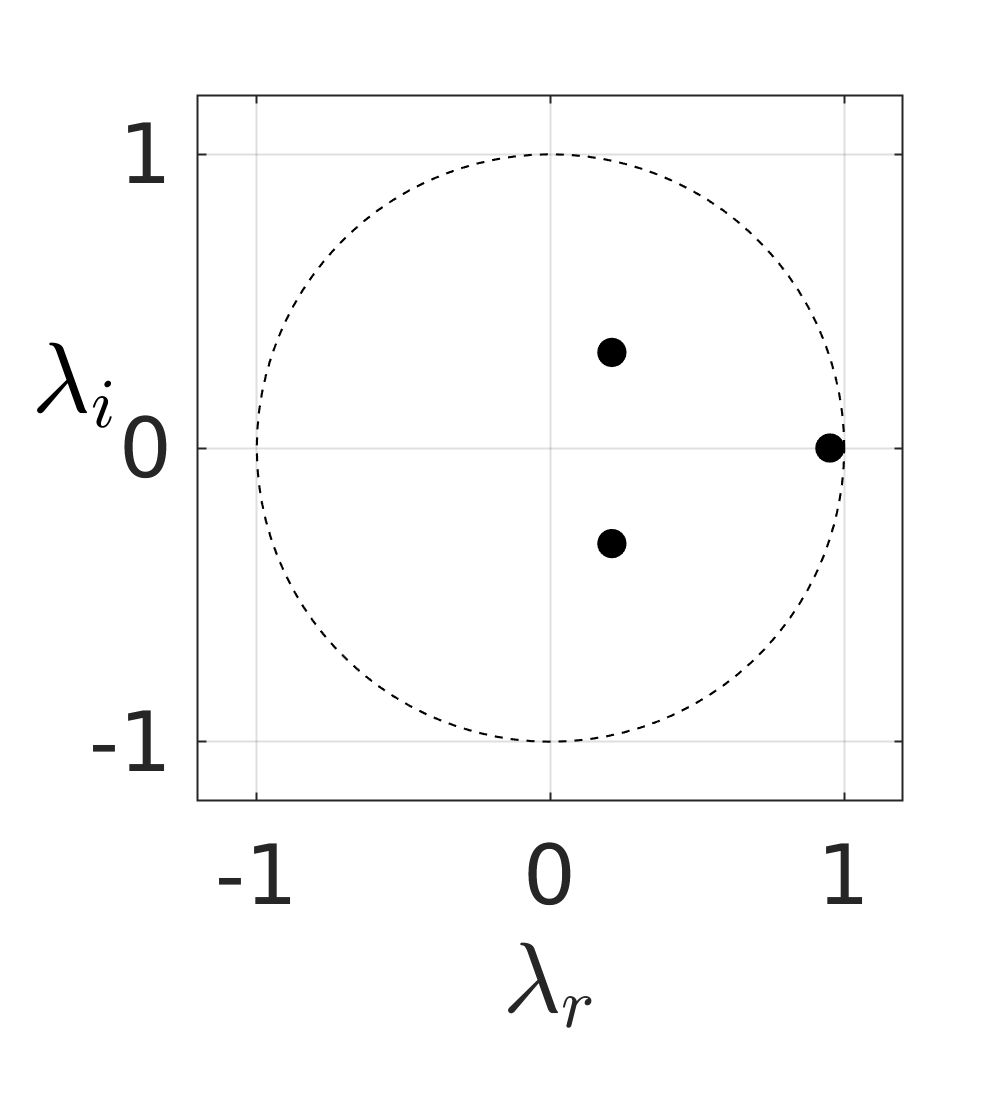}   &
\includegraphics[width=0.25\textwidth]{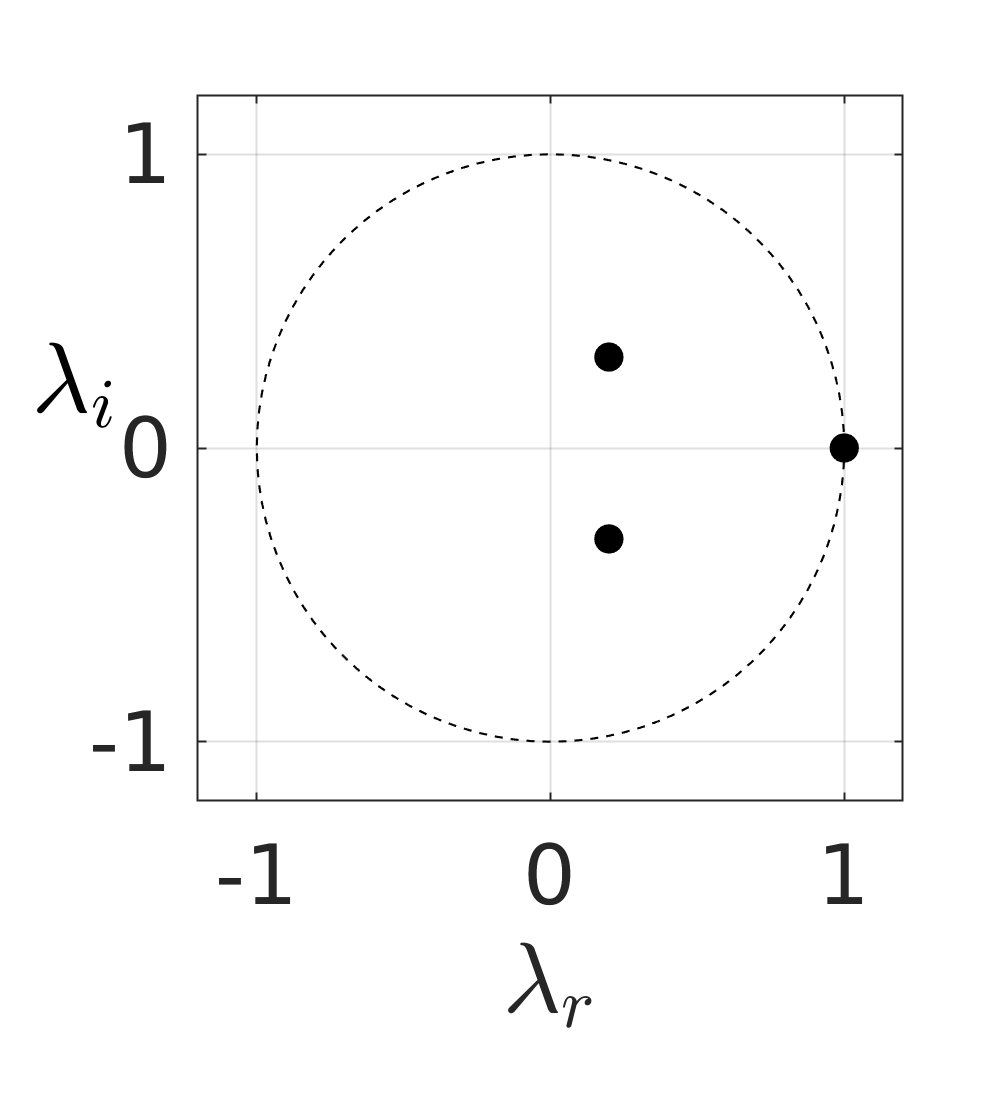}   &
\includegraphics[width=0.25\textwidth]{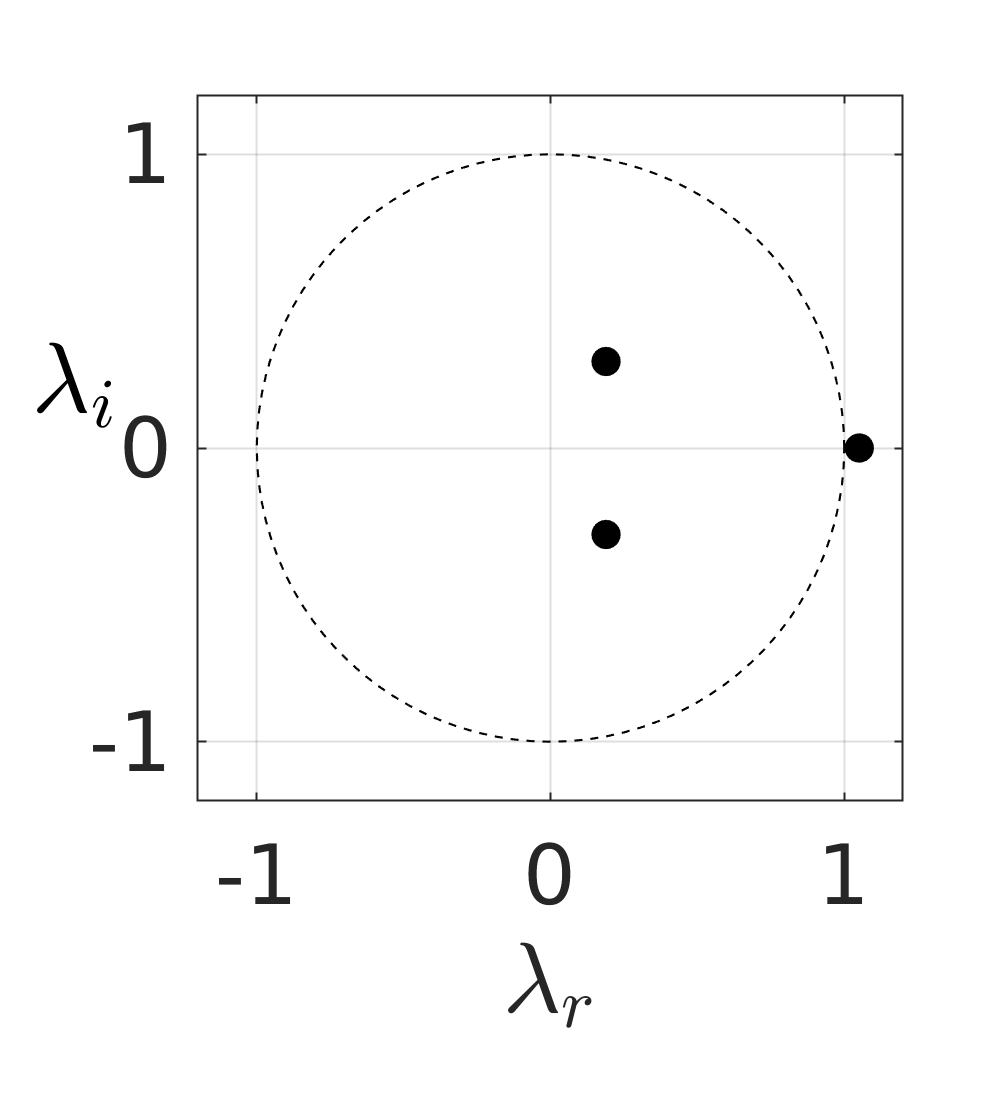} \\
 (a) & (b) & (c) 
\end{tabular}
\caption{Floquet stability analysis of the $z=0$ periodic solutions at different $\mu$ = (a)1.95, (b)2.0, and (c)2.05}
\label{fig:FSA_SS}
\end{figure}

The stability of this periodic solution is determined by the multipliers $\lambda=\exp( (\sigma_F+i\omega_F)/T )$, being the eigenvalues of the monodromy matrix. As shown in figure~\ref{fig:FSA_SS}, the periodic solution $\bm{q}_{p}^0$: $(\sqrt{\mu-\mu_1}\cos(\theta),\sqrt{\mu-\mu_1}\sin(\theta),0)$ for $\mu \ge \mu_2$ becomes unstable. A real multiplier $\lambda$ crosses the unit cycle at $+1$. The Floquet exponent of this real multiplier is $\mu-\mu_2$, equal to the growth rate at the fixed point $\bm{q}_s^0$.

The multipliers of the other two periodic solutions $\bm{q}_{p}^\pm$: $(\sqrt{\mu-\mu_1}\cos(\theta),\sqrt{\mu-\mu_1}\sin(\theta),$ $\pm \sqrt{\mu-\mu_2})$ is shown in figure~\ref{fig:LSA_SS_asym}(right). The Floquet exponent of the leading multiplier is $-2(\mu-\mu_2)$, equal to the growth rate at the fixed point $\bm{q}_s^\pm$. 

\begin{figure}
\centering
\begin{tabular}{cc}
\includegraphics[width=0.25\textwidth]{./Figures/FSA205.png}  &
\includegraphics[width=0.25\textwidth]{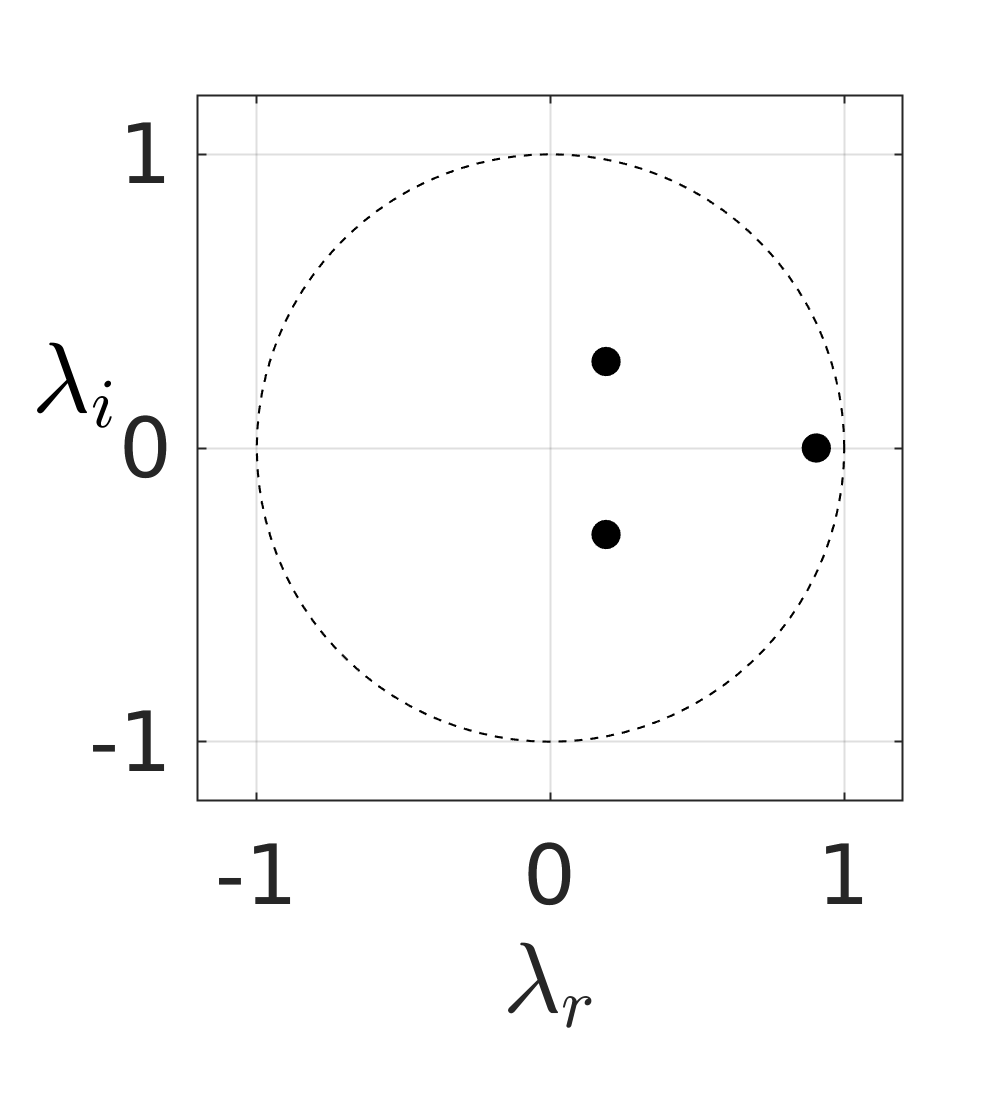}  \\
\end{tabular}
\caption{Floquet stability analysis of the periodic solutions $\bm{q}_{p}^0$(left) and $\bm{q}_{p}^\pm$(right), at $\mu=2.05$}
\label{fig:LSA_SS_asym}
\end{figure}

As a consequence, if the additional degree of freedom $z$, introduced by the pitchfork bifurcation, do not couple, at the onset of the bifurcation, to the primary degrees of freedom $x,y$ associated with the Hopf bifurcation (see \S~\ref{Sec:MultipleBifurcation}), then it is easy to understand that both the symmetric steady solution $(x_s,y_s,z_s)\equiv  (0,0,0)$ and the statistically symmetric periodic solution $(x_c(t+T),y_c(t+T),0) = (x_c(t),y_c(t),0)$, $T$ being the period of the limit cycle, will both undergo an instability with respect to the symmetry breaking provoked by the pitchfork bifurcation ($z\neq 0$). This simultaneous instability of both the symmetric fixed point and the statistically symmetric limit cycle looks like an instability of the subspace ($x,y$) with respect to the transverse direction $z$, associated with the active degree of freedom introduced by the pitchfork bifurcation. 

\end{appendix}

\bibliographystyle{jfm}
\bibliography{Main}

\begin{thebibliography}{71}
\expandafter\ifx\csname natexlab\endcsname\relax\def\natexlab#1{#1}\fi
\def\au#1{#1} \def\ed#1{#1} \def\yr#1{#1}\def\at#1{#1}\def\jt#1{\textit{#1}}
  \def\bt#1{#1}\def\bvol#1{\textbf{#1}} \def\vol#1{#1} \def\pg#1{#1}
  \def\publ#1{#1}\def\arxiv#1{#1}\def\org#1{#1}\def\st#1{\textit{#1}}

\bibitem[Bansal \& Yarusevych(2017)]{bansal2017experimental}
{\sc \au{Bansal, M.~S.} \& \au{Yarusevych, S.}} \yr{2017}  \at{Experimental
  study of flow through a cluster of three equally spaced cylinders}.  \jt{Exp.
  Thermal Fluid Sci.}  \bvol{80},  \pg{203--217}.

\bibitem[Barkley(2006)]{barkley2006europhys}
{\sc \au{Barkley, D.}} \yr{2006}  \at{Linear analysis of the cylinder wake mean
  flow}.  \jt{EPL (Europhysics Letters)}  \bvol{75}~(5),  \pg{750}.

\bibitem[Barkley \& Henderson(1996)]{barkley_henderson_1996}
{\sc \au{Barkley, D.} \& \au{Henderson, R.}} \yr{1996}  \at{Three-dimensional
  floquet stability analysis of the wake of a circular cylinder}.  \jt{J.~Fluid
  Mech.}  \bvol{322},  \pg{215–241}.

\bibitem[Bonnavion \& Cadot(2018)]{bonnavion_JFM2018}
{\sc \au{Bonnavion, G.} \& \au{Cadot, O.}} \yr{2018}  \at{Unstable wake
  dynamics of rectangular flat-backed bluff bodies with inclination and ground
  proximity}.  \jt{J.~Fluid Mech.}  \bvol{854},  \pg{196--232}.

\bibitem[Bourgeois {\em et~al.\/}(2013)Bourgeois, Noack \&
  Martinuzzi]{Bourgeois2013jfm}
{\sc \au{Bourgeois, J.~A.}, \au{Noack, B.~R.} \& \au{Martinuzzi, R.~J.}}
  \yr{2013}  \at{Generalised phase average with applications to sensor-based
  flow estimation of the wall-mounted square cylinder wake}.  \jt{J.~Fluid
  Mech.}  \bvol{736},  \pg{316--350}.

\bibitem[Brunton \& Noack(2015)]{Brunton2015amr}
{\sc \au{Brunton, S.~L.} \& \au{Noack, B.~R.}} \yr{2015}  \at{Closed-loop
  turbulence control: Progress and challenges}.  \jt{Appl. Mech. Rev.}
  \bvol{67}~(5),  \pg{050801:01--48}.

\bibitem[Brunton {\em et~al.\/}(2016)Brunton, Proctor \& Kutz]{brunton2016pnas}
{\sc \au{Brunton, S.~L.}, \au{Proctor, J.~L.} \& \au{Kutz, J.~N.}} \yr{2016}
  \at{Discovering governing equations from data by sparse identification of
  nonlinear dynamical systems}.  \jt{Proc. Natl. Acad. Sci.}  \bvol{113}~(5),
  \pg{3932--3937}.

\bibitem[Cadot {\em et~al.\/}(2015)Cadot, Evrard \& Pastur]{cadot_PRE2015}
{\sc \au{Cadot, O.}, \au{Evrard, A.} \& \au{Pastur, L.}} \yr{2015}
  \at{Imperfect supercritical bifurcation in a three-dimensional turbulent
  wake}.  \jt{Phys. Rev. E}  \bvol{91}~(6),  \pg{063005}.

\bibitem[Cornejo~Maceda(2017)]{Cornejo2017limsi}
{\sc \au{Cornejo~Maceda, G.~Y.}} \yr{2017} Machine learning control applied to
  wake stabilization. MS2 Internship Report, LIMSI and ENSAM, Paris, France.

\bibitem[Cross \& Hohenberg(1993)]{Cross1993rpm}
{\sc \au{Cross, M.~C.} \& \au{Hohenberg, P.~C.}} \yr{1993}  \at{Pattern
  formation outside of equilibrium}.  \jt{Rev. Mod. Phys.}  \bvol{65}~(3),
  \pg{851}.

\bibitem[Ding \& Kawahara(1999)]{ding1999three}
{\sc \au{Ding, Y.} \& \au{Kawahara, M.}} \yr{1999}  \at{Three-dimensional
  linear stability analysis of incompressible viscous flows using the finite
  element method}.  \jt{Int. J. Num. Meth. Fluids}  \bvol{31}~(2),
  \pg{451--479}.

\bibitem[Fabre {\em et~al.\/}(2008)Fabre, Auguste \& Magnaudet]{Fabre2008pof}
{\sc \au{Fabre, D.}, \au{Auguste, F.} \& \au{Magnaudet, J.}} \yr{2008}
  \at{Bifurcations and symmetry breaking in the wake of axisymmetric bodies}.
  \jt{Phys. Fluids}  \bvol{20}~(5),  \pg{051702}.

\bibitem[Fletcher(1984)]{Fletcher1984book}
{\sc \au{Fletcher, C.~A.}} \yr{1984} {\em Computational Galerkin Methods\/},
  1st edn.  \publ{New York: Springer}.

\bibitem[Gomez {\em et~al.\/}(2016)Gomez, Blackburn, Rudman, Sharma \&
  McKeon]{Gomez2016jfm}
{\sc \au{Gomez, F.}, \au{Blackburn, H.~M.}, \au{Rudman, M.}, \au{Sharma, A.~S.}
  \& \au{McKeon, B.~J.}} \yr{2016}  \at{A reduced-order model of
  three-dimensional unsteady flow in a cavity based on the resolvent operator}.
   \jt{J.~Fluid Mech.}  \bvol{798},  \pg{R2--1..14}.

\bibitem[Gorban \& Karlin(2005)]{Gorban2005book}
{\sc \au{Gorban, A.~N.} \& \au{Karlin, I.~V.}} \yr{2005} {\em Invariant
  Manifolds for Physical and Chemical Kinetics\/}. {\em Lecture Notes in
  Physics\/} Vol.\ 660.  \publ{Berlin: Springer-Verlag}.

\bibitem[Grandemange {\em et~al.\/}(2012)Grandemange, Cadot \&
  Gohlke]{grandemange_PRE2012}
{\sc \au{Grandemange, M.}, \au{Cadot, O.} \& \au{Gohlke, M.}} \yr{2012}
  \at{Reflectional symmetry breaking of the separated flow over
  three-dimensional bluff bodies}.  \jt{Phys. Rev. E}  \bvol{86}~(3),
  \pg{035302}.

\bibitem[Grandemange {\em et~al.\/}(2013)Grandemange, Gohlke \&
  Cadot]{Grandemange2013jfm}
{\sc \au{Grandemange, M.}, \au{Gohlke, M.} \& \au{Cadot, O.}} \yr{2013}
  \at{Turbulent wake past a three-dimensional blunt body. part 1. global modes
  and bi-stability}.  \jt{J.~Fluid Mech.}  \bvol{722},  \pg{51--84}.

\bibitem[Grandemange {\em et~al.\/}(2014)Grandemange, Gohlke \&
  Cadot]{Grandemange_EXIF2014}
{\sc \au{Grandemange, M.}, \au{Gohlke, M.} \& \au{Cadot, O.}} \yr{2014}
  \at{Statistical axisymmetry of the turbulent sphere wake}.  \jt{Exp.~Fluids}
  \bvol{55}~(11),  \pg{1838}.

\bibitem[Gumowski {\em et~al.\/}(2008)Gumowski, Miedzik, Goujon-Durand, Jenffer
  \& Wesfreid]{gumowski_PRE2008}
{\sc \au{Gumowski, K.}, \au{Miedzik, J.}, \au{Goujon-Durand, S.}, \au{Jenffer,
  P.} \& \au{Wesfreid, J.~E.}} \yr{2008}  \at{Transition to a time-dependent
  state of fluid flow in the wake of a sphere}.  \jt{Phys. Rev. E}
  \bvol{77}~(5),  \pg{055308}.

\bibitem[Hopf(1948)]{Hopf1948}
{\sc \au{Hopf, E.}} \yr{1948}  \at{A mathematical example displaying features
  of turbulence}.  \jt{Commun. Pure Appl. Math.}  \bvol{1},  \pg{303--322}.

\bibitem[Ishar {\em et~al.\/}(2019)Ishar, Kaiser, Morzynski, Albers, Meysonnat,
  Schr\"{o}der \& Noack]{Ishar2019jfm}
{\sc \au{Ishar, R.}, \au{Kaiser, E.}, \au{Morzynski, M.}, \au{Albers, M.},
  \au{Meysonnat, P.}, \au{Schr\"{o}der, W.} \& \au{Noack, B.~R.}} \yr{2019}
  \at{Metric for attractor overlap}.  \jt{J.~Fluid Mech.}  \bvol{(in print)}.

\bibitem[Jim{\'e}nez(2018)]{jimenez2018jfm}
{\sc \au{Jim{\'e}nez, J.}} \yr{2018}  \at{Coherent structures in wall-bounded
  turbulence}.  \jt{J.~Fluid Mech.}  \bvol{842}.

\bibitem[Jordan \& Smith(1999)]{jordan1999nonlinear}
{\sc \au{Jordan, D.~W.} \& \au{Smith, P.}} \yr{1999} {\em Nonlinear ordinary
  differential equations: an introduction to dynamical systems\/}, ,
  \vol{vol.~2}.  \publ{Oxford University Press, USA}.

\bibitem[Lam \& Cheung(1988)]{lam1988phenomena}
{\sc \au{Lam, K.} \& \au{Cheung, W.~C.}} \yr{1988}  \at{Phenomena of vortex
  shedding and flow interference of three cylinders in different equilateral
  arrangements}.  \jt{J.~Fluid Mech.}  \bvol{196},  \pg{1--26}.

\bibitem[Landau(1944)]{Landau1944}
{\sc \au{Landau, L.~D.}} \yr{1944}  \at{On the problem of turbulence}.
  \jt{C.R. Acad. Sci. USSR}  \bvol{44},  \pg{311--314}.

\bibitem[Landau \& Lifshitz(1987)]{Landau1987book}
{\sc \au{Landau, L.~D.} \& \au{Lifshitz, E.~M.}} \yr{1987} {\em Fluid
  Mechanics\/}, 2nd edn. {\em Course of Theoretical Physics\/} Vol. 6.
  \publ{Oxford: Pergamon Press}.

\bibitem[Loiseau {\em et~al.\/}(2018)Loiseau, Noack \& Brunton]{Loiseau2018jfm}
{\sc \au{Loiseau, J.~C.}, \au{Noack, B.~R.} \& \au{Brunton, S.~L.}} \yr{2018}
  \at{Sparse reduced-order modeling: Sensor-based dynamics to full-state
  estimation}.  \jt{J.~Fluid Mech.}  \bvol{844},  \pg{459--490}.

\bibitem[Luchtenburg {\em et~al.\/}(2009)Luchtenburg, G\"{u}nter, Noack, King
  \& Tadmor]{Luchtenburg2009jfm}
{\sc \au{Luchtenburg, D.~M.}, \au{G\"{u}nter, B.}, \au{Noack, B.~R.}, \au{King,
  R.} \& \au{Tadmor, G.}} \yr{2009}  \at{A generalized mean-field model of the
  natural and actuated flows around a high-lift configuration}.  \jt{J.~Fluid
  Mech.}  \bvol{623},  \pg{283--316}.

\bibitem[Malkus(1956)]{Malkus1956jfm}
{\sc \au{Malkus, W.~V.~R.}} \yr{1956}  \at{Outline of a theory of turbulent
  shear flow}.  \jt{J.~Fluid Mech.}  \bvol{1},  \pg{521--539}.

\bibitem[Manneville(2010)]{Manneville2010book}
{\sc \au{Manneville, P.}} \yr{2010} {\em Instabilities, chaos and
  turbulence\/}, ,  \vol{vol.~1}.  \publ{World Scientific}.

\bibitem[Meliga {\em et~al.\/}(2009)Meliga, Chomaz \& Sipp]{Meliga2009jfm}
{\sc \au{Meliga, P.}, \au{Chomaz, J.-M.} \& \au{Sipp, D.}} \yr{2009}
  \at{Global mode interaction and pattern selection in the wake of a disk: a
  weakly nonlinear expansion}.  \jt{J.~Fluid Mech.}  \bvol{633},
  \pg{159--189}.

\bibitem[Mittal(1999)]{mittal_AIAA1999}
{\sc \au{Mittal, R.}} \yr{1999}  \at{Planar symmetry in the unsteady wake of a
  sphere}.  \jt{AIAA J.}  \bvol{37}~(3),  \pg{388--390}.

\bibitem[Morzynski {\em et~al.\/}(1999)Morzynski, Afanasiev \&
  Thiele]{MORZYNSKI1999161}
{\sc \au{Morzynski, M.}, \au{Afanasiev, K.} \& \au{Thiele, F.}} \yr{1999}
  \at{Solution of the eigenvalue problems resulting from global non-parallel
  flow stability analysis}.  \jt{Comput. Methods. Appl. Mech. Engrg}
  \bvol{169}~(1),  \pg{161 -- 176}.

\bibitem[Newhouse {\em et~al.\/}(1978)Newhouse, Ruelle \& Takens]{newhouse1978}
{\sc \au{Newhouse, S.}, \au{Ruelle, D.} \& \au{Takens, F.}} \yr{1978}
  \at{Occurrence of strange axiom {A} attractors near quasi periodic flows on
  ${T}^m$, m $\geq$ 3}.  \jt{Commun. Math. Phys.}  \bvol{64}~(1),  \pg{35--40}.

\bibitem[Noack(2016)]{Noack2016jfm2}
{\sc \au{Noack, B.~R.}} \yr{2016}  \at{From snapshots to modal expansions --
  bridging low residuals and pure frequencies}.  \jt{J.~Fluid Mech.}
  \bvol{802},  \pg{1--4}.

\bibitem[Noack {\em et~al.\/}(2003)Noack, Afanasiev, Morzy\'nski, Tadmor \&
  Thiele]{Noack2003jfm}
{\sc \au{Noack, B.~R.}, \au{Afanasiev, K.}, \au{Morzy\'nski, M.}, \au{Tadmor,
  G.} \& \au{Thiele, F.}} \yr{2003}  \at{A hierarchy of low-dimensional models
  for the transient and post-transient cylinder wake}.  \jt{J.~Fluid Mech.}
  \bvol{497},  \pg{335--363}.

\bibitem[Noack \& Eckelmann(1994{\natexlab{{\em a\/}}})]{Noack1994jfm}
{\sc \au{Noack, B.~R.} \& \au{Eckelmann, H.}} \yr{1994{\natexlab{{\em a\/}}}}
  \at{A global stability analysis of the steady and periodic cylinder wake}.
  \jt{J.~Fluid Mech.}  \bvol{270},  \pg{297--330}.

\bibitem[Noack \& Eckelmann(1994{\natexlab{{\em b\/}}})]{Noack1994zamm}
{\sc \au{Noack, B.~R.} \& \au{Eckelmann, H.}} \yr{1994{\natexlab{{\em b\/}}}}
  \at{Theoretical investigation of the bifurcations and the turbulence
  attractor of the cylinder wake}.  \jt{Z.\ angew.\ Math.\ Mech.}
  \bvol{74}~(5),  \pg{T396--T397}.

\bibitem[Noack \& Morzy\'nski(2017)]{Noack2017put}
{\sc \au{Noack, B.~R.} \& \au{Morzy\'nski, M.}} \yr{2017}  \bt{The fluidic
  pinball --- a toolkit for multiple-input multiple-output flow control
  (version 1.0)}. {\em Tech. Rep.\/} 02/2017.  \org{Chair of Virtual
  Engineering, Poznan University of Technology, Poland}.

\bibitem[Noack {\em et~al.\/}(2008)Noack, Schlegel, Ahlborn, Mutschke,
  Morzy\'nski, Comte \& Tadmor]{Noack2008jnet}
{\sc \au{Noack, B.~R.}, \au{Schlegel, M.}, \au{Ahlborn, B.}, \au{Mutschke, G.},
  \au{Morzy\'nski, M.}, \au{Comte, P.} \& \au{Tadmor, G.}} \yr{2008}  \at{A
  finite-time thermodynamics of unsteady fluid flows}.  \jt{J. Non-Equilibr.
  Thermodyn.}  \bvol{33},  \pg{103--148}.

\bibitem[Noack {\em et~al.\/}(2016)Noack, Stankiewicz, Morzyński \&
  Schmid]{Noack2016jfm}
{\sc \au{Noack, B.~R.}, \au{Stankiewicz, W.}, \au{Morzyński, M.} \&
  \au{Schmid, P.~J.}} \yr{2016}  \at{Recursive dynamic mode decomposition of
  transient and post-transient wake flows}.  \jt{J.~Fluid Mech.}  \bvol{809},
  \pg{843--872}.

\bibitem[Price \& Paidoussis(1984)]{price1984aerodynamic}
{\sc \au{Price, S.~J.} \& \au{Paidoussis, M.~P.}} \yr{1984}  \at{The
  aerodynamic forces acting on groups of two and three circular cylinders when
  subject to a cross-flow}.  \jt{J.~Wind Eng. Indust. Aero.}  \bvol{17}~(3),
  \pg{329--347}.

\bibitem[Rempfer(1994)]{Rempfer1994pf}
{\sc \au{Rempfer, D.}} \yr{1994}  \at{On the structure of dynamical systems
  describing the evolution of coherent structures in a convective boundary
  layer}.  \jt{Phys. Fluids}  \bvol{6}~(3),  \pg{1402--4}.

\bibitem[Rempfer \& Fasel(1994)]{Rempfer1994jfm}
{\sc \au{Rempfer, D.} \& \au{Fasel, H.~F.}} \yr{1994}  \at{Evolution of
  three-dimensional coherent structures in a flat-plate boundary-layer}.
  \jt{J.~Fluid Mech.}  \bvol{260},  \pg{351--375}.

\bibitem[Reynolds \& Hussain(1972)]{Reynolds1972jfm}
{\sc \au{Reynolds, W.~C.} \& \au{Hussain, A.~K.~M.~F.}} \yr{1972}  \at{The
  mechanics of an organized wave in turbulent shear flow. {P}art 3.
  {T}heoretical model and comparisons with experiments.}  \jt{J.~Fluid Mech.}
  \bvol{54},  \pg{263--288}.

\bibitem[Rigas {\em et~al.\/}(2014)Rigas, Oxlade, Morgans \&
  Morrison]{rigas_JFM2014}
{\sc \au{Rigas, G.}, \au{Oxlade, A.~R.}, \au{Morgans, A.~S.} \& \au{Morrison,
  J.~F.}} \yr{2014}  \at{Low-dimensional dynamics of a turbulent axisymmetric
  wake}.  \jt{J.~Fluid Mech.}  \bvol{755}.

\bibitem[Rigas {\em et~al.\/}(2017)Rigas, Schmidt, Colonius \&
  Bres]{Rigas2017aiaa}
{\sc \au{Rigas, G.}, \au{Schmidt, O.~T.}, \au{Colonius, T.} \& \au{Bres,
  G.~A.}} \yr{2017} One-way navier-stokes and resolvent analysis for modeling
  coherent structures in a supersonic turbulent jet.  \bt{In {\em 23rd
  AIAA/CEAS Aeroacoustics Conference\/}},  \pg{p.~11}.  \publ{Reston, VA, USA:
  AIAA - American Institute of Aeronautics and Astronautics}, 23rd AIAA/CEAS
  Aeroacoustics Conference, Denver, CO, USA.

\bibitem[Roweis \& Saul(2000)]{Roweis2000s}
{\sc \au{Roweis, S.~T.} \& \au{Saul, L.~K.}} \yr{2000}  \at{Nonlinear
  dimensionality reduction by locally linear embedding}.  \jt{Science}
  \bvol{290}~(5500),  \pg{2323--2326}.

\bibitem[Rowley \& Dawson(2017)]{rowley2017arfm}
{\sc \au{Rowley, C.~W.} \& \au{Dawson, S.~T.}} \yr{2017}  \at{Model reduction
  for flow analysis and control}.  \jt{Ann. Rev. Fluid Mech.}  \bvol{49},
  \pg{387--417}.

\bibitem[Rowley {\em et~al.\/}(2009)Rowley, Mezi\'c, Bagheri, Schlatter \&
  Henningson]{Rowley2009jfm}
{\sc \au{Rowley, C.~W.}, \au{Mezi\'c, I.}, \au{Bagheri, S.}, \au{Schlatter, P.}
  \& \au{Henningson, D.~S.}} \yr{2009}  \at{Spectral analysis of nonlinear
  flows}.  \jt{J.~Fluid Mech.}  \bvol{645},  \pg{115--127}.

\bibitem[Ruelle \& Takens(1971)]{Ruelle1971nature}
{\sc \au{Ruelle, D.} \& \au{Takens, F.}} \yr{1971}  \at{On the nature of
  turbulence}.  \jt{Les rencontres physiciens-math{\'e}maticiens de
  Strasbourg-RCP25}  \bvol{12},  \pg{1--44}.

\bibitem[Sayers(1987)]{sayers1987flow}
{\sc \au{Sayers, A.~T.}} \yr{1987}  \at{Flow interference between three
  equispaced cylinders when subjected to a cross flow}.  \jt{J.~Wind Eng.
  Indust. Aero.}  \bvol{26}~(1),  \pg{1--19}.

\bibitem[Schatz {\em et~al.\/}(1995)Schatz, Barkley \&
  Swinney]{schatz1995instability}
{\sc \au{Schatz, M.~F.}, \au{Barkley, D.} \& \au{Swinney, H.~L.}} \yr{1995}
  \at{Instability in a spatially periodic open flow}.  \jt{Phys. Fluids}
  \bvol{7}~(2),  \pg{344--358}.

\bibitem[Schewe(1983)]{Schewe1983jfm}
{\sc \au{Schewe, G.}} \yr{1983}  \at{On the force fluctuations acting on a
  circular cylinder in crossflow from subcritical up to transcritical
  {R}eynolds numbers}.  \jt{J.~Fluid Mech.}  \bvol{133},  \pg{265--285}.

\bibitem[Schlegel \& Noack(2015)]{schlegel2015jfm}
{\sc \au{Schlegel, M.} \& \au{Noack, B.~R.}} \yr{2015}  \at{On long-term
  boundedness of galerkin models}.  \jt{J.~Fluid Mech.}  \bvol{765},
  \pg{325--352}.

\bibitem[Schmid(2010)]{Schmid2010jfm}
{\sc \au{Schmid, P.~J.}} \yr{2010}  \at{Dynamic mode decomposition for
  numerical and experimental data}.  \jt{J.~Fluid Mech.}  \bvol{656},
  \pg{5--28}.

\bibitem[Schumm {\em et~al.\/}(1994)Schumm, Berger \& Monkewitz]{Schumm1994jfm}
{\sc \au{Schumm, M.}, \au{Berger, E.} \& \au{Monkewitz, P.~A.}} \yr{1994}
  \at{Self-excited oscillations in the wake of two-dimensional bluff bodies and
  their control}.  \jt{J.~Fluid Mech.}  \bvol{271},  \pg{17--53}.

\bibitem[Sipp \& Lebedev(2007)]{Sipp2007jfm}
{\sc \au{Sipp, D.} \& \au{Lebedev, A.}} \yr{2007}  \at{Global stability of base
  and mean flows: a general approach and its applications to cylinder and open
  cavity flows}.  \jt{J.~Fluid Mech.}  \bvol{593}~(1),  \pg{333--358}.

\bibitem[Strogatz {\em et~al.\/}(1994)Strogatz, Friedman, Mallinckrodt \&
  McKay]{strogatz1994book}
{\sc \au{Strogatz, S.}, \au{Friedman, M.}, \au{Mallinckrodt, A.~J.} \&
  \au{McKay, S.}} \yr{1994}  \at{Nonlinear dynamics and chaos: with
  applications to physics, biology, chemistry, and engineering}.  \jt{Computers
  Phys.}  \bvol{8}~(5),  \pg{532--532}.

\bibitem[Strykowski \& Sreenivasan(1990)]{Strykowski1990jfm}
{\sc \au{Strykowski, P.~J.} \& \au{Sreenivasan, K.~R.}} \yr{1990}  \at{On the
  formation and suppression of vortex `shedding' at low reynolds numbers}.
  \jt{J.~Fluid Mech.}  \bvol{218},  \pg{71--107}.

\bibitem[Stuart(1958)]{Stuart1958jfm}
{\sc \au{Stuart, J.~T.}} \yr{1958}  \at{On the non-linear mechanics of
  hydrodynamic stability}.  \jt{J.~Fluid Mech.}  \bvol{4},  \pg{1--21}.

\bibitem[Swift \& Hohenberg(1977)]{Swift1977pra}
{\sc \au{Swift, J.} \& \au{Hohenberg, P.~C.}} \yr{1977}  \at{Hydrodynamic
  fluctuations at the convective instability}.  \jt{Phys. Rev. A}
  \bvol{15}~(1),  \pg{319}.

\bibitem[Szaltys {\em et~al.\/}(2012)Szaltys, Chrust, Przadka, Goujon-Durand,
  Tuckerman \& Wesfreid]{szaltys_JFS2012}
{\sc \au{Szaltys, P.}, \au{Chrust, M.}, \au{Przadka, A.}, \au{Goujon-Durand,
  S.}, \au{Tuckerman, L.~S.} \& \au{Wesfreid, J.~E.}} \yr{2012}  \at{Nonlinear
  evolution of instabilities behind spheres and disks}.  \jt{J.~Fluids~Struct.}
   \bvol{28},  \pg{483--487}.

\bibitem[Tadmor {\em et~al.\/}(2011)Tadmor, Lehmann, Noack, Cordier, Delville,
  Bonnet \& Morzy\'nski]{Tadmor2011ptrsa}
{\sc \au{Tadmor, G.}, \au{Lehmann, O.}, \au{Noack, B.~R.}, \au{Cordier, L.},
  \au{Delville, J.}, \au{Bonnet, J.-P.} \& \au{Morzy\'nski, M.}} \yr{2011}
  \at{Reduced order models for closed-loop wake control}.  \jt{Philos. Trans.
  R. S. A}  \bvol{369}~(1940),  \pg{1513--1524}.

\bibitem[Taira {\em et~al.\/}(2017)Taira, Brunton, Dawson, Rowley, Colonius,
  McKeon, Schmidt, Gordeyev, Theofilis \& Ukeiley]{taira_AIAA2017}
{\sc \au{Taira, K.}, \au{Brunton, S.~L.}, \au{Dawson, S.~T.}, \au{Rowley,
  C.~W.}, \au{Colonius, T.}, \au{McKeon, B.~J.}, \au{Schmidt, O.~T.},
  \au{Gordeyev, S.}, \au{Theofilis, V.} \& \au{Ukeiley, L.~S.}} \yr{2017}
  \at{Modal analysis of fluid flows: An overview}.  \jt{AIAA J.}  \pg{pp.
  4013--4041}.

\bibitem[Tatsuno {\em et~al.\/}(1998)Tatsuno, Amamoto \&
  Ishi-i]{tatsuno1998effects}
{\sc \au{Tatsuno, M.}, \au{Amamoto, H.} \& \au{Ishi-i, K.}} \yr{1998}
  \at{Effects of interference among three equidistantly arranged cylinders in a
  uniform flow}.  \jt{Fluid Dyn. Res.}  \bvol{22}~(5),  \pg{297}.

\bibitem[Taylor \& Hood(1973)]{Taylor1973ef}
{\sc \au{Taylor, C.} \& \au{Hood, P.}} \yr{1973}  \at{A numerical solution of
  the navier-stokes equations using the finite element technique}.  \jt{Comput.
  Fluids}  \bvol{1},  \pg{73--100}.

\bibitem[Turton {\em et~al.\/}(2015)Turton, Tuckerman \&
  Barkley]{turton2015pre}
{\sc \au{Turton, S.~E.}, \au{Tuckerman, L.~S.} \& \au{Barkley, D.}} \yr{2015}
  \at{Prediction of frequencies in thermosolutal convection from mean flows}.
  \jt{Phys. Rev. E}  \bvol{91}~(4),  \pg{043009}.

\bibitem[Watson(1960)]{Watson1960jfm}
{\sc \au{Watson, J.}} \yr{1960}  \at{On the non-linear mechanics of wave
  disturbances in stable and unstable parallel flows. {P}art 2. the development
  of a solution for plane {P}oiseuille flow and for plane {C}ouette flow}.
  \jt{J.~Fluid Mech.}  \bvol{9},  \pg{371--389}.

\bibitem[Zaitsev \& Shliomis(1971)]{Zaitsev1971}
{\sc \au{Zaitsev, V.~M.} \& \au{Shliomis, M.~I.}} \yr{1971}  \at{Hydrodynamic
  fluctuations near convection threshold}.  \jt{Sov. Phys. JETP}  \bvol{32},
  \pg{866}.

\bibitem[Zhang {\em et~al.\/}(1994)Zhang, Noack \& Eckelmann]{Zhang1994mpisf}
{\sc \au{Zhang, H.-Q.}, \au{Noack, B.~R.} \& \au{Eckelmann, H.}} \yr{1994}
  \bt{Numerical computation of the 3-d cylinder wake}. {\em Tech. Rep.\/}
  3/1994.  \org{Max-Planck-Institut f\"{u}r Str\"{o}mungsforschung,
  G\"{o}ttingen, Germany}.

\end{thebibliography}

\end{document}